\documentclass[a4paper,12pt]{article}

\usepackage{jheppub}
\usepackage[english]{babel}
\pdfoutput=1
\usepackage[utf8]{inputenc}
\usepackage[T1]{fontenc}
\usepackage{amsmath}
\usepackage{float}
\usepackage{graphicx}
\usepackage{comment}
\usepackage[dvipsnames]{xcolor}
\usepackage{color}
\usepackage{epsfig}
\usepackage{dcolumn}
\usepackage{hyperref}
\usepackage{bm}
\usepackage{amssymb}
\usepackage{slashed}
\usepackage[caption=false]{subfig}
\usepackage{epstopdf}
\usepackage{setspace}
\usepackage[normalem]{ulem}	
\usepackage{multicol}
\usepackage{mathrsfs}
\usepackage{fancybox}
\usepackage{cancel}
\usepackage{empheq}
\pdfoutput=1

\newcommand*\widefbox[1]{\fbox{\hspace{2em}#1\hspace{2em}}}

\usepackage{tikz,pgfplots}
\usetikzlibrary{arrows,shapes,trees,calc,positioning,patterns}

\definecolor{corlinks}{RGB}{0,0,128}
\definecolor{cormenu}{RGB}{0,0,128}
\definecolor{corurl}{RGB}{0,0,128}

\definecolor{colRed0}{rgb}{0.85, 0.05, 0.12}
\definecolor{colRed1}{rgb}{0.92, 0.1, 0.05}
\definecolor{colRed2}{rgb}{0.95, 0.35, 0.05}
\definecolor{colYellow1}{rgb}{1., 0.82, 0.}
\definecolor{colBlue1}{rgb}{0.0, 0., 0.4}
\definecolor{colBlue2}{rgb}{0.1, 0.3, 0.9}
\definecolor{colBlue3}{rgb}{0.15, 0.4, 0.75}
\definecolor{colBlue4}{rgb}{0.3, 0.8, 0.93}
\definecolor{colGreen0}{rgb}{0.0, 0.15, 0.05}
\definecolor{colGreen1}{rgb}{0.0, 0.35, 0.1}
\definecolor{colGreen2}{rgb}{0.1, 0.65, 0.2}
\definecolor{colGreen3}{rgb}{0.3, 0.85, 0.5}
\definecolor{colBrown1}{rgb}{0.3, 0.18, 0.12}
\definecolor{colBrown2}{rgb}{0.5, 0.3, 0.20}
\definecolor{colViolet1}{rgb}{0.4, 0.18, 0.42}
\definecolor{colViolet2}{rgb}{0.5, 0.3, 0.70}

\hypersetup{
colorlinks=true,
urlcolor=corlinks,
linkcolor=corlinks,
menucolor=cormenu,
citecolor=corlinks,
pdfborder= 0 0 0
}

\DeclareRobustCommand{\Eq}[1]{Eq.~(\ref{#1})}
\DeclareRobustCommand{\Fig}[1]{Fig.~\ref{#1}}

\DeclareRobustCommand{\Sec}[1]{Sec.~\ref{#1}}

\title{Axion Fragmentation}

\date{\today}

\author[a,b]{Nayara Fonseca,}
\author[a,c]{Enrico Morgante,}
\author[a]{Ryosuke Sato,}
\author[a,d]{G\'eraldine Servant}

\affiliation[a\,]{DESY, Notkestrasse 85, D-22607 Hamburg, Germany}
\affiliation[b\,]{Abdus Salam International Centre for Theoretical Physics, Strada Costiera 11, 34151, Trieste, Italy}
\affiliation[c\,]{PRISMA$^+$  Cluster  of  Excellence  and  Mainz  Institute  for  Theoretical  Physics, Johannes  Gutenberg-Universit\"at  Mainz,  D-55099  Mainz,  Germany}
\affiliation[d\,]{II. Institute of Theoretical Physics, Univ. Hamburg, D-22761 Hamburg, Germany}

\emailAdd{nfonseca@ictp.it}
\emailAdd{emorgant@uni-mainz.de}
\emailAdd{ryosuke.sato@desy.de}
\emailAdd{geraldine.servant@desy.de}

\abstract{
We investigate the production of axion quanta during the early universe evolution of an axion-like field rolling down a wiggly potential.
We compute the growth of quantum fluctuations and  their back-reaction on the homogeneous zero-mode. 
 We evaluate the transfer of kinetic energy from the zero mode to the quantum fluctuations and  the conditions to decelerate the axion zero-mode  as a function of the Hubble rate, the slope of the potential, the size of the barriers and the initial field velocity.
We discuss how these effects  impact the relaxion mechanism. 
}

\begin{document}

\begin{flushright} 
DESY 19-202 \\
MITP/19-079
\end{flushright}

\maketitle

\clearpage

\section{Introduction}

Axion-like particles (ALPs) are ubiquitous in a large number of high-energy completions of the Standard Model (SM). 
They are for instance generic predictions in the low-energy spectrum of string compactifications \cite{Svrcek:2006yi,Arvanitaki:2009fg}. 
ALPs denote any particles which are a pseudo-Nambu Goldstone Bosons enjoying  a discrete symmetry. They are light and have weak self-interactions.
Interestingly, they can be used to solve open problems of the SM as they are well-motivated candidates to explain \textit{e.g.} dark matter~\cite{Preskill:1982cy, Abbott:1982af, Dine:1982ah}, inflation~\cite{Pajer:2013fsa, Adshead:2015pva, Domcke:2019qmm, Adshead:2019lbr}, and baryogenesis~\cite{Servant:2014bla,Domcke:2019mnd}. They were originally introduced to solve the Strong CP problem  \cite{Wilczek:1977pj,Peccei:1977hh} and still remain the most popular explanation to this puzzle. Independently of their virtues in providing solutions to some major open problems in the SM, they are also interesting purely from their specific phenomenological properties, notably in cosmology. They can leave imprints in many different ways such as in the Cosmic Microwave Background, in Large Scale Structures, in Gravitational Waves, in stellar astrophysics and they can be searched for directly in dedicated laboratory experiments such as haloscopes (see e.g \cite{Marsh:2015xka, Irastorza:2018dyq} for recent reviews).

The role of axions in cosmology has been the subject of a large number of studies.
Specifically, zero-mode axion oscillations around the minimum of the axion potential can provide a large component of the energy density of the universe and mimic the effect of dark matter.
Surprisingly, the  effect of axion quantum  fluctuations in the early universe has been mostly overlooked so far in the literature where only the zero mode homogeneous mode has been considered. In this paper, we investigate in details  the effect of axion particle production during the evolution of the homogeneous zero-mode. 

The production of axion quanta when the zero-mode field oscillates around one minimum of the potential is generally suppressed unless the initial position of the field is extremely close to the maximum of the potential~\cite{Greene:1998pb}.%
\footnote{As this work was being completed, Ref.~\cite{Arvanitaki:2019rax} appeared, which considers the production of axion quantum fluctuations during oscillations about the minimum of the potential, as in~\cite{Greene:1998pb}.
This effect is important only if the initial position of the axion field is tuned very close to the top of the barrier of the axion potential, at the level of $\sim 10^{-7}$.  Such peculiar initial position was recently motivated by some inflation dynamics in \cite{Co:2018mho}  and by anthropic arguments in \cite{Freivogel:2008qc}.}
In this work, we are instead investigating the exponential production of axion quanta when the axion is rolling down its potential with a large velocity and the axion is crossing a large number of minima/maxima during its evolution. Such situation was somehow rarely investigated before, although it is quite generic. We are interested in the resulting friction force on the zero-mode. Our findings do not require any tuning of initial conditions.
Our only assumption is the existence of a small slope together with  some wiggles, and an initial kinetic energy larger than the barriers potential energy. 
In fact, this is precisely the type of potential introduced in the relaxion mechanism proposed in 
 \cite{Graham:2015cka}
 to resolve the electroweak scale hierarchy problem.
 This type of axion potential, with a linear term plus a cosine, was first considered in the context of string cosmology, in models of axion 
 monodromy~\cite{McAllister:2008hb, Flauger:2009ab}%
 \footnote{In  Ref.~\cite{McAllister:2008hb}, particle production was not alluded to. In the follow-up paper~\cite{Flauger:2009ab}, 
the axion is the inflaton 
and axion particle production was mentioned,
but in  the case of small wiggles,
which does not lead to exponential particle production.
Friction from backreaction on the zero mode was  therefore not discussed.
Later, Ref.~\cite{Jaeckel:2016qjp} studied a quadratic potential with wiggles.
The particle production, from the zero-mode oscillations around the global minimum with large amplitude, was discussed at the  linearized level.
However, the backreaction on the zero mode was not.
In a follow-up study \cite{Berges:2019dgr}, the particle production was discussed  by non-perturbative numerical analysis. The aim is not to exploit this effect as a stopping condition (the field eventually stops at a global minimum) but as a dark matter production mechanism. 
 It is there suggested that axions quanta produced during the rolling stage could be (warm) dark matter candidates.}.

In this paper, we show that the  production of  axion particles generates a friction that decelerates the rolling of the field.
This occurs if the field velocity is large enough to overcome the sinusoidal term and the field goes over a large number of wiggles.
As we will see, the equation of motion for the axion fluctuation can be described by the Mathieu equation and parametric resonance gives an exponential production of the particles with specific wave numbers.
We denote this phenomenon as \textit{axion fragmentation.}
This provides a novel mechanism to stop axion rolling.
We focus our attention on the most dramatic effect of fragmentation, \textit{i.e.}, the regime in which the axion stops its motion by transferring all of its kinetic energy to the fluctuations. Other effects can arise from fragmentation. For example, in situation in which a field rolls down a steep potential, fragmentation can be a way of generating a slow-roll regime not sustained by the sole Hubble friction. At the same time, the effect discussed in this paper is mathematically very similar to the amplification of fluctuations in oscillons~\cite{ Hertzberg:2010yz, Amin:2010xe, Amin:2010dc, Antusch:2017flz, Olle:2019kbo}%
\footnote{A recent analysis \cite{Olle:2019kbo}, although not related to axions, suggests that quanta produced by oscillons could be the dark matter candidates.}. 

The phenomenon of scalar field fragmentation has been studied in the context of preheating \cite{Dolgov:1989us, Traschen:1990sw, Kofman:1994rk, Shtanov:1994ce, Kofman:1997yn}.
The excitation of gauge field quanta from an axion has been studied extensively in different contexts, mainly in axion inflation models where the axion has a coupling to gauge fields, see \textit{e.g.}~\cite{Pajer:2013fsa,Domcke:2019qmm}.
 We provide a model-independent detailed analytical treatment of axion fragmentation that can be applied to various setups.
We discuss the precise conditions for fragmentation to stop the field (even far away from the global minimum), which we check over a numerical analysis. 
We discuss the implications of these findings for the relaxation mechanisms of the electroweak scale.
In this case, the {\it relaxion} scans the Higgs mass-squared term in the early universe,
and dynamically realizes an electroweak scale which is suppressed compared to the cutoff scale.
A key ingredient of this scenario is the friction that slows down the relaxion field rolling.
In the original paper~\cite{Graham:2015cka}, Hubble friction is responsible for the slow roll of the relaxion.
Alternatively, a coupling to SM gauge bosons $\phi F_{\mu\nu} \tilde F^{\mu\nu}$ can provide the necessary friction, through tachyonic  particle production~\cite{Hook:2016mqo, Choi:2016kke, Tangarife:2017rgl, Matsedonskyi:2017rkq, Fonseca:2018xzp, Fonseca:2018kqf}. In~\cite{Ibe:2019udh} the necessary friction is provided by parametric resonance of the Higgs zero mode. In~\cite{Wang:2018ddr} the field is stopped thanks to a potential instability. Finally, in~\cite{Kadota:2019wyz} the relaxion is slowed down via the production of dark fermions.
So far, the excitation of axion particle themselves, was not considered, although they are present in the most minimal models where the relaxion has no extra couplings to gauge fields.
We analyse in details how this can be used as a stopping mechanism for the relaxion in our companion paper~\cite{Fonseca:2019lmc}, the results of which are summarised here.

The paper is organized as follows.
In section \ref{sec:intuitive derivation},
we describe our setup and derive the equation of motion  of the fluctuations at the leading order.
We also give an intuitive discussion on the particle production effect.
Then, we derive semi-analytic formula in section \ref{sec:analytical discussion},
and show numerical results in section \ref{sec:numeircal analysis}.
We discuss the analysis beyond the leading order in section \ref{sec:beyond linear}.
In Sec.~\ref{sec:Consequences} we discuss how our results apply to the relaxion mechanism.
Our conclusions are drawn in Sec.~\ref{sec:conclusions}. In the appendices \ref{sec:technical details} and \ref{sec:detail on phiddot} we provide further details on the derivation of our results.

\section{Axion fragmentation in a nutshell} \label{sec:intuitive derivation}
\begin{figure}[t]
\centering
\includegraphics[width=0.6\hsize]{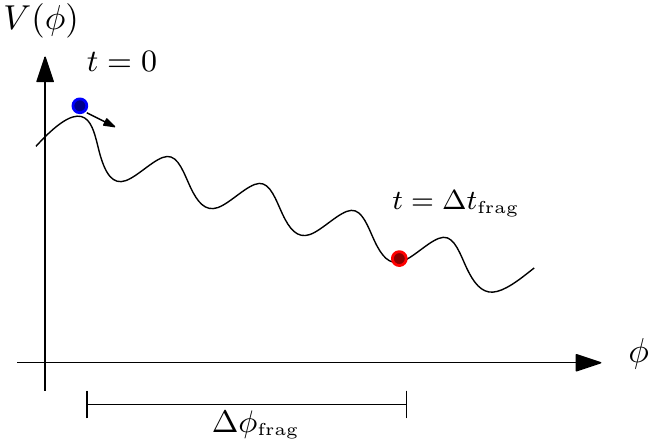}
\caption{
Sketch of the axion potential given in \Eq{eq:potential}.  Once \emph{axion fragmentation} starts, the field $\phi$ takes a time $\Delta t_{\rm frag}$ and moves a distance   $\Delta\phi_{\rm frag}$ until it gets trapped in one of the wiggles.
}\label{fig:potential}
\end{figure}

In this section we discuss how the axion field $\phi$ evolution is affected by  the axion fragmentation phenomenon.  
The dynamics of axion quantum fluctuation is  described by a Mathieu equation with time varying coefficients, from which we can estimate how the fluctuations back-react on the zero mode.  Our goal is to study the dynamics of axion particle production when the axion field is rolling down a wiggly potential.
For concreteness, we consider the following potential
\begin{align}
V(\phi) = \Lambda_b^4 \cos \frac{\phi}{f} - \mu^3 \phi, \label{eq:potential}
\end{align}
and we assume the height of the barrier $\Lambda_b$ as a constant for simplicity (see \Fig{fig:potential} for a sketch of the potential).
For simplicity, we do not include any cosmological constant term in Eq.~(\ref{eq:potential}), and we assume that the vacuum energy is cancelled in the late universe by some other mechanism, about which we remain agnostic.
We are interested in the case in which the barriers are large, \textit{i.e.}
\begin{equation} \label{eq:local minima}
\frac{\Lambda_b^4}{f} > \mu^3
\end{equation}
which corresponds to say that the potential has local minima. Additionally, we assume that the kinetic energy of $\phi$ is large enough to overcome the barriers, $\dot\phi>\Lambda_b^2$.

The equation of motion of the axion $\phi(x,t)$ is given by
\begin{align}
\ddot\phi + 3H\dot\phi - \frac{1}{a^2} \nabla^2\phi + V'(\phi)=0, \label{eq:EOM}
\end{align}
where $a$ is the scale factor of the Friedmann-Lema\^itre-Robertson-Walker metric and $H=\dot a/a$ is the Hubble expansion rate.
Let us decompose $\phi(x,t)$ into a classical homogeneous mode $\phi(t)$ and small fluctuations $\delta\phi$
(with no risk of confusion, we will denote the homogeneous mode as $\phi(t)$):
\begin{align}
\phi(x,t)
= \phi(t) + \delta\phi(x,t) 
= \phi(t) + \left( \int\frac{d^3 k}{(2\pi)^3} a_k u_{k}(t) e^{ikx} + h.c. \right), \label{eq:phi mode expansion}
\end{align}
with  $a_k$, $a_k^\dagger$ being  respectively the annihilation and creation operators which satisfy
\begin{align}
[a_k, a_{k'}^\dagger] = (2\pi)^3 \delta^{(3)}(k-k'),
\end{align}
and the initial condition of the mode function $u_k$ at $t \to -\infty$ is given by
\begin{align}
u_k(t) = \frac{e^{-i(k/a) t}}{a\sqrt{2k}}. \label{eq:initial}
\end{align}
In the analysis of this work, we treat $\delta\phi$ as small perturbation.
We discuss the validity of this approximation in Sec.~\ref{sec:beyond linear}.
By using this approximation, we expand the last term of the LHS of Eq.~(\ref{eq:EOM}) as $V'(\phi) + V''(\phi) \delta\phi + (1/2) V'''(\phi)\delta\phi^2 + \cdots$.
The third term of this expansion gives the dominant source of the backreaction to the zero mode from the particle production.
The equations of motion of $\phi$ and $\delta\phi$ are given by
\begin{align}
\ddot{\phi} + 3H\dot{\phi} + V'(\phi) + \frac{1}{2} V'''(\phi) \int \frac{d^3 x}{V_{\rm vol}} \langle \delta\phi(x)^2 \rangle &= 0, \label{eq:EOM1}\\
\ddot{\delta\phi} + 3H\dot{\delta\phi} + \left[ -\frac{1}{a^2} \nabla^2 + V''(\phi) \right] \delta\phi &= 0, \label{eq:EOM2}
\end{align}
where $\langle\dots\rangle$ indicates a quantum expectation value. Equations~(\ref{eq:EOM1}) and~(\ref{eq:EOM2}) can be rewritten in terms of the mode functions $u_k$ as
\begin{align}
\ddot{\phi} + 3H\dot{\phi} + V'(\phi) + \frac{1}{2}V'''(\phi) \int\frac{d^3 k}{(2\pi)^3} |u_k|^2 &=0, \label{eq:zeromode}\\
\ddot{u_k} + 3H\dot{u}_k + \left[ \frac{k^2}{a^2} + V''(\phi)\right] u_k &= 0. \label{eq:fluctuation}
\end{align}
Let us denote with $\dot\phi_0$ the initial velocity of the field $\phi$. We assume that $\dot{\phi}_0$ is large enough to overcome the barrier,  \textit{i.e.}, $$\dot\phi_0 \gtrsim \Lambda_b^2,$$ otherwise $\phi$ is trapped in the first valley.
This marks a crucial difference with the well-studied case of parametric resonance due to a scalar field which oscillates coherently at the minimum of its potential~\cite{Dolgov:1989us, Traschen:1990sw, Kofman:1994rk, Shtanov:1994ce, Kofman:1997yn, Greene:1998pb, Hertzberg:2010yz, Amin:2010xe, Amin:2010dc, Antusch:2017flz, Olle:2019kbo}.
Our case of study is sketched in \Fig{fig:potential}. The field $\phi$ rolls over many wiggles until it gets trapped.

\begin{figure}[t]\label{fig:Mathieu instability}
\centering
\includegraphics[width=.6\textwidth]{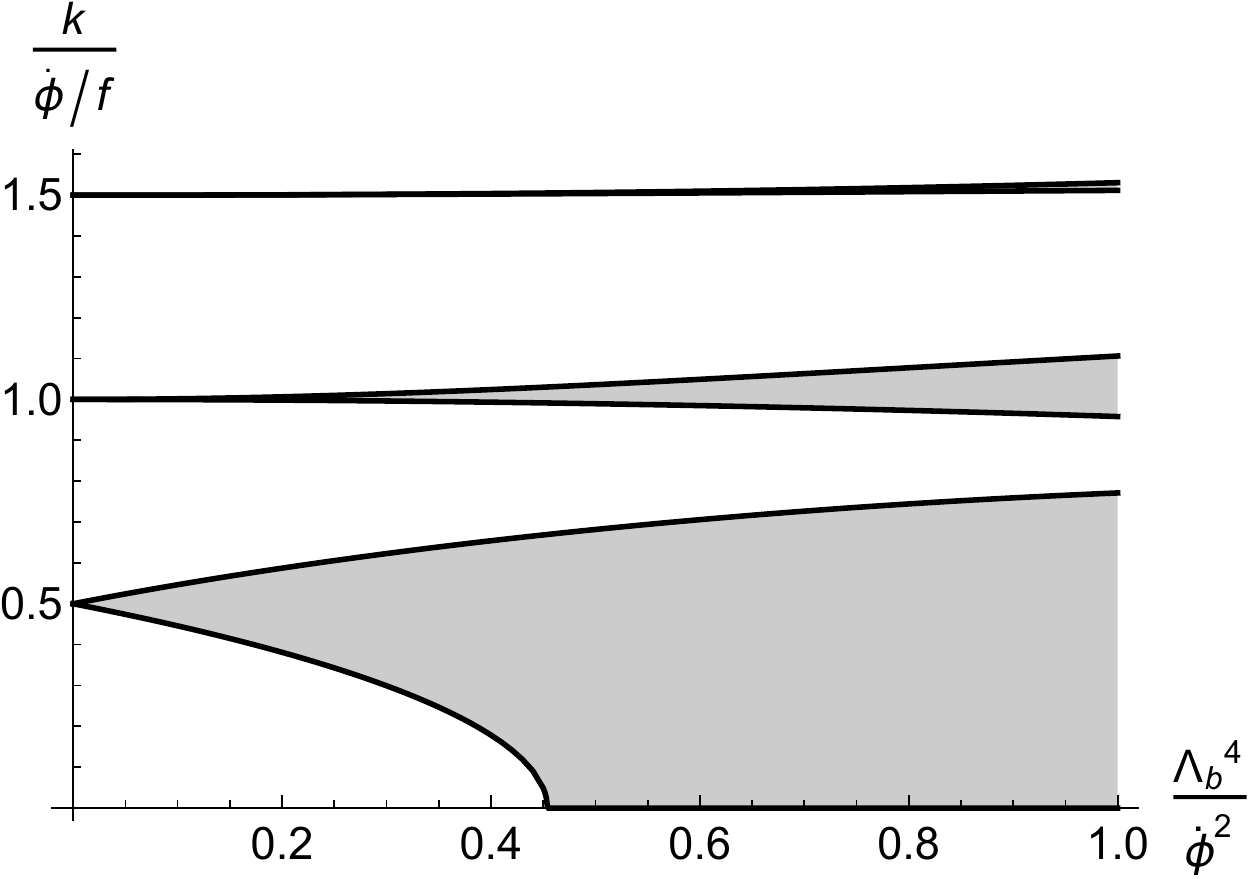}
\caption{In grey, the instability bands of \Eq{eq:Mathieu uk constant phidot}. Inside those bands, the solution grows exponentially.
For the calculation of the boundaries of the instability bands, see \textit{e.g.}, Ref.~\cite{kovacic2018mathieu}.
}
\end{figure}

To get insights about the axion particle production, let us estimate the effect of  friction.
In the limit of constant $\dot\phi$ and $H=0$, Eq.~(\ref{eq:fluctuation}) is simplified to the Mathieu equation \cite{mclachlan}:
\begin{align}\label{eq:Mathieu uk constant phidot}
\ddot u_k + \left( k^2 - \frac{\Lambda_b^4}{f^2} \cos \frac{\dot\phi}{f}t \right) u_k = 0.
\end{align}
Solutions to the Mathieu equation have instability and grow exponentially if the parameters are in some specific regions.%
\footnote{See \textit{e.g.}, figure 8 (A) of \cite{mclachlan} and section~IV of \cite{Kofman:1997yn}}
For $\dot\phi^2_0 > \Lambda_b^4$, the instability region presents a band structure, as it is shown in Fig.~\ref{fig:Mathieu instability}. 
In the limit $\dot\phi^2_0 \gg \Lambda_b^4$, the solution has an instability if the momentum $k$ is close to $n \dot\phi_0/2f$ with integer $n\geq 1$.
For $n\geq 2$, the speed of the growth is slow and the size of the instability band is small.
Hence, the instability band with $n=1$ gives the most important source of the friction to decelerate the axion rolling, which is given by
\begin{align}
\frac{\dot\phi^2}{4f^2} - \frac{\Lambda_b^4}{2f^2}
< k^2 < 
\frac{\dot\phi^2}{4f^2} + \frac{\Lambda_b^4}{2f^2}. \label{eq:instability band}
\end{align}
Equivalently,  one can write the instability band as $|k-k_{\rm cr}| < \delta k_{\rm cr}$ for $\dot\phi^2 \gg \Lambda_b^4$, where $k_{\rm cr}$ and $\delta k_{\rm cr}$ are defined as
\begin{align}\label{eq:k crit}
k_{\rm cr} = \frac{\dot\phi}{2f}, \quad
\delta k_{\rm cr} = \frac{\Lambda_b^4}{2f\dot\phi}\,,
\end{align}
where, initially, $\dot\phi=\dot\phi_0$. Inside the instability band, the asymptotic behavior of $u_k$ at large $t$ is given by
\begin{align}
u_k \propto \exp\left( \sqrt{ (\delta k_{\rm cr} )^2 - \left( k- k_{\rm cr}  \right)^2 } t \right) \sin \left(k_{\rm cr} t + \arctan\sqrt{ \frac{\delta k_{\rm cr} + (k - k_{\rm cr} ) }{ \delta k_{\rm cr} - (k - k_{\rm cr} )  } } \right). \label{eq:asymptotic uk}
\end{align}

Let us now estimate the energy of the growing modes.
The number of modes which exponentially grow per unit volume is $\sim (k_{\rm cr}^0)^2\, \delta k_{\rm cr}^0$.
The energy density of the fluctuations is
\begin{equation}
\delta\rho = \frac{1}{2} \langle(\dot{\delta\phi})^2\rangle + \frac{1}{2} \langle(\vec\nabla \,\delta\phi)^2\rangle \,.
\end{equation}
As long as $\dot\phi$ is constant, this grows as
\begin{align} \label{eq:rho fluctuations}
\rho_{\rm fluc}(t) \sim (k_{\rm cr}^0)^3 \,\delta k_{\rm cr}^0 \exp(2 \, \delta k_{\rm cr}^0 \, t ).
\end{align}
\begin{figure}[t]
\centering
\includegraphics[width=0.9\hsize]{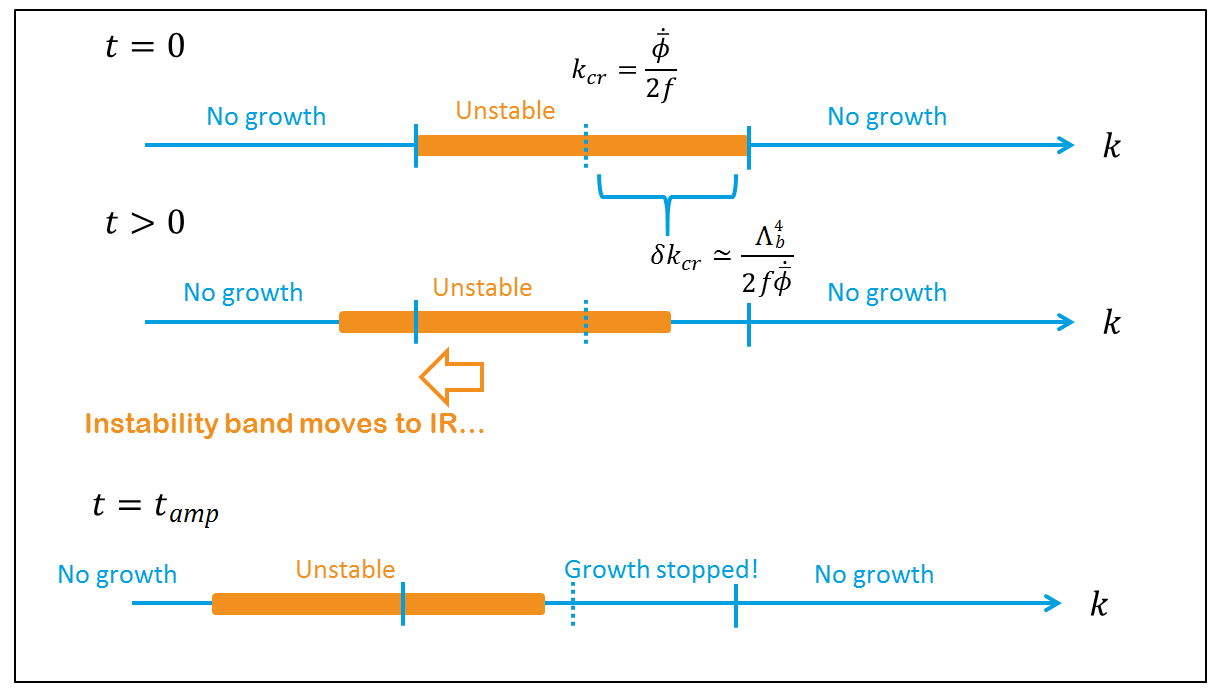}
\caption{Sketch of different time frames showing  how the instability band moves due to  particle production. For $t=0$, $|k-k_{\rm cr}| \lesssim \delta k_{\rm cr}$ is satisfied so there is an exponential growth of the fluctuations for the mode $k_{\textrm{cr}}$. Subsequently, the zero mode  looses  kinetic energy and the instability band starts moving towards smaller values of momentum. The exponential growth of  mode with $k_{\textrm{cr}}$  stops when it moves beyond the instability band.
}\label{fig:sketch}
\end{figure}
The homogeneous mode gradually looses its kinetic energy because of back-reaction,
and the instability band moves towards the region of small values of $k$ (see \Fig{fig:sketch}).
The exponential growth of the modes with the wave number $k_{\rm cr}^0$ stops when this mode goes out from the instability band.
At that time, the critical wave number has changed by $\delta k_{\rm cr}^0$. Using the definition of $k_{\rm cr}$, the kinetic energy of the zero mode decreases by
\begin{equation}
\delta K \approx \frac{1}{2}4 f^2 [(k_{\rm cr}^0-\delta k_{\rm cr}^0)^2 - (k_{\rm cr}^0)^2] \approx - 4 f^2 k_{\rm cr}^0 \delta k_{\rm cr}^0 = - \dot\phi_0^2 \frac{\delta k_{\rm cr}^0}{k_{\rm cr}^0}\, ,
\end{equation}
with the new $\dot\phi$  given by
\begin{equation} \label{eq:phidot variation}
\dot\phi \sim \dot\phi_0 \left(1-\frac{\delta k_{\textrm{cr}}^0}{k_{\textrm{cr}}^0} \right) \,. 
\end{equation}
The energy density of the fluctuations $\rho_{\rm fluc}$ is
\begin{align}\label{eq:rhofluc estimate}
\rho_{\rm fluc} \approx - \delta K \approx \dot\phi_0^2 \times \frac{\delta k_{\rm cr}^0}{k_{\rm cr}^0}\,.
\end{align}
The timescale that the mode with wave number $k_{\rm cr}^0$ spends inside the instability band can be estimated combining Eqs.~(\ref{eq:rhofluc estimate}) and~(\ref{eq:rho fluctuations}):
\begin{align} \label{eq:deltatime ampl}
\delta t_{\rm amp} = \frac{1}{2\,\delta k_{\rm cr}^0} \log \frac{\dot\phi_0^2}{(k_{\rm cr}^0)^4}.
\end{align}
The time evolution of kinetic energy $d\rho/dt$ is roughly given as $\sim \delta K / \delta t_{\rm amp} $:
\begin{align}\label{eq:phi double dot approx}
\frac{d}{dt} \left(\frac{1}{2}\dot\phi^2\right)
\sim -\dot\phi^2 \frac{2(\delta k_{\rm cr})^2}{k_{\rm cr}} \left( \log\frac{\dot\phi^2}{k_{\rm cr}^4} \right)^{-1}
\sim -\frac{\Lambda_b^8}{f \dot\phi} \left( \log\frac{16f^4}{\dot\phi^2} \right)^{-1} \,,  
\end{align}
where we have dropped the subscript $0$ because this equation of motion is now valid at any velocity. Eq.~(\ref{eq:phi double dot approx}) can be integrated exactly from $\dot\phi_0$ to $0$, giving the time $\Delta t_{\rm frag}$ and the field excursion $\Delta\phi_{\rm frag}$ from the beginning of particle production until the field stops:
\begin{align}
\Delta t_{\rm frag} &\sim \frac{f \dot\phi_0^3}{\Lambda_b^8} \log\frac{f^4}{\dot\phi_0^2}, \label{eq:Delta t frag approx} \\
\Delta \phi_{\rm frag} &\sim \frac{f \dot\phi_0^4}{\Lambda_b^8} \log\frac{f^4}{\dot\phi_0^2} . \label{eq:Delta phi frag approx}
\end{align}
Here $\dot\phi_0$ is the velocity of $\phi$ at the beginning of the particle production. 
In the equations above we neglected $\mathcal{O}(1)$ factors as the calculation above was only approximate. In the next section, we will derive these expressions using  a more precise treatment. The correct numerical factors are the ones of Eqs.~(\ref{eq:dtppapprox}) and~(\ref{eq:dphippapprox}) below, which reproduce the parametric dependence of Eqs.~(\ref{eq:Delta t frag approx}) and ~(\ref{eq:Delta phi frag approx}).

\section{Analytical discussion} \label{sec:analytical discussion}
In what follows we   discuss in detail the axion fragmentation dynamics introduced in the previous section.  
We   establish the   conditions to  decelerate   an axion field  uniquely due  to particle production friction from the axion field itself.
The approximate analytical formulae derived here  will be compared with the numerical solutions in the next section. 

In the intuitive discussion in \Sec{sec:intuitive derivation} we considered the limit in which the Hubble expansion is negligible, \textit{i.e.}, $H=0$.
Before deriving the conditions to make the field decelerate,  let us  consider the effect of the Hubble friction in the equation of motion for the fluctuation in Eq.~(\ref{eq:fluctuation}).
We can anticipate two additional effects once we consider the cosmic expansion.  Most importantly,  the growth of the modes is suppressed by the friction term $3H\dot u_k$ in Eq.~(\ref{eq:fluctuation}).
In addition, since the instability band moves towards smaller values of momentum when the zero-mode decelerates,  Hubble expansion makes a given mode to spend more time inside the instability band due to the red-shift of the physical momentum $k/a(t)$.

Let us assume for the moment $H$ to be constant.
By defining $\tilde u_k \equiv e^{3Ht/2} u_k$, Eq.~(\ref{eq:fluctuation}) can be rewritten as
\begin{align} \label{eq:eomfluctuations}
{\ddot{\tilde u}}_k + \left( e^{-2Ht} k^2 + V''(\phi) + \frac{9}{4}H^2 \right) \tilde u_k = 0.
\end{align}
For $H=0$, the last equation simply reproduces~\Eq{eq:Mathieu uk constant phidot}.
According to Eq.~(\ref{eq:asymptotic uk}), the exponential growth of $ u_k$ is at most $\exp( \Lambda_b^4 t/2f\dot\phi )$.
Thus, in order for $\tilde u_k$ to grow, the Hubble expansion rate should be bounded by
\begin{align}
H < \frac{\Lambda_b^4}{3f\dot\phi}~. \label{eq:condition 0}
\end{align}
Equivalently, the last equation can be rewritten as
\begin{align}
\dot\phi < \frac{\Lambda_b^4}{\mu^3 f} \dot\phi_{\mathrm{SR}}, \label{eq:condition 0'}
\end{align}
where
\begin{equation}
\dot\phi_{\mathrm{SR}} \equiv \frac{\mu^3}{3H}
\end{equation}
is the slow-roll velocity of the field in the linear potential $-\mu^3 \phi$ for a constant Hubble rate $H$. With a slight abuse of notation, in the following we will use this definition also in the case in which $H$ is not constant or the potential is not linear, and this quantity will turn out to be useful, even without representing a proper slow-roll velocity.
In addition, to have particle production active, the field should go over the barriers, $H < \dot\phi / 2\pi f$.%
\footnote{
This condition guarantees that the time needed to go over one wiggle is shorter than one Hubble time.
If this were not the case, the effect of Hubble friction would be dominant with respect to fragmentation.
In particular, if the wiggles are large enough, instead of rolling over many of them the field would stopped as soon as $V'=0$, just due to cosmic expansion. For a more detailed discussion about this point, we refer the reader to Ref.~\cite{Fonseca:2019lmc}.}
However, this condition is trivially satisfied when both of Eq.~(\ref{eq:condition 0}) and $\dot\phi > \Lambda_b^2$ are satisfied.
Hence, we assume
\begin{equation}\label{eq:assumptions}
\dot\phi_0^2 > \Lambda_b^4 \qquad \text{and} \qquad H \ll \frac{\Lambda_b^4 }{3 f\dot\phi_0 } \,.
\end{equation}
The first of this assumption is valid until the field keeps rolling. As $\dot\phi\approx\Lambda_b^2$, the field stops.
Contrary to the former, the second becomes easier to satisfy as the velocity decreases, therefore it is enough to assume that it is valid for the initial conditions.
The assumptions above allow us to simplify the analysis  due to the following three reasons.
First, we can regard $\dot\phi$ as a smooth function of the time $t$.
The numerical solution of $\dot\phi$ has a smooth component and a rapidly oscillating component with frequency $\dot\phi / 2\pi f$.
This oscillating component is caused by the wiggles and its relative size compared to the smooth component is $\sim \Lambda_b^4/\dot\phi^2$. We then neglect this oscillating term in this section.
Second, we can assume that $\ddot\phi$ is constant during the  amplification time 
$\delta t_{\rm amp}$, which we defined before as the time it takes for the mode $k_\mathrm{cr}$ to exit the instability band. This can be calculated, using \Eq{eq:phidot variation}, as
\begin{equation}\label{eq:delta time ampl 2}
\delta t_{\rm amp} \sim \frac{\dot\phi(\delta t_{\rm amp}) - \dot\phi}{\ddot\phi} \sim \frac{\Lambda_b^4}{\dot\phi|\ddot\phi|}.
\end{equation}
Using the result in (\ref{eq:delta time ampl 2}), we   impose that  $|\dddot\phi| \delta t_{\rm amp} \ll |\ddot\phi|$,  which we will justify later in \Eq{eq:phi double dot 2}. This condition can be rewritten as
\begin{align}
\left| \frac{\Lambda_b^4}{\dot\phi} \frac{d \log |\ddot\phi|}{d\dot\phi} \right|   \ll 1\,, \label{eq:constant ddot phi}
\end{align}
which shows that if $\ddot\phi$ depends on $\dot\phi$ polynomially, this condition is satisfied if $\Lambda_b^4 \ll \dot\phi^2$.
As a third simplification,  we can drop the friction term $3H \dot{u}_k$ in Eq.~(\ref{eq:fluctuation}) without changing the physical momentum  by ${\cal O}(1)$ fraction during the amplification.

\medskip

Now that we have presented our simplifying assumptions, let us go back to discuss the equation of motion.
For a given velocity $\dot\phi$, once cosmic expansion is taken into account, the critical mode $k_\mathrm{cr}$ and the width $\delta k_\mathrm{cr}$ are obtained dividing the left hand side of Eq.~(\ref{eq:k crit}) by the scale factor $a$:
\begin{align}\label{eq:k crit expansion}
\frac{k_{\rm cr}}{a} = \frac{\dot\phi}{2f}, \quad
\frac{\delta k_{\rm cr}}{a} = \frac{\Lambda_b^4}{2f\dot\phi}\,.
\end{align}
The condition $H<\dot\phi/(2\pi f)$, which was discussed above Eq.~(\ref{eq:assumptions}), ensures that $k_\mathrm{cr}> a \,H$. In other words, the amplification process takes place and ends when the modes are well inside the horizon. This marks an important difference with the case in which quantum fluctuations grow until they exit the horizon, as it happens, for example, with the gauge fields generated at the end of axion inflation~\cite{Pajer:2013fsa}.
For our discussion, we do not need to specify when the axion fragmentation dynamics takes place. The latter can be embedded in the cosmological history of the universe at different epochs. The usual dynamics of the modes crossing the horizon during inflation and then re-entering after the Big Bang does not affect our discussion.

By taking the initial condition Eq.~(\ref{eq:initial}) and assuming constant $\ddot\phi$,
the asymptotic behaviour of $u_k$ with $k/a = k_\mathrm{cr}/a = \dot\phi/2f$ after the amplification is
\begin{align}
u_{k_\mathrm{cr}}(t) \simeq \frac{1}{a} \sqrt{\frac{2}{k_\mathrm{cr}}} \exp\left( \frac{\pi \Lambda_b^8}{4f\dot\phi^2\,|\ddot\phi + H\dot\phi|} \right) \sin\left( \frac{1}{a} k_\mathrm{cr} t + \delta \right)\,. \label{eq:asymptotic uk 2}
\end{align}
Equation~(\ref{eq:asymptotic uk 2}) is derived in Appendix \ref{sec:technical details} in the following way. The equation of motion Eq.~(\ref{eq:eomfluctuations}) is solved by means of a WKB approximation in three separate time intervals, before the mode $k_\mathrm{cr}$ enters the instability band, when the mode is well inside the instability band and after it has left it. In the two transition regions, when the mode enters and exit the instability band, the solution is found in terms of Airy functions. Finally, the five intervals are matched by imposing the continuity of the solution.
From Eq.~(\ref{eq:asymptotic uk 2}) we see that for exponential particle production occurs, one needs:
\begin{align}
\frac{\pi \Lambda_b^8 }{ 2f\dot\phi^2\,|\ddot\phi + H\dot\phi|} > 1 \,, \label{eq:condition exponent}
\end{align}
where the factor of $2$ difference compared to Eq.~(\ref{eq:asymptotic uk 2}) depends on the fact that the energy density scales with $|u_k|^2$.
We discuss the validity of this assumption later, around Eq.~(\ref{eq:exponential pp condition phidot}).
Using \Eq{eq:phi mode expansion} and \Eq{eq:asymptotic uk 2},
we can estimate the energy density per volume in comoving momentum space right after the end of the amplification as
\begin{align}
\frac{d^3 \rho_{\textrm{fluc}}}{d k^3} \biggr|_{k=k_\mathrm{cr}} =
\frac{1}{(2\pi)^3} \frac{1}{2 a^2} k_\mathrm{cr}^2 \left| u_{k_\mathrm{cr}} \right|^2 \approx 
\frac{1}{(2\pi)^3} \frac{k_\mathrm{cr}}{a^4} \exp\left( \frac{\pi \Lambda_b^8}{2f\dot\phi^2\,|\ddot\phi + H\dot\phi|} \right).
\label{eq:d3rhodk3}
\end{align}
Using the initial condition for $u_k$ in \Eq{eq:initial}, the energy density before the amplification is  $d^3\rho_{\textrm{fluc}}/dk^3 = k/[2 a^4 (2\pi)^3]$. Then, 
the energy of the fluctuation is amplified by a factor of
\begin{equation} \label{eq:amplification factor}
2\exp\left( \frac{\pi \Lambda_b^8} { 2 f\dot\phi^2 |\ddot\phi + H\dot\phi| }\right).
\end{equation}
When this factor is much larger than $1$,  particle production is efficient to provide the friction for the homogeneous mode.
The increment of the energy density of the fluctuations because of the particle production is
\begin{align}
\frac{d\rho_{\textrm{fluc}}}{dt} = \left|\frac{dk_{cr}}{dt}\right| \times 4\pi k_\mathrm{cr}^2 \, \frac{d^3 \rho_{\textrm{fluc}}}{dk^3} \biggr|_{k = k_\mathrm{cr}},
\end{align}
where $dk_{\rm cr}/dt$ is the velocity of the instability band in the momentum space.
For non-zero $H$, $k_{\mathrm{cr}}$ is determined as
$$k_{\mathrm{cr}}/a = \dot\phi / (2f).$$
Therefore, the velocity of the instability band is given as $dk_{\mathrm{cr}}/dt = a(\ddot\phi + H \dot\phi)/(2f)$.
Thus, 
\begin{align}
\frac{d\rho_{\textrm{fluc}}}{dt} 
&= a \left| \frac{\ddot\phi + H\dot\phi}{2f} \right| \times 4\pi k_\mathrm{cr}^2 \times \frac{1}{(2\pi)^3} \frac{k_\mathrm{cr}}{a^4} \exp\left( \frac{\pi \Lambda_b^8}{2f\dot\phi^2|\ddot\phi + H\dot\phi|} \right) \nonumber\\
&= \frac{1}{32\pi^2 f^4} \dot\phi^3 |\ddot\phi + H\dot\phi| \exp\left( \frac{\pi \Lambda_b^8}{2f\dot\phi^2|\ddot\phi + H\dot\phi|} \right).
\label{eq:fluctuation production}
\end{align}
The kinetic energy of the homogeneous mode is $\dot\phi^2/2$ and its time derivative is $\dot\phi\ddot\phi$.
As a result, from conservation of energy, we obtain the following equation:
\begin{align}
\boxed{
\dot\phi\ddot\phi
=
- 3 H \dot\phi^2
+ \mu^3 \dot\phi
- \frac{1}{32\pi^2 f^4} \dot\phi^3 |\ddot\phi + H\dot\phi| \exp\left( \frac{\pi \Lambda_b^8}{2f\dot\phi^2|\ddot\phi + H\dot\phi|} \right).}
\label{eq:eq for phi double dot}
\end{align}
Compared to Eqs.~(\ref{eq:EOM1}) and~(\ref{eq:zeromode}), Eq.(\ref{eq:eq for phi double dot}) describes the effect of fragmentation after averaging over many oscillations of the sinusoidal potential, and is more suitable for our analysis.
Using this equation we will determine the general conditions to stop the field due to axion fragmentation, which are obtained in  the next sections.
The first and second term in the right-handed side of \Eq{eq:eq for phi double dot} are the effect of Hubble friction and acceleration by the slope, respectively.
This equation can be regarded as a consistency condition for $\ddot\phi$ during the fragmentation phase.
By solving the above equation, $\ddot\phi$ can be calculated as a function of $\dot\phi$, $\Lambda_b$, $f$, $\mu^3$, and $H$.

\subsection{General condition to stop the axion}
Let us discuss conditions to stop the axion field.
Here we summarize the results, while the details of the derivation are given in Appendix \ref{sec:detail on phiddot}.

\begin{figure}
\centering
\includegraphics[width=.55\textwidth]{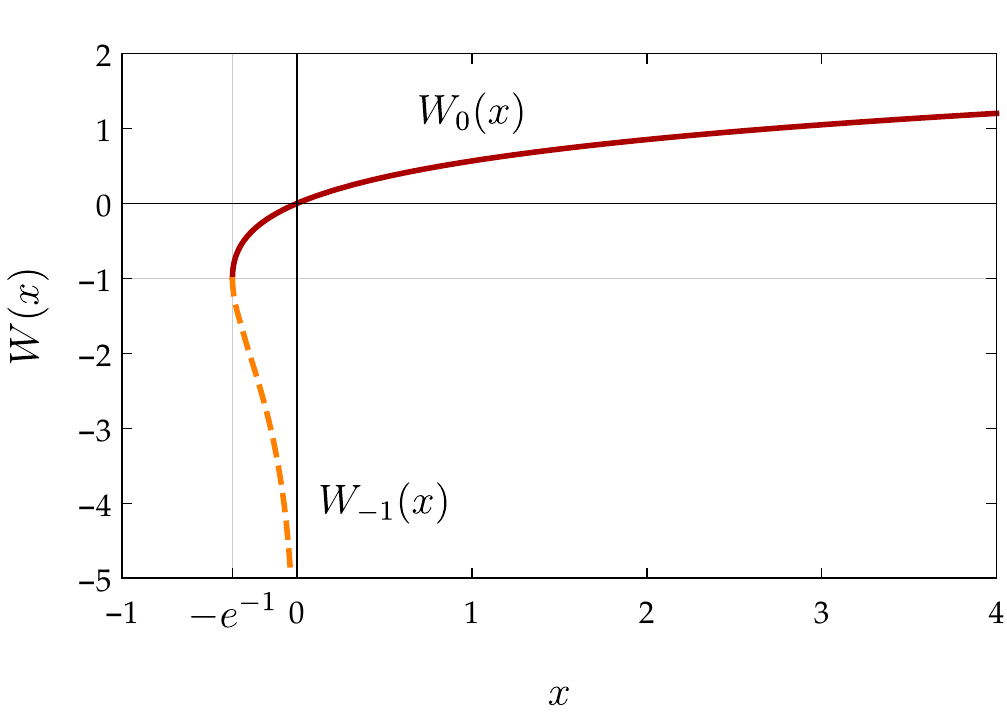}
\caption{\label{fig:Lambert}The two real branches of the product logarithm function.}
\end{figure}

$\ddot\phi<0$ must hold from the initial time until the field has come to a complete stop. As detailed in Appendix~\ref{sec:stopping condition}, this is realized if (and only if) the following condition  holds for the initial velocity:
\begin{empheq}[box=\widefbox]{align}
\mu^3 < 
2H\dot\phi_0 + \displaystyle\frac{\pi\Lambda_b^8}{2f\dot\phi_0^2} \left( W_0\left( \displaystyle\frac{32\pi^2f^4}{e\dot\phi_0^2} \right) \right)^{-1}\,.
\label{eq:condition for no positive phiddot}
\end{empheq}
Here $W_n(x)$ is the product logarithm function (also known as Lambert $W$ function),%
\footnote{
The product logarithm function is the inverse function of $W e^W = x$.
In general, there exist infinite number of solutions for this equation, and $W_0$ and $W_{-1}$ are the two real ones.
In particular, $W_0(x)$ is real for $-e^{-1} \leq x$ and $W_{-1}$ is real for $-e^{-1} \leq x < 0$.
Also, $W_0(x) = \log x - \log\log x + \cdots$ for large $x$.
A plot of $W_0$ and $W_{-1}$ for small values of $x$ is shown in Fig.~\ref{fig:Lambert}.
This function is available \texttt{ProductLog} in \texttt{Mathematica} or \texttt{special.lambertw} in \texttt{SciPy}.
See \textit{e.g.}, \url{http://mathworld.wolfram.com/LambertW-Function.html}. 
}
whose real branches are plotted in Fig.~\ref{fig:Lambert}.
\Eq{eq:condition for no positive phiddot} expresses an equilibrium between the slope and the Hubble expansion which allows for efficient fragmentation: if the slope increases, in order to avoid the field acceleration, the friction due to  cosmic expansion should compensate this effect. Alternatively, for $\dot\phi_0 < \mu^3/(2 H) = 3/2 \dot\phi_\mathrm{SR}$, one can see \Eq{eq:condition for no positive phiddot} as a lower bound on $\Lambda_b$ that expresses, for given $\mu^3$ and $H$, the necessary amount of fragmentation needed in order to slow down the field.

If Eq.~(\ref{eq:condition for no positive phiddot}) is satisfied, Eq.~(\ref{eq:eq for phi double dot}) has only one solution, with negative $\ddot\phi$, which is given by%
\footnote{Before fragmentation is active, the field is only subject to its potential and to Hubble friction, and its equation of motion is simply  $\ddot\phi = \mu^3 - 3 H \dot\phi$, where we neglected a small oscillating term.
This cannot be obtained as the $\Lambda_b\to0$ limit of Eq.~(\ref{eq:solution phiddot}). The reason is that the equation of motion Eq.~(\ref{eq:eq for phi double dot}) was derived assuming fragmentation is active. In particular, we assumed $\ddot\phi\ll\Lambda_b^4/f$. The acceleration is initially $\ddot\phi = \mu^3 - 3 H \dot\phi$, and asymptotes to Eq.~(\ref{eq:solution phiddot}) as fragmentation starts. }
\begin{align}
\ddot\phi &=
\begin{cases}
- H\dot\phi + \displaystyle\frac{\pi \Lambda_b^8}{2f\dot\phi^2}\left[ b + W_{-1}(-ab e^{-b} ) \right]^{-1} & (b>0) \\
- H\dot\phi + \displaystyle\frac{\pi \Lambda_b^8}{2f\dot\phi^2}\left[ b + W_0(-ab e^{-b} ) \right]^{-1} & (b<0) \\
\end{cases}
.
\label{eq:solution phiddot}
\end{align}
Here $a$ and $b$ are dimensionless parameters which are defined as
\begin{align}
a \equiv \frac{\dot\phi^2}{32\pi^2 f^4},\quad
b \equiv \frac{\pi\Lambda_b^8}{2f\dot\phi^2(\mu^3-2H\dot\phi)}.
\label{eq:a b definition}
\end{align}

\begin{figure}[ht]
\centering
\includegraphics[width=0.69\hsize]{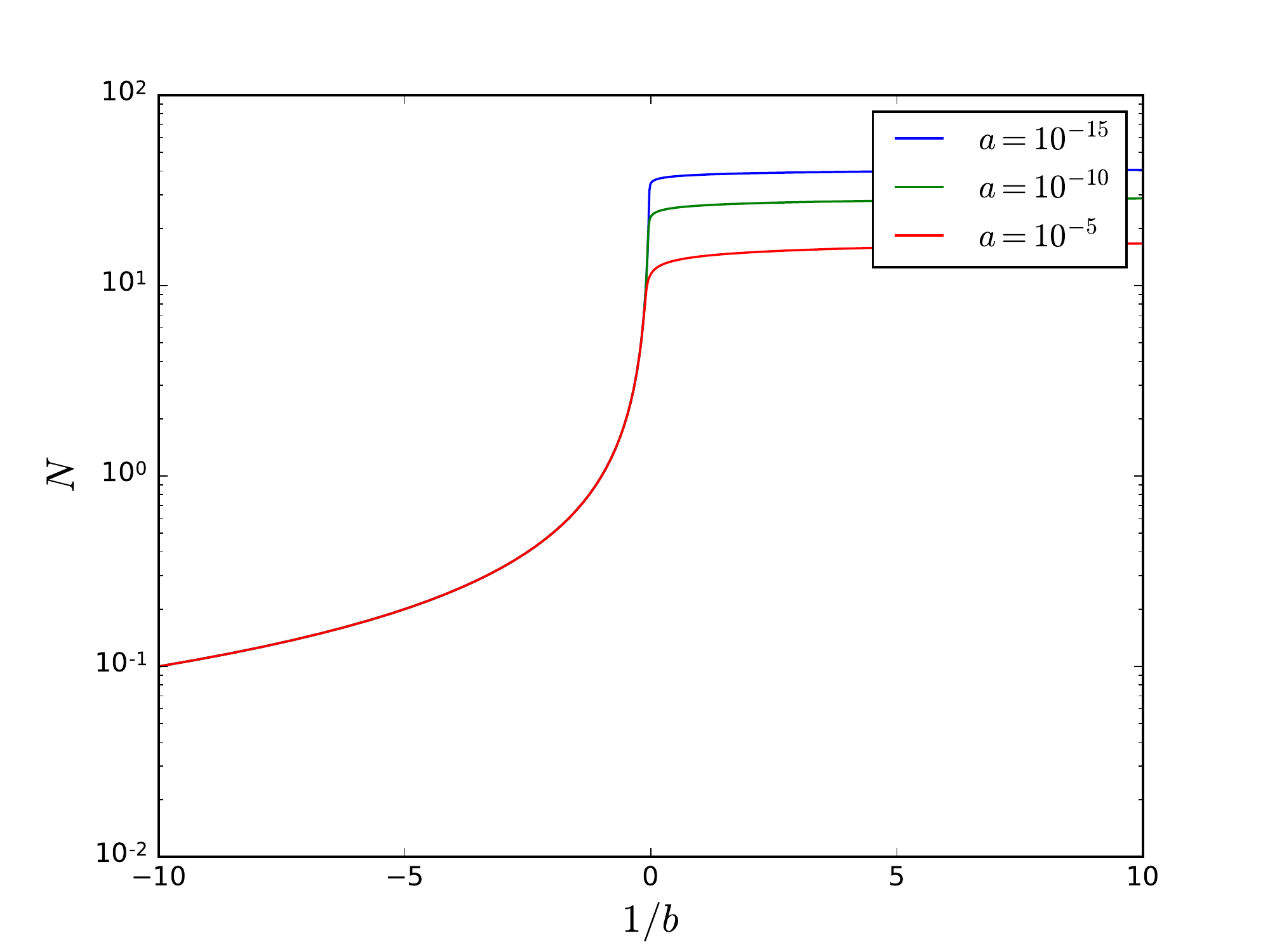}
\caption{The exponent $N$ given in Eq.~(\ref{eq:exponent}) as a function of $1/b$.}\label{fig:exponent}
\end{figure}

Let us discuss the validity of the assumption Eq.~(\ref{eq:condition exponent}).
The effect of the axion fragmentaion in Eq.~(\ref{eq:eq for phi double dot}) has
a exponential factor with a exponent $N \equiv  \pi \Lambda_b^8 / 2f\dot\phi^2 |\ddot\phi + H \dot\phi|$.
As we have discussed, for exponential particle production to occur, $N$ should be larger than 1.
By using Eq.~(\ref{eq:solution phiddot}), we obtain
\begin{align}
N =
\begin{cases}
|b + W_{-1}(-abe^{-b})| & (b>0) \\
|b + W_{ 0}(-abe^{-b})| & (b<0) 
\end{cases}. \label{eq:exponent}
\end{align}
As we can see in Fig.~\ref{fig:exponent}, for fixed $a$, $N$ is an monotonously increasing function of $1/b$.
For $a\ll 1$, as we will always assume, $N$ becomes small at $1/b\lesssim 0$, and is well approximated as $N \simeq -b$.
Thus, by requiring $N > 1$, we obtain 
\begin{align}
\frac{2f\dot\phi^2 (\mu^3 - 2H\dot\phi)}{\pi\Lambda_b^8} > -1.
\label{eq:exponential pp condition phidot}
\end{align}
If $\dot\phi < (3/2)\dot\phi_{\rm SR}$, the LHS of Eq.~(\ref{eq:exponential pp condition phidot}) is positive and the inequality is satisfied.
Thus, the above condition can be rewritten as
\begin{align}
\dot\phi < \frac{3}{2}\dot\phi_{\rm SR}
\qquad{\rm or}\qquad
\mu^3 > 2H\dot\phi - \frac{\pi \Lambda_b^8}{2f\dot\phi^2}.
\label{eq:exponential pp condition phidot 2}
\end{align}
As long as Eq.~(\ref{eq:exponential pp condition phidot}) (or equivalently Eq.~(\ref{eq:exponential pp condition phidot 2})) is satisfied,
we can safely use $\ddot\phi$ given in Eq.~(\ref{eq:solution phiddot}).
Note that $2H\dot\phi - \pi \Lambda_b^8/(2f\dot\phi^2)$ is a monotonously increasing function of $\dot\phi$.
Thus, we can see that Eq.~(\ref{eq:exponential pp condition phidot 2}) is satisfied for any $\dot\phi < \dot\phi_0$ if (and only if)
\begin{align}
\dot\phi_0 < \frac{3}{2}\dot\phi_{\rm SR}
\qquad{\rm or}\qquad
\mu^3 > 2H\dot\phi_0 - \frac{\pi \Lambda_b^8}{2f\dot\phi_0^2}.
\label{eq:exponential pp condition}
\end{align}
If this condition is satisfied, Eq.~(\ref{eq:solution phiddot}) for the acceleration $\ddot\phi$ can be used to describe the fragmentation process from its begininng until the end of that.
In phenomenologically interesting applications $\dot\phi \leq \dot\phi_{\rm SR}$, and Eq.~(\ref{eq:exponential pp condition}) is always satisfied.

\medskip

To summarize,  in order for fragmentation to stop the rolling of the axion field we need to simultaneously impose  the conditions of \Eq{eq:condition 0'} and \Eq{eq:condition for no positive phiddot}  to stop the axion. We are interested in the case in which the potential has local minima (\textit{i.e.} $\Lambda_b^4/f > \mu^3$), and we will assume $\dot\phi_0 \leq \dot\phi_{\mathrm{SR}}$. In this case, \Eq{eq:condition 0'} is trivially satisfied, and the only condition that must hold is \Eq{eq:condition for no positive phiddot}.

\begin{figure}[t]
\centering
\includegraphics[width=0.69\hsize]{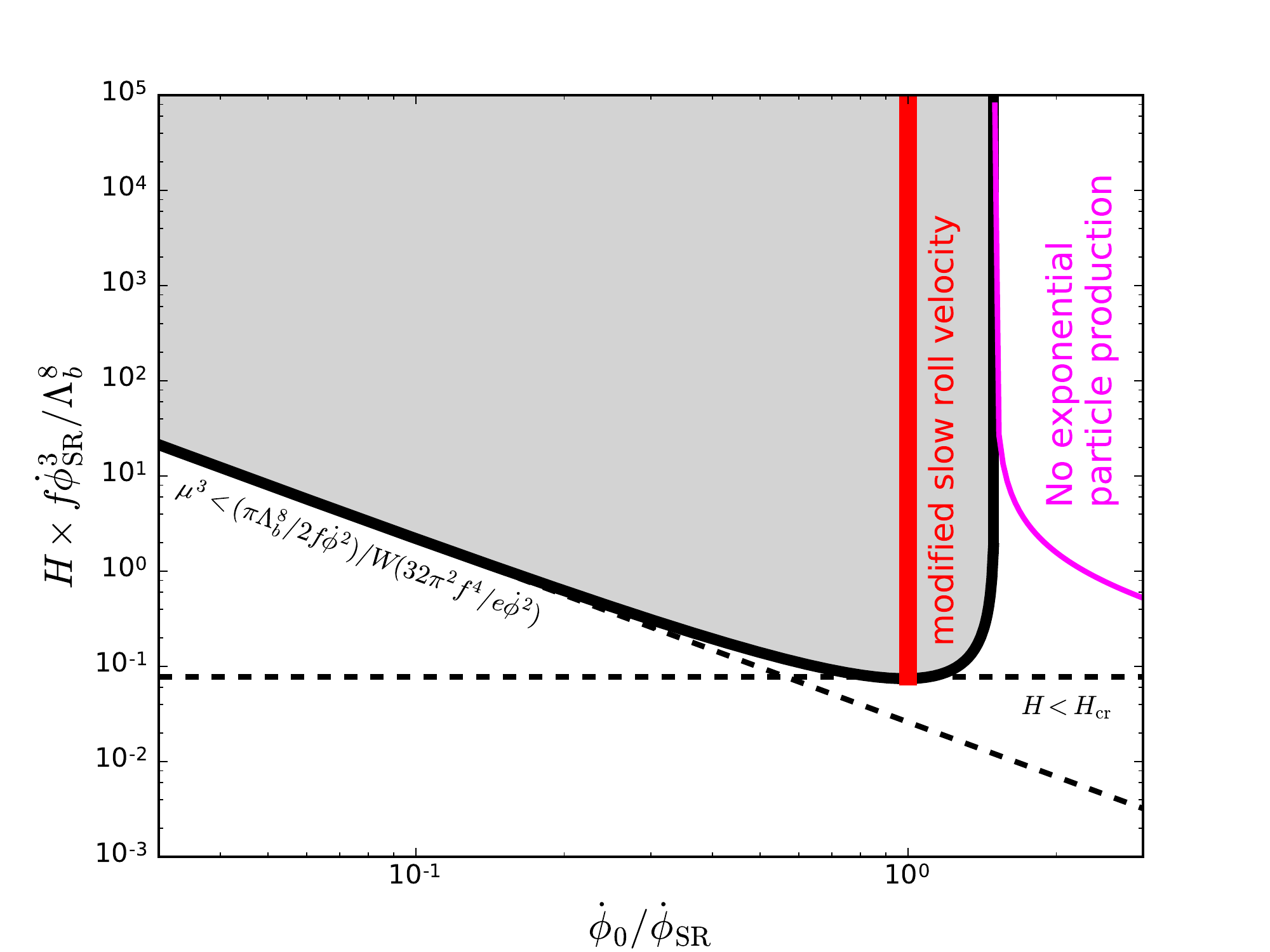}
\caption{
The parameter space which is excluded by Eq.~(\ref{eq:condition for no positive phiddot}).
The ratio between the decay constant and the slow roll velocity is fixed to $f^2/\dot\phi_{\textrm{SR}} = 10^4$.
The coefficient of the slope is fixed to $\mu^3 = 3 H \dot\phi_{\textrm{SR}}$.
Eq.~(\ref{eq:condition for no positive phiddot}) is \textit{not} satisfied in the gray regions.
The red solid line shows the modified slow roll velocity $\dot\phi_{\rm SR(frag)}$ given in Eq.~(\ref{eq:modifield slow roll velocity}).
$H_{\rm cr}$ is defined in Eq.~(\ref{eq:Hcr}).
The magenta line shows a condition given in Eq.~(\ref{eq:exponential pp condition}).
The axion can be successfully stopped in the white region.
}\label{fig:conditions and parameter space 1}
\end{figure}

\begin{figure}[t]
\centering
\includegraphics[width=0.69\hsize]{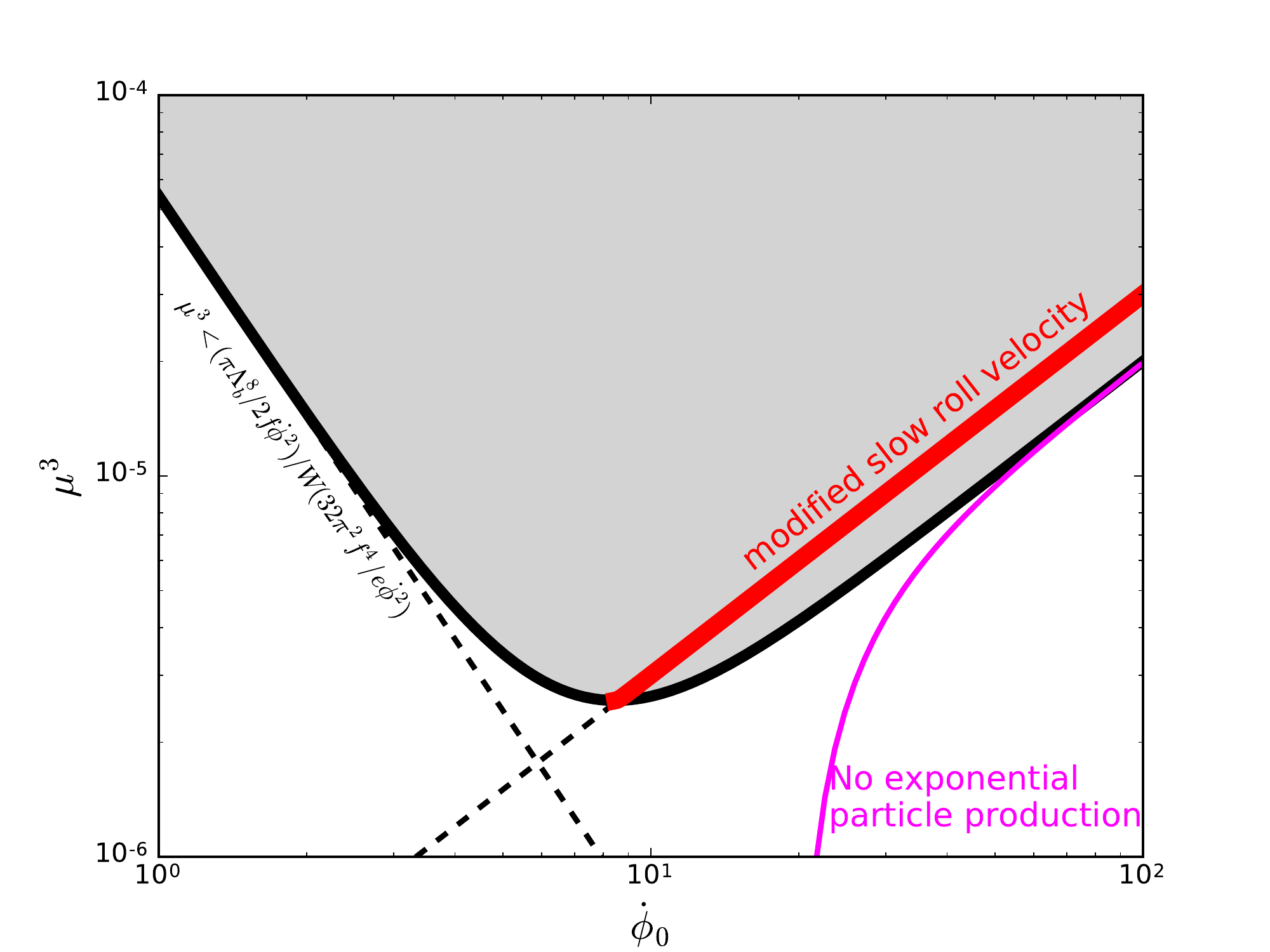}
\caption{
Same as Fig.~\ref{fig:conditions and parameter space 1}.
The decay constant and Hubble expansion rate are fixed to $f/\Lambda_b = 10^3$, $H/\Lambda_b = 10^{-7}$.
}\label{fig:conditions and parameter space 2}
\end{figure}

Figure~\ref{fig:conditions and parameter space 1} and Fig.~\ref{fig:conditions and parameter space 2} show
examples of the parameter space in which Eq.~(\ref{eq:condition for no positive phiddot}) is satisfied.
In the case of $\dot\phi_0 \lesssim \dot\phi_{\mathrm{SR}}$, there is an upperbound on $\mu^3$ because the acceleration effect by the slope should be weaker than the particle production effect.
On the other hand, the bound on $\mu^3$ is very weak for $\dot\phi \gtrsim \dot\phi_{\mathrm{SR}}$
and it becomes trivial for $\dot\phi_0 \geq (3/2)\dot\phi_{\mathrm{SR}}$.
Indeed, for $\dot\phi_0 \geq (3/2)\dot\phi_{\mathrm{SR}}$, Eq.~(\ref{eq:condition for no positive phiddot}) is always satisfied, and the field is slowed down by Hubble friction without the need of fragmentation. In this region, the only bound comes from imposing that the exponential amplification of the fluctuations is active, as we do in Eq.~(\ref{eq:exponential pp condition}).

It is interesting to investigate whether Eq.~(\ref{eq:eq for phi double dot}) admits constant velocity solutions. If such solutions exist the field can reach a steady state, and fragmentation can not stop the evolution.
Equation~(\ref{eq:eq for phi double dot}) with $\ddot\phi = 0$ has no solution for $\dot\phi$ if (and only if):
\begin{align}
H < H_{\rm cr} \simeq \frac{\pi \Lambda_b^8}{2f\dot\phi_{\rm SR}^3} \left( W_0\left( \frac{32\pi^2f^4}{e\dot\phi_{\rm SR}^2} \right) \right)^{-1}
\label{eq:Hcr}
\end{align}
is satisfied. This means that the axion zero mode cannot roll with constant velocity in such cases. For details, see the Appendix \ref{sec:modified slow roll velocity}. Note that \Eq{eq:Hcr} represents an upper bound on $H$ for fixed $\dot\phi_\mathrm{SR}$, but for fixed $\mu^3$ it can be rewritten as a lower bound on $H$.
On the other hand, for $H > H_{\rm cr}$, Eq.~(\ref{eq:eq for phi double dot}) with $\ddot\phi=0$ admits solutions. To distinguish it from the slow roll velocity $\dot\phi_{\rm SR} = \mu^3/3H$, we denote this velocity as the modified slow roll velocity $\dot\phi_{\rm SR(frag)}$. We show this velocity in red in Fig.~\ref{fig:conditions and parameter space 1} and Fig.~\ref{fig:conditions and parameter space 2}. As long as $\dot\phi_{\rm SR} \ll f^2$, the modification to the slow roll velocity is small, \textit{i.e.}, $\dot\phi_{\rm SR(frag)} \simeq \dot\phi_{\rm SR}$.

\subsection{Stopping conditions in several limits}
In the previous section, we described the generic negative solution of $\ddot\phi$ and the stopping conditions.
Here we discuss several cases in which the conditions can be simplified by taking some limits of the parameters.

\subsubsection{$H=0$ and $\mu^3=0$}
First, let us discuss the limit of $H=0$ and $\mu^3=0$.
In this limit, Eqs.~(\ref{eq:condition 0}) and (\ref{eq:condition for no positive phiddot}) are satisfied automatically.
The expression for $\ddot\phi$ given in Eq.~(\ref{eq:solution phiddot}) is simplified to%
\footnote{
Eq.~(\ref{eq:phi double dot 2}) can be obtained by separately taking the limits $\mu^3-2H\dot\phi \to 0^+$ and $\mu^3-2H\dot\phi \to 0^-$ of the two branches of Eq.~(\ref{eq:solution phiddot}), respectively, and using the expansions $W_0(x)\approx \log x -\log\log x$ for $x\to+\infty$ and $W_1(x)\approx \log(-x) - \log(-\log(-x))$ for $x\to 0^-$, from which we get 
\[
\lim_{b\to+\infty} b + W_{-1}(-abe^{-b}) =
\lim_{b\to-\infty} b + W_{0}(-abe^{-b}) = \log a
\]
}
\begin{align}
\ddot\phi = -\frac{\pi\Lambda_b^8}{2\dot\phi^2 f}\left( \log\frac{32\pi^2 f^4}{\dot\phi^2} \right)^{-1}. \label{eq:phi double dot 2}
\end{align}
This can be regarded as a refinement of Eq.~(\ref{eq:phi double dot approx}).
In this case, the amplification factor of the fluctuation energy in \Eq{eq:amplification factor} is given by
\begin{align}\label{eq:amplification mu=0 H=0}
2\exp\left( \frac{\pi \Lambda_b^8}{2 f\dot\phi^2 |\ddot\phi|} \right) = \frac{64\pi^2 f^4}{\dot\phi^2}.
\end{align}
In the derivation of Eq.~(\ref{eq:asymptotic uk 2}) we assumed that $\dot\phi\ll f^2$, which is necessary for the validity of the low-energy EFT of the axion field, thus the amplification factor is much larger than $1$, enhancing the efficiency of fragmentation.
The time scale and the field excursion during the axion particle production are
\begin{align}
\Delta t_{\rm frag} &\equiv \int^0_{\dot\phi_0}\frac{d\dot\phi}{\ddot\phi}
\simeq \frac{2f \dot\phi_0^3}{3\pi \Lambda_b^8} \log\frac{32\pi^2f^4}{\dot\phi_0^2}, 
\label{eq:dtppapprox}\\
\Delta \phi_{\rm frag} &\equiv \int^0_{\dot\phi_0}\frac{\dot\phi\, d\dot\phi}{\ddot\phi}
\simeq \frac{f \dot\phi_0^4}{2\pi \Lambda_b^8} \log\frac{32\pi^2f^4}{\dot\phi_0^2} ,
\label{eq:dphippapprox}
\end{align}
where we dropped subleading terms.
The number of wiggles which the axion travels until it stops is
\begin{align}
\frac{\Delta\phi_{\rm frag}}{2\pi f}
=
\frac{\dot\phi_0^4}{4\pi^2 \Lambda_b^8} \log\frac{32\pi^2f^4}{\dot\phi_0^2}.
\end{align}
For example, for $\dot\phi_0 / \Lambda_b^2 = 10^2$ and $\log \,(32\pi^2 f^4/\dot\phi_0^2) = 10$, this number is $\sim 3 \times 10^7$.

\subsubsection{$H\simeq 0$ and $\mu^3 \neq 0$}

\begin{figure}[t]
\centering
\includegraphics[width=0.7\hsize]{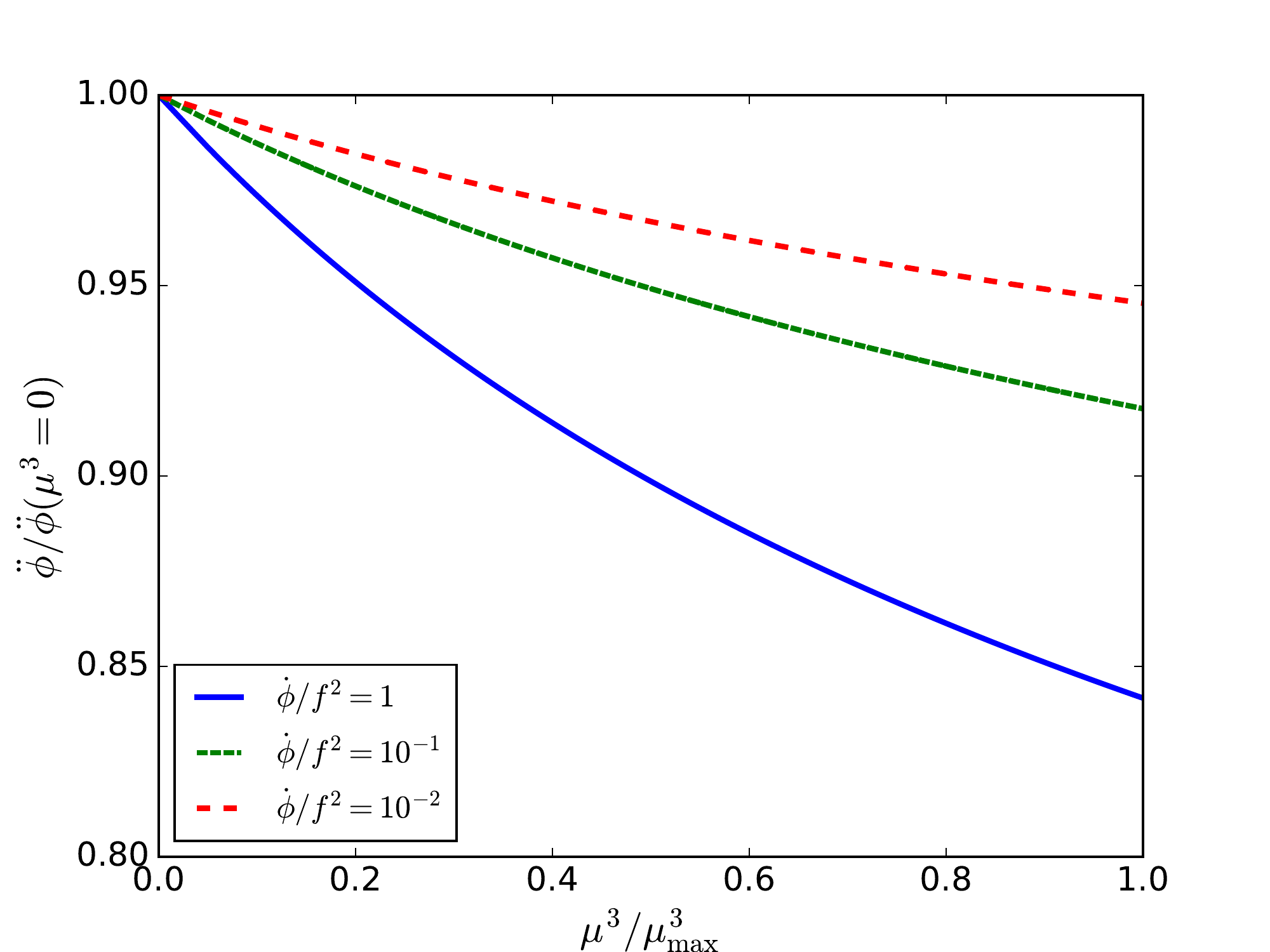}
\caption{
The ratio between $\ddot\phi$ with non-zero slope $\mu^3$ (Eq.~(\ref{eq:phi double dot with H=0}))
and $\ddot\phi$ with $\mu^3=0$ (Eq.~(\ref{eq:phi double dot 2})).
We take $H=0$ in this figure.
}\label{eq:phiddot ratio}
\end{figure}

Next, let us discuss the case in which $\mu^3\neq0$ and  $H$ is small enough to be neglected.
For $\dot\phi_0 \ll \dot\phi_\mathrm{SR}$, Eq.~(\ref{eq:condition for no positive phiddot}) can be simplified as
\begin{align}
\mu^3 < \mu_\mathrm{max}^3\equiv \frac{\pi\Lambda_b^8}{2f\dot\phi_0^2} \left( W_0\left( \frac{32\pi^2 f^4}{e\dot\phi^2_0} \right) \right)^{-1}. \label{eq:slope bound (H=0)}
\end{align}
This corresponds to the left part of Figs.~\ref{fig:conditions and parameter space 1} and \ref{fig:conditions and parameter space 2}, which show how $H$ is irrelevant if $\dot\phi_0 \ll \dot\phi_{\textrm{SR(frag)}} \approx \dot\phi_{\textrm{SR}}$.
In the limit of $H=0$, the negative $\ddot\phi$ solution Eq.~(\ref{eq:solution phiddot}) is simplified as
\begin{align}
\ddot\phi \simeq \frac{\pi \Lambda_b^8}{2f\dot\phi^2} [ b + W_{-1}(-abe^{-b}) ]^{-1}\,, \label{eq:phi double dot with H=0}
\end{align}
where $a,b$ are defined in \Eq{eq:a b definition}. In the limit $\mu^3 =0$ we recover Eq.~(\ref{eq:phi double dot 2}).
The acceleration $|\ddot\phi|$ monotonically decreases as a function of $\mu^3$.
In Fig.~\ref{eq:phiddot ratio}, we show the relative variation of $\ddot\phi$ when the slope $\mu^3$ goes from $0$ to $\mu_\mathrm{max}$ defined in Eq.~(\ref{eq:slope bound (H=0)}).
As long as $\dot\phi < f$, the decrements of $\ddot\phi$ is at most $\sim 15$ \%.
Hence, the estimates for the fragmentation time and the total field excursion of Eqs.~(\ref{eq:dtppapprox}, \ref{eq:dphippapprox}) still provide a reliable approximation.

\subsubsection{$\dot\phi = \dot\phi_{\textrm{SR}}$}
Let us discuss the case in which the initial velocity is equal to the slow roll velocity, \textit{i.e.}, $\dot\phi_0 = \dot\phi_{\textrm{SR}}$. This is the case if the dynamics takes place during inflation, since the velocity is exponentially driven to the attractor slow-roll velocity irrespectively of the initial conditions.
In this case, the condition Eq.~(\ref{eq:condition for no positive phiddot}) can be simplified to the following equation:
\begin{align}
H < \frac{\pi \Lambda_b^8}{2f\dot\phi_{\textrm{SR}}^3} \left( W_0\left( \frac{32\pi^2 f^4}{e \dot\phi_{\textrm{SR}}^2} \right) \right)^{-1}.
\label{eq:condition 3}
\end{align}
By replacing $\dot\phi_{\textrm{SR}} = \mu^3/(3 H)$, the above condition Eq.~(\ref{eq:condition 3}) can be equivalently rewritten as a condition for $\mu^3$ and $H$ for given $\Lambda_b$ and $f$:
\begin{align}
\mu^3 < \mu_{th}^3 \equiv \frac{3e \Lambda_b^8}{64\pi f^5}
\quad
{\rm or}
\quad
H > H_{min} \equiv \sqrt{ \frac{2}{27\pi} } \frac{f^{1/2} \mu^{9/2}}{\Lambda_b^4} \sqrt{ \log\frac{64\pi f^5 \mu^3}{3 e\Lambda_b^8} }.
 \label{eq:Hmin mu3th}
\end{align}
This condition tells us that
the Hubble friction is required to prevent acceleration in order to work the fragmentation mechanism
if the slope is steeper than the threshold value $\mu^3_{th}$.
Alternatively, one can think of Eq.~(\ref{eq:condition 3}) as a lower bound on $\Lambda_b$ to have enough fragmentation to stop the field, for given $\mu^3$ and $H$.

\begin{figure}[t]
\centering
\includegraphics[width=0.7\hsize]{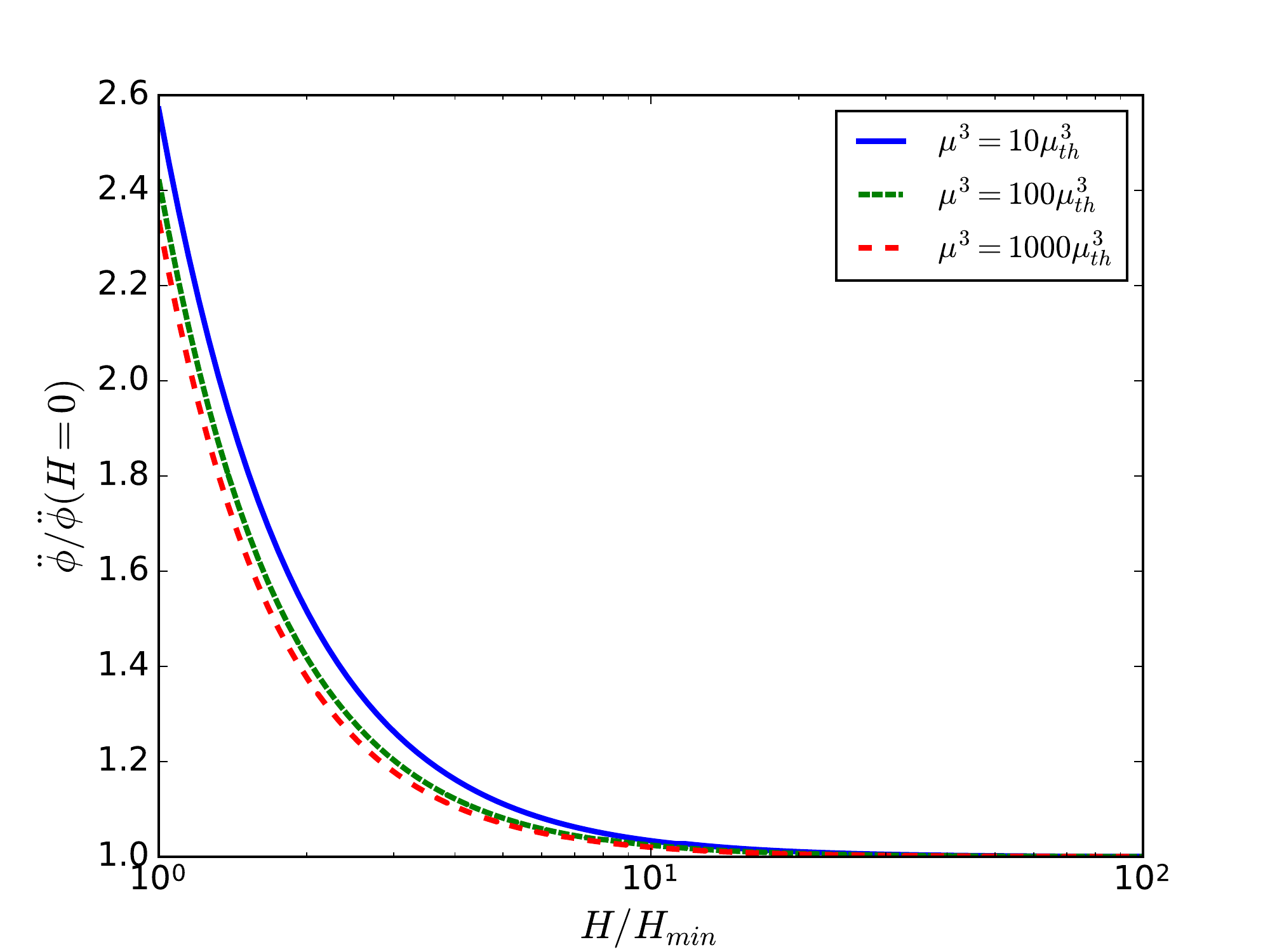}
\caption{
The ratio between $\ddot\phi$ by Eq.~(\ref{eq:phi double dot SR})
and $\ddot\phi$ by Eq.~(\ref{eq:phi double dot 2}) with $\dot\phi = \mu^3/3H$.
$H_{min}$ and $\mu^3_{th}$ are defined in Eq.~(\ref{eq:Hmin mu3th}).
This figure is independent on the choice of $f$.
}
\label{eq:phiddot ratio slowroll}
\end{figure}

The negative $\ddot\phi$ solution Eq.~(\ref{eq:solution phiddot}) for $\dot\phi = \dot\phi_\mathrm{SR}$ is given as
\begin{align}\label{eq:phi double dot SR}
\ddot\phi = \frac{\pi \Lambda_b^8}{2f\dot\phi_{\textrm{SR}}^2} \left( -\frac{1}{\tilde b} + \frac{1}{\tilde b + W_{-1}(- a\tilde b e^{-\tilde b} ) }\right),\quad
\tilde b \equiv \frac{\pi\Lambda_b^8}{2f H \dot\phi_{\textrm{SR}}^3}.
\end{align}
In Fig.~\ref{eq:phiddot ratio slowroll}, we show the ratio between
$\ddot\phi$ evaluated with Eq.~(\ref{eq:phi double dot SR}) and $\ddot\phi$ computed in Eq.~(\ref{eq:phi double dot 2}) with $H,\mu^3=0$, with the same velocity $\dot\phi = \dot\phi_{\rm SR}$.
Again, as long as Eq.~(\ref{eq:Hmin mu3th}) (or equivalently Eq.~(\ref{eq:condition 3})) is satisfied, the acceleration is well described by the $H=0$ equation~(\ref{eq:phi double dot 2}), thus the estimates for the fragmentation time and the total field excursion of Eqs.~(\ref{eq:dtppapprox}, \ref{eq:dphippapprox}) are reliable also in this case.

\section{Numerical analysis of the equations of motion} \label{sec:numeircal analysis}

In this section, we test the analytical understanding developed in Sec.~\ref{sec:analytical discussion} against a numerical solution of the equations of motion for the homogeneous mode $\phi(t)$, Eq.~(\ref{eq:zeromode}), and for the fluctuations $\delta \phi(x,t)$, Eq.~(\ref{eq:fluctuation}). We still limit ourselves to a linear level analysis, the validity of which will be further discussed in Sec.~\ref{sec:beyond linear}.
For this calculation, we discretize the integral in Eq.~(\ref{eq:zeromode})
and take 10000 modes whose momentum is between  $10^{-4} k_{\rm cr}^0 $ and $k_{\rm cr}^0 + 10 \delta k_{\rm cr}^0$.
The momenta are evenly spaced in logarithmic scale.
The differential equations are solved numerically by the fourth order Runge-Kutta method.

\begin{figure}[t]
\centering
\includegraphics[width=0.7\hsize]{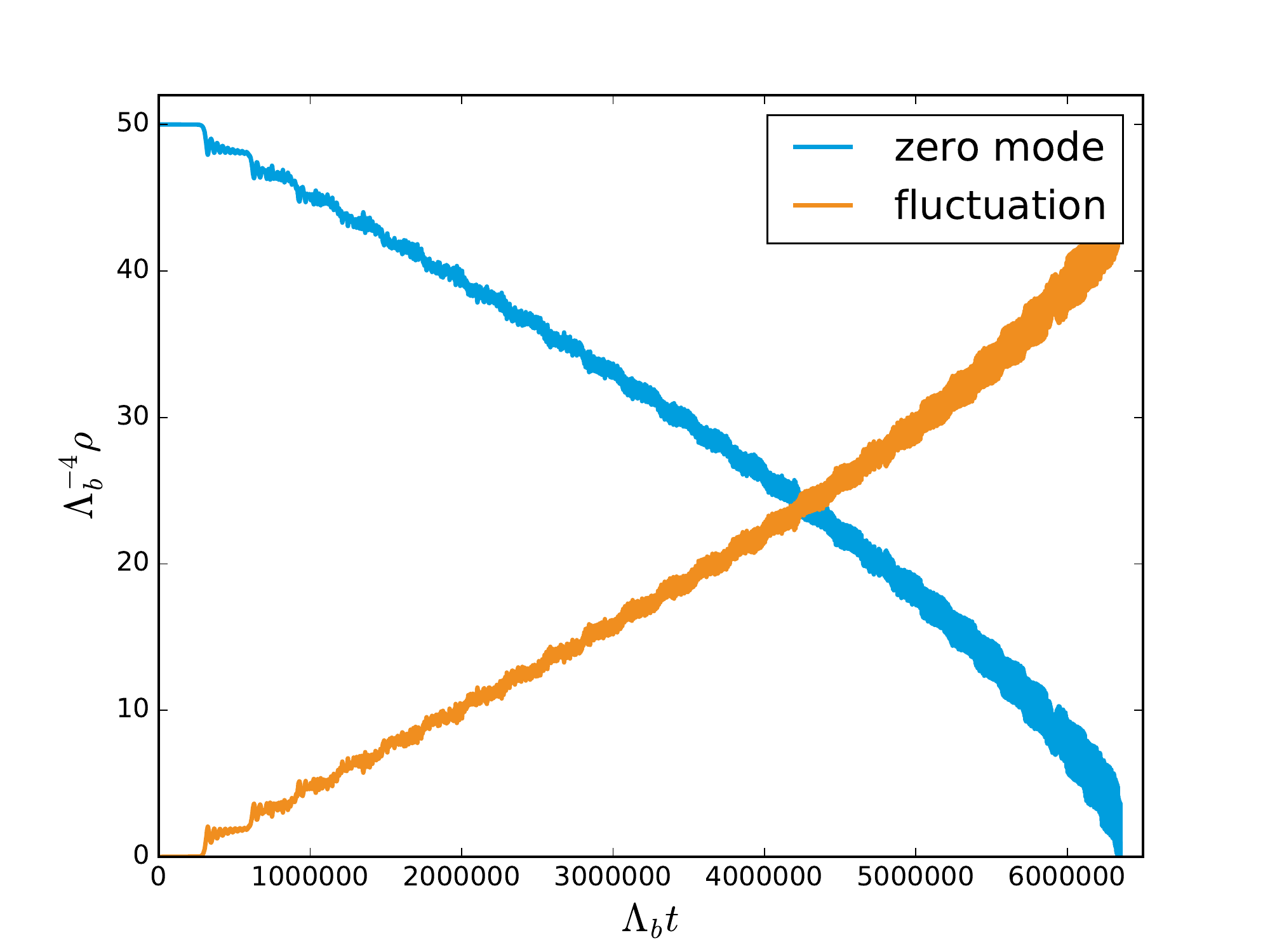}
\caption{
Time evolution of energy of zero-mode and fluctuations.
We take $f = 1000 \Lambda_b$, $\dot\phi_0 = 10 \Lambda_b^2$, $\mu^3=0$, and $H=0$.
}\label{fig:time evolution of energy}
\end{figure}
\begin{figure}[t]
\centering
\includegraphics[width=0.7\hsize]{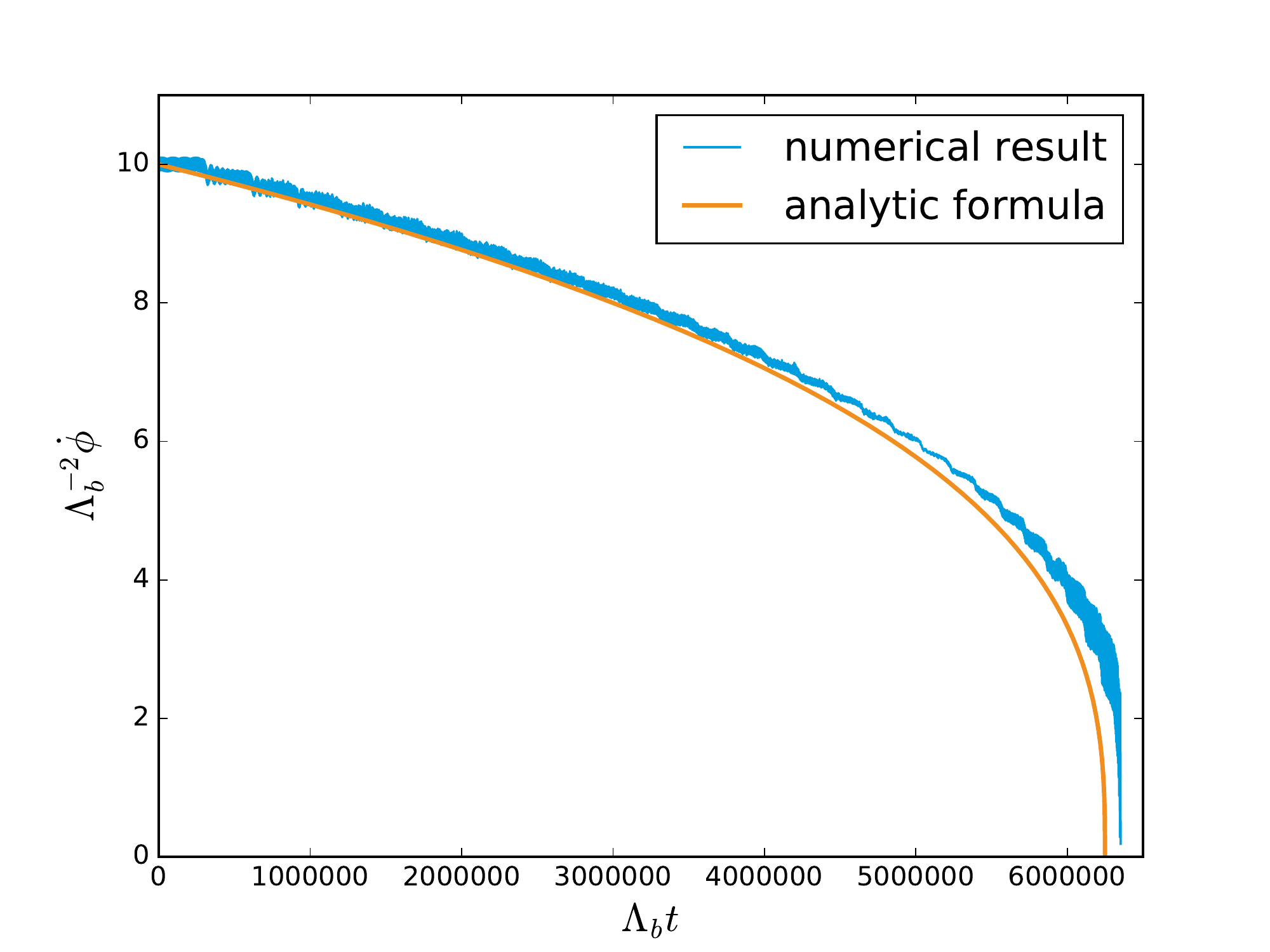}
\caption{
Time evolution of $\dot\phi$ with $f = 1000 \Lambda_b$, $\dot\phi = 10 \Lambda_b^2$, $\mu^3=0$, and $H=0$.
The blue line is obtained by solving Eqs.~(\ref{eq:zeromode}) and (\ref{eq:fluctuation}),
and the orange line is the solution of Eq.~(\ref{eq:phi double dot 2}).
We take the same parameters as Fig.~\ref{fig:time evolution of energy}.
}\label{fig:time evolution of phidot}
\end{figure}
\begin{figure}[t]
\centering
\includegraphics[width=0.7\hsize]{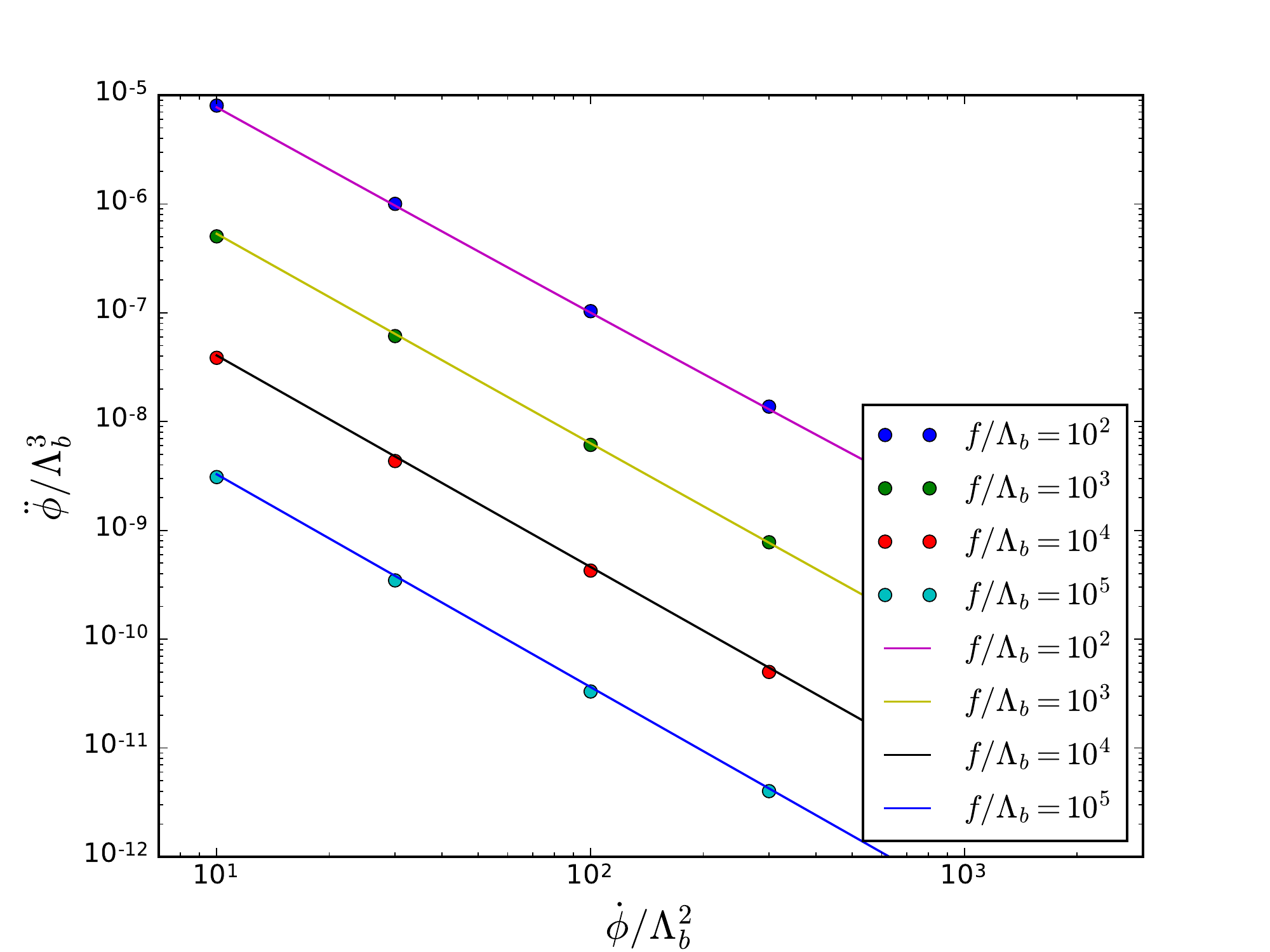}
\caption{
$\ddot\phi$ as a function of $\dot\phi$ with $\mu^3=0$ and $H=0$.
The lines are obtained from Eq.~(\ref{eq:phi double dot 2}).
The dots are calculated from the numerical solution of Eqs.~(\ref{eq:zeromode}) and (\ref{eq:fluctuation}).
}\label{fig:phi doubledot as a function of phidot}
\end{figure}
\begin{figure}[t]
\centering
\includegraphics[width=0.7\hsize]{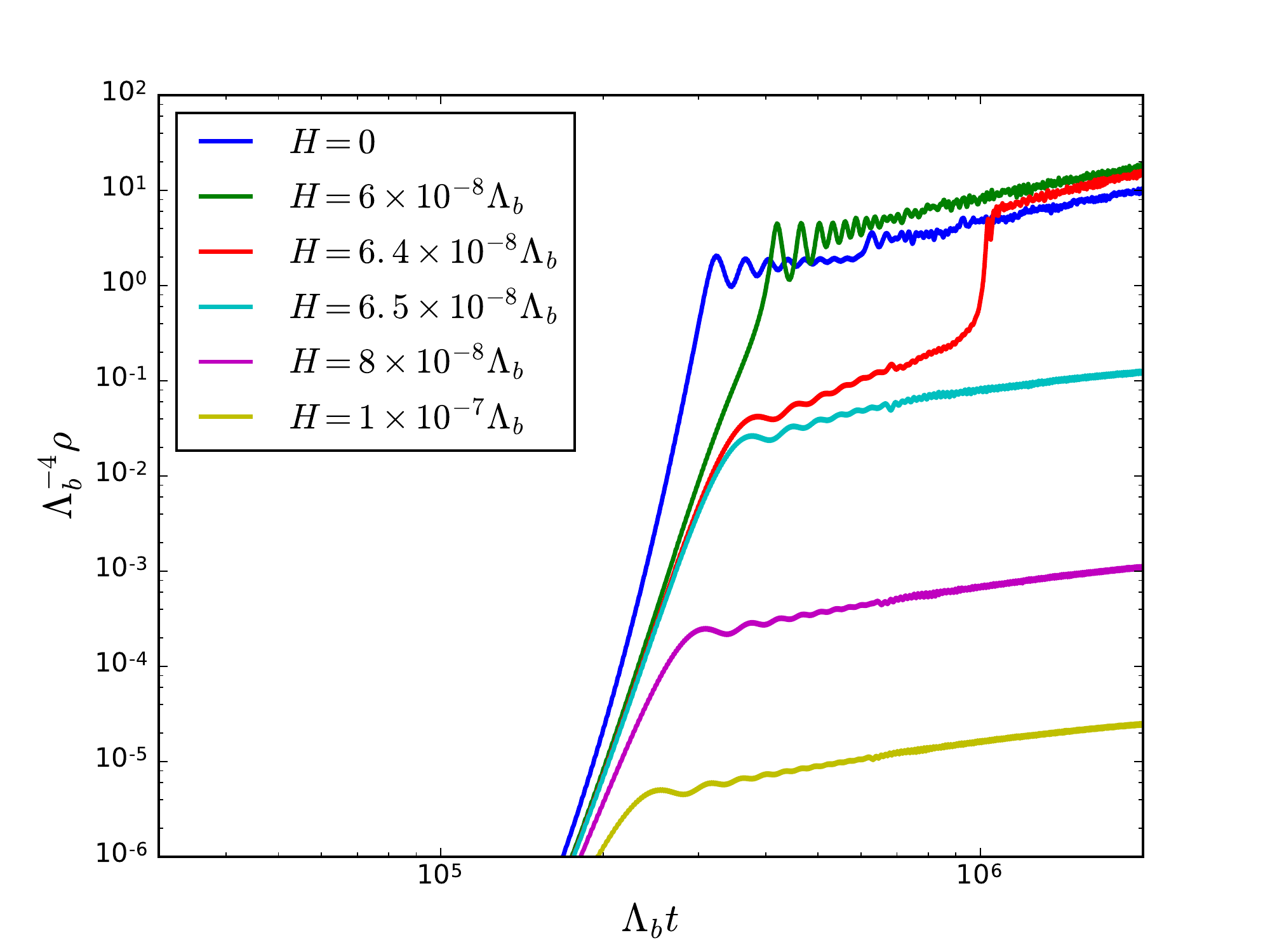}
\caption{
Time evolution of the energy of the fluctuations with $f = 1000 \Lambda_b$, $\dot\phi_0 = 10 \Lambda_b^2$, and $\mu^3 = 3H\dot\phi_0$. For small $H$, the evolution is fairly insensitive to it, while for larger values Hubble friction progressively reduces the efficiency of fragmentation. A transition is clearly visible around $H\simeq 6.4\times 10^{-8} \Lambda_b$.
}\label{fig:time evolution of fluctuations}
\end{figure}
In Fig.~\ref{fig:time evolution of energy}, we show the time evolution of the energy of zero-mode and of the fluctuations, for $\mu^3 = 0$ and $H=0$.
The total energy is conserved and the figure shows that the energy of the zero-mode is successfully transferred to the fluctuations.
In Figs.~\ref{fig:time evolution of phidot} and \ref{fig:phi doubledot as a function of phidot} we show the time evolution of $\dot\phi$ and $\ddot\phi$, again with $\mu^3=0$ and $H=0$ The numerical solution of Eqs.~(\ref{eq:zeromode}), (\ref{eq:fluctuation}) is compared to the result of Eq.~(\ref{eq:phi double dot 2}).
In Fig.~ \ref{fig:phi doubledot as a function of phidot}, we show $\ddot\phi$ as a function of $\dot\phi$ and $f$, again comparing the numerical solution with the analytical results.
Both Figs.~\ref{fig:time evolution of phidot} and \ref{fig:phi doubledot as a function of phidot} show that Eq.~(\ref{eq:phi double dot 2}) is consistent with the direct numerical calculation with Eqs.~(\ref{eq:zeromode}, \ref{eq:fluctuation}).

The effect of Hubble is shown in Fig.~\ref{fig:time evolution of fluctuations}, where we plot the time evolution of the energy of fluctuations for several value of $H$, with $\dot\phi_0 = \dot\phi_{\rm SR}$.
As long as Eq.~(\ref{eq:condition 3}) is satisfied, Eq.~(\ref{eq:eq for phi double dot}) has only one solution and $\ddot\phi$ mildly depends on $H$ in this regime.
However, when $H$ becomes larger than the critical value, the additional solutions given in Eqs.~(\ref{eq:phiddot 2}), (\ref{eq:phiddot 3}) appear.
Fig.~\ref{fig:time evolution of fluctuations} shows this transition behavior
and the fragmentation process becomes slower for large value of $H$.

\begin{figure}[ht]
\centering
\includegraphics[width=0.65\hsize]{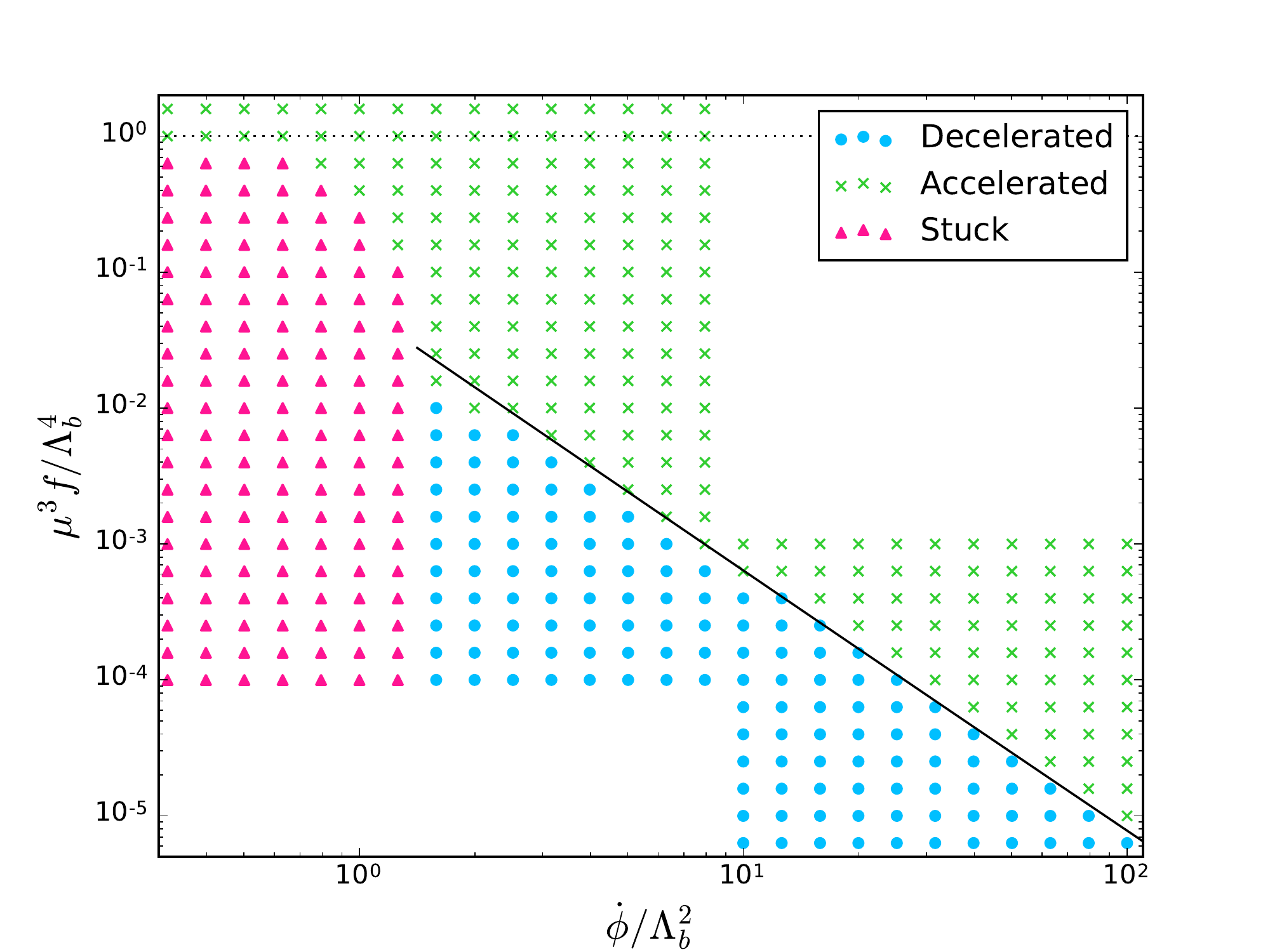}
\caption{
Phase diagram of axion fragmentation with $H=0$.
Blue: Particle production is efficient enough to stop the axion rolling.
Green: The axion is accelerated by the slope and  particle production is not efficient enough to stop its rolling.
Red: The initial kinetic energy is not large enough to overcome the first barrier.
The black line shows the condition in Eq.~(\ref{eq:slope bound (H=0)}), which reproduces the boundary between the blue and the green regions.
}\label{fig:phase diagram H=0}
\end{figure}
\begin{figure}[ht]
\centering
\includegraphics[width=0.65\hsize]{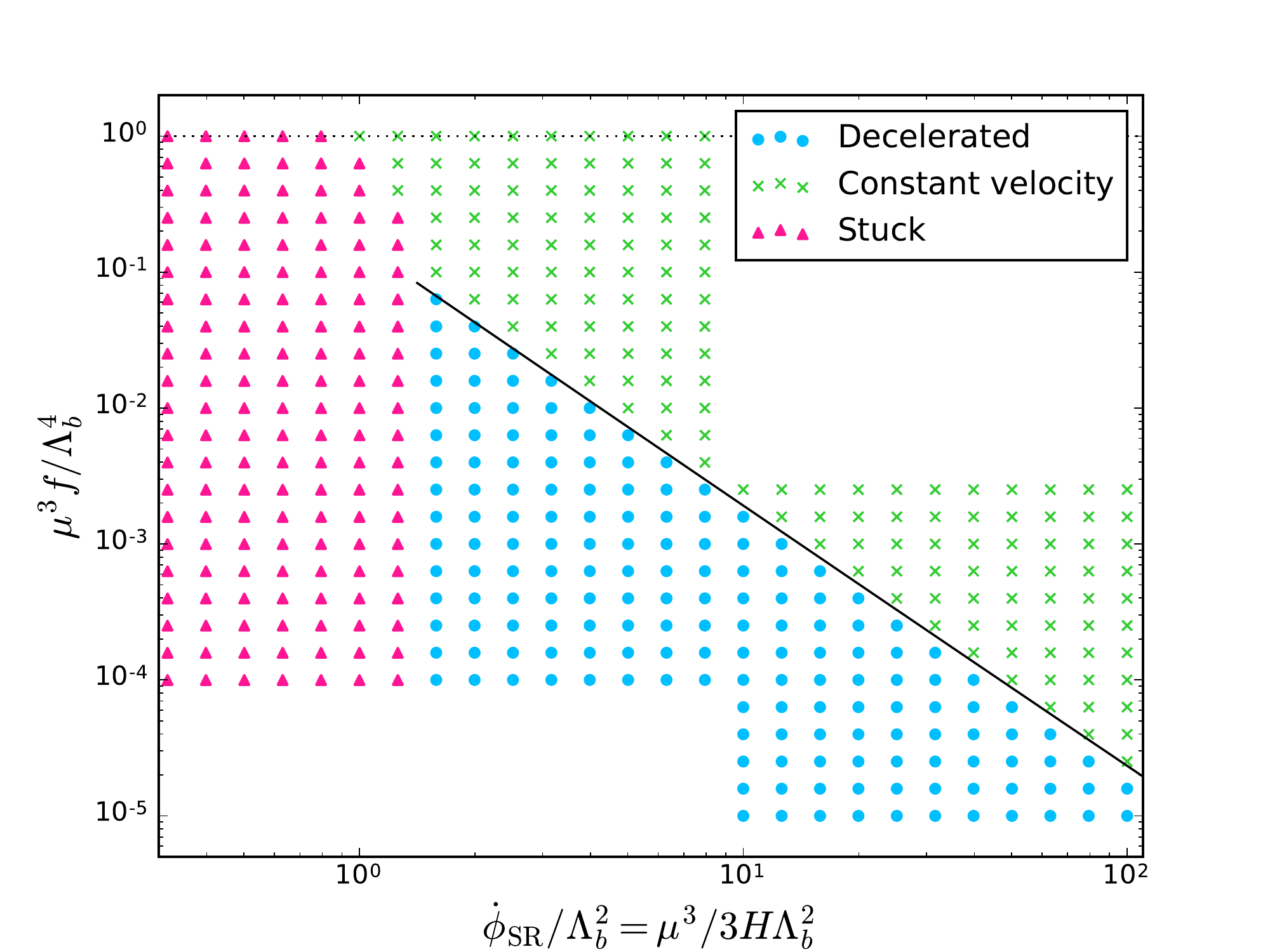}
\caption{
Phase diagram of axion particle production effect with $H\neq 0$.
The initial velocity is taken to be the slow roll velocity $\mu^3 / 3H$.
Blue: Particle production is efficient enough to stop the axion rolling.
Green: The velocity of the axion is fixed to the slow roll velocity.
Red: The initial kinetic energy is not large enough to overcome the first barrier.
The black line shows the condition in Eq.~(\ref{eq:condition 3}), which reproduces the boundary between the blue and the green regions.
}\label{fig:phase diagram H!=0}
\end{figure}
The phase diagrams of the axion particle production are shown in Figs.~\ref{fig:phase diagram H=0} and \ref{fig:phase diagram H!=0}
for general values of $\mu^3$ and $H$.
In these figures, we take $\phi = -\pi f/2$ as the initial condition so that $\Lambda_b^4 \cos \phi/f = 0$ at the beginning.
In Fig.~\ref{fig:phase diagram H=0}, we take $H=0$ and show the parameter region
in which the particle production is efficient in $\dot\phi$-$\mu^3$ plane.
The figure shows that the condition Eq.~(\ref{eq:slope bound (H=0)}) successfully reproduces the numerical result for the maximal slope $\mu^3$ that allows stopping due to fragmentation.
In Fig.~\ref{fig:phase diagram H!=0}, we take nonzero $H$ and show the parameter region
in which the particle production is efficient in the $\dot\phi_\mathrm{SR}-\mu^3$ plane.
The figure shows the excellent agreement between Eq.~(\ref{eq:condition 3}) and the numerical results.

\begin{figure}
\centering
\includegraphics[width=0.7\hsize]{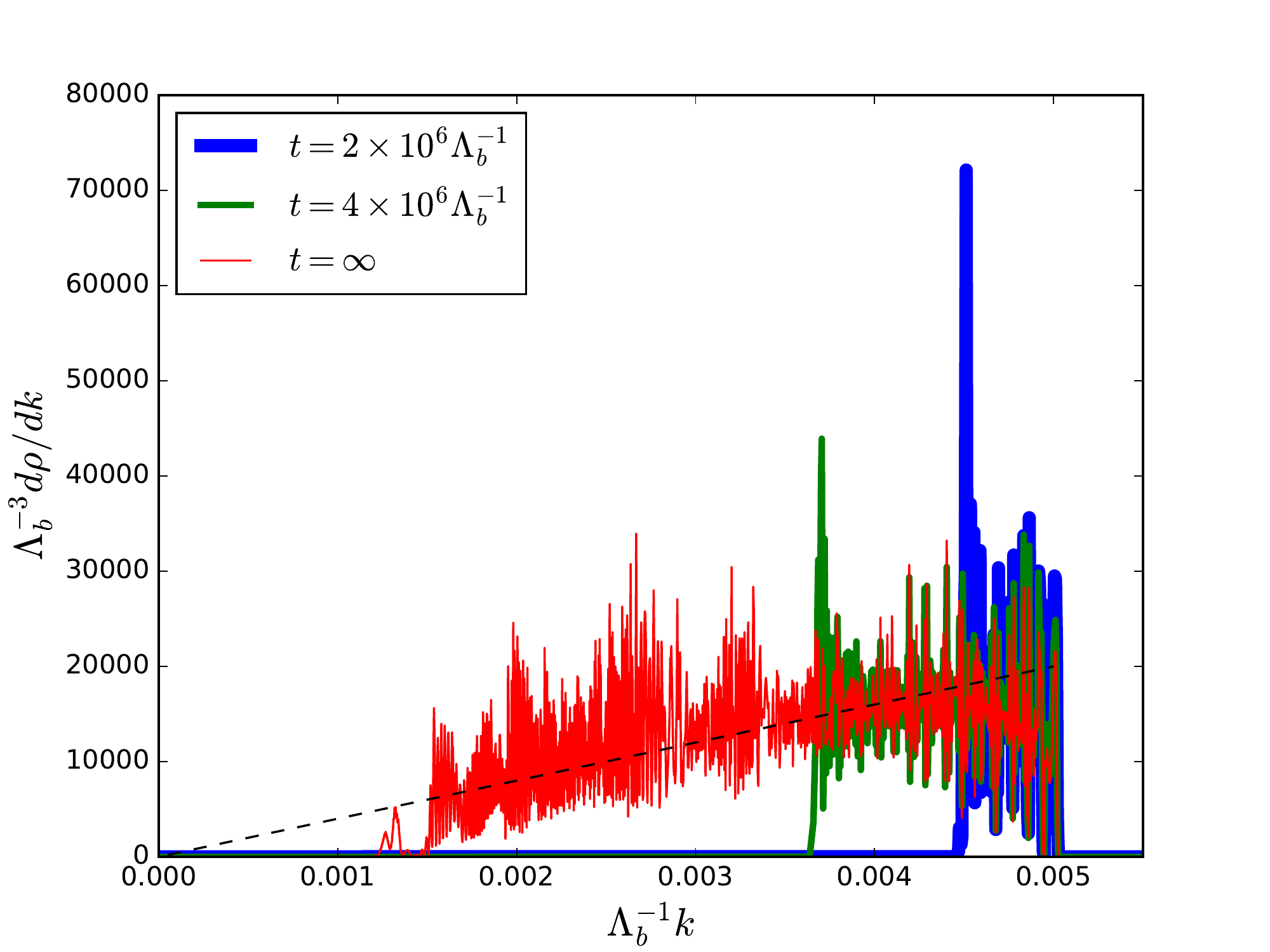}
\caption{
Energy spectrum of the fluctuations.
We took the same parameters as Fig.~\ref{fig:time evolution of phidot}.
The black line shows Eq.~(\ref{eq:drhodk}).
}\label{fig:fluctuation spectrum}
\end{figure}
In Fig.~\ref{fig:fluctuation spectrum}, we show the time evolution of the energy spectrum of the fluctuations. This quantity can be easily estimated as follows.
For $H=0$ and $\mu^3 = 0$, by using Eqs.~(\ref{eq:asymptotic uk 2}) and~(\ref{eq:phi double dot 2}), we get
\begin{align}
u_{k_\mathrm{cr}}(t) = \frac{4 \pi f}{k_\mathrm{cr}^{3/2}} \sin(k_\mathrm{cr} t+\delta).
\label{eq:uk H=0 mu3=0}
\end{align}
Then, we can calculate the energy spectrum after fragmentation as
\begin{align}
\frac{d\rho}{dk} = 4 k f^2 \,, \label{eq:drhodk}
\end{align}
where we dropped the oscillating terms.
Fig.~\ref{fig:fluctuation spectrum} shows that this estimation agrees with the result of the numerical calculation.
From Eq.~(\ref{eq:drhodk}), we see that the peak frequency coincides with the initial position of the instability band, $k_\mathrm{cr}^0 = \dot\phi_0/(2 f)$. Since $\dot\phi_0> \Lambda_b^2$, the emitted particles are typically relativistic.
We expect that non-perturbative effects will broaden this spectrum (see the discussion in the next section).

\section{Beyond the perturbative analysis}\label{sec:beyond linear}
\begin{figure}[t]
\centering
\includegraphics[width=0.8\hsize]{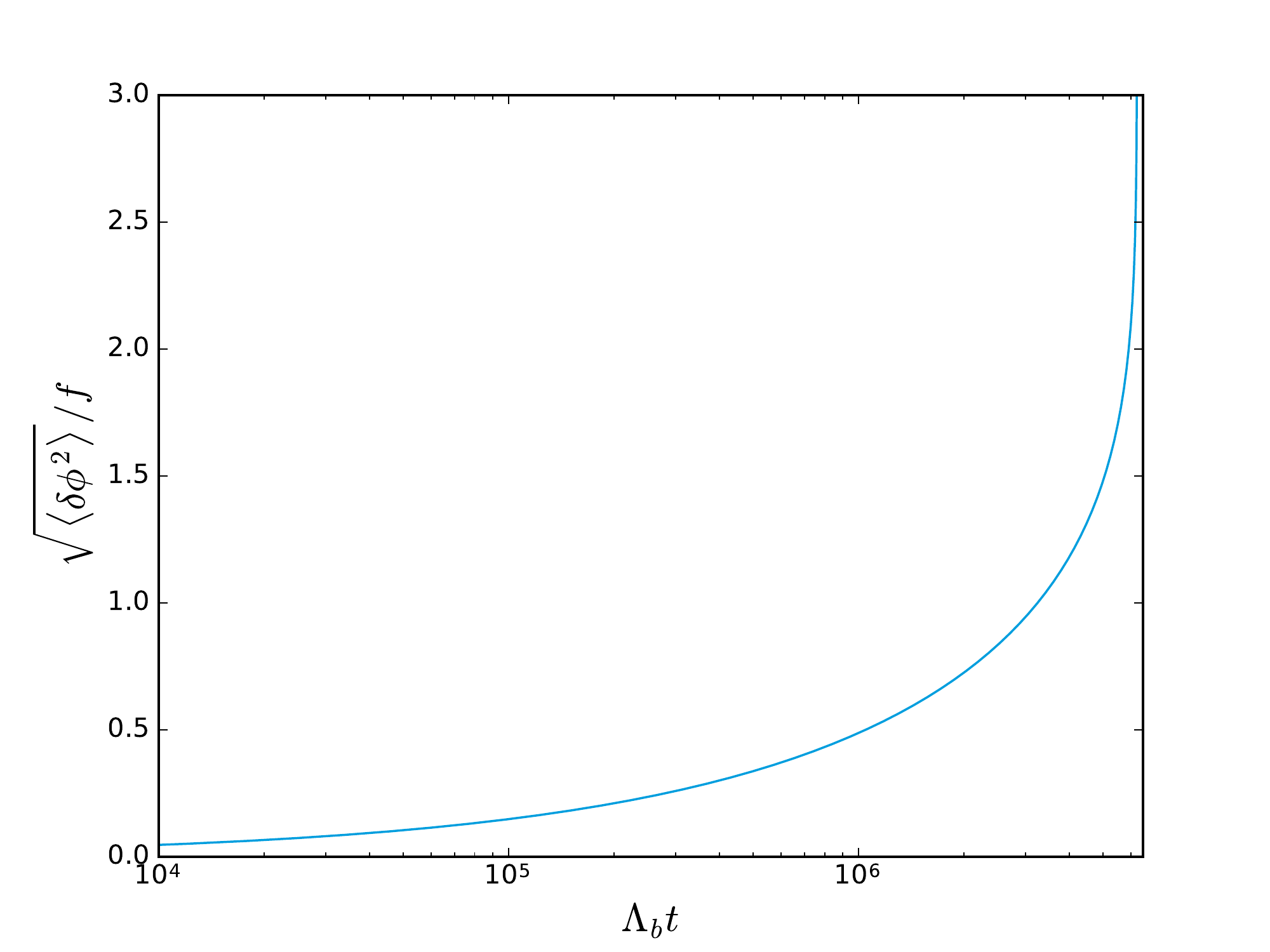}
\caption{Time evolution of $\sqrt{\langle \delta\phi^2\rangle}/f$.
The parameters are same as Fig.~\ref{fig:time evolution of phidot}.}\label{fig:deltaphi}
\end{figure}

Let us comment on the validity of the leading order expansion of $\delta\phi$.
For $H=0$ and $\mu^3=0$,
the size of fluctuation which is generated by the axion particle production
can be estimated by using
Eq.~(\ref{eq:uk H=0 mu3=0}) as
\begin{align}
\langle \delta\phi^2 \rangle
= \int^{\dot\phi_0/2f}_{\dot\phi/2f} dk \frac{4\pi k^2}{(2\pi)^3} |u_k|^2
\approx 4 f^2 \log \frac{\dot\phi_0}{\dot\phi} \,,
\end{align}
where we averaged out the oscillating term in the integral.
The time evolution of the ratio between $\sqrt{\langle\delta\phi^2\rangle}$ and $f$ is shown in Fig.~\ref{fig:deltaphi}.
We can see  that $\sqrt{\langle\delta\phi^2\rangle}$ at late times becomes large and 
we need to use non-perturbative methods for a concrete analysis in this regime.
However, Fig.~\ref{fig:deltaphi} shows that $\sqrt{\langle\delta\phi^2\rangle} / f \lesssim {\cal O}(1)$ is satisfied during most of fragmentation process.
Thus, we can expect that the estimation on the time scale Eq.~(\ref{eq:dtppapprox}) and field excursion Eq.~(\ref{eq:dphippapprox})
during fragmentation do not considerably vary from the ones obtained using a non-perturbative analysis, unless non-linear effects lead to a significant suppression of the growth of the field perturbations.
This suppression is not expected to happen in our scenario due to the periodic potential, as we elaborate in the next paragraph.

A non-linear analysis would require a dedicated lattice study, on which we are currently working~\cite{lattice}. Preliminary results indicate that our main quantitative result, namely the estimate of the total field excursion before the axion is stopped Eq.~(\ref{eq:dphippapprox}), is correct up to a factor of order 1. The main difference that we expect to emerge from a non-linear analysis is the spectrum of excited fluctuations, which will be broadened by interactions among different modes.
Moreover, the final field configuration will be inhomogeneous, with the scalar field populating more than one minima of the potential. This observation does not change significantly the picture we described so far, since the spread $\sqrt{\langle\delta\phi^2\rangle}$ will always be much smaller compared to the total field excursion $\Delta\phi_\mathrm{frag}$. Interestingly, formation of domain walls can occur in this scenario, depending on the non-linear evolution of the system and on other model dependent inputs such as the axion lifetime. We postpone this discussion to a future publication~\cite{lattice}.

A thorough comparison with other resonant systems is non-trivial, and it can hardly provide any insight on the non-linear behaviour of our model. Many models which were studied in the context of preheating feature a strong suppression of the growth at the non-linear level, generically due to the appearance of large effective mass terms. As an example, in Ref.~\cite{Prokopec:1996rr}, the resonant production of scalar particles $\chi$ from the oscillations of the inflaton $\varphi$ is studied, and it is shown that both the perturbations of the inflaton field $\delta\varphi$ (which are generated non-linearly) and the inclusion of the quartic coupling $\chi^4$, suppress the further growth of $\chi$ modes through a mass term $(\langle\delta\phi^2\rangle + \langle\chi^2\rangle)\chi^2$ (with appropriate coupling constants). Moreover, as fluctuations grow and drain energy from the zero-mode, the amplitude of the latter decreases, and hence the force driving the growth also progressively decreases.
The case under study here has two peculiarities with respect to the one above. First, since the field traverses many periods of the periodic potential, the oscillating term that stimulates the growth of fluctuations has effectively a constant amplitude. Secondly, because of the approximate shift symmetry, no effective mass scale is generated in our case. Instead, all corrections enter the equation of motions only through the cosine potential, and thus there is no reason to expect any suppression.
A setup similar to ours is discussed in~\cite{Berges:2019dgr}, in which a monodromy potential $m^2\phi^2 + \Lambda_b^4\cos\phi/f$ is studied. The paper shows that the evolution of the zero-mode stops shortly after the fluctuations have entered the non-linear regime, in accordance with our expectations.

\section{Consequences: Relaxation of the electroweak scale}\label{sec:Consequences}

The axion fragmentation dynamics explored in this work  should be taken into account in the evolution of any axion field which rolls down a wiggly potential. This phenomenon can 
fundamentally impact on a broad range of models, such as axion monodromy constructions  and relaxion scenarios.  
 In this section, we consider  the effects of  axion fragmentation on the relaxation mechanisms of the electroweak scale.

The relaxion mechanism is a solution to the electroweak hierarchy problem in which the Higgs mass term is controlled by the evolution of an axion-like field, the relaxion \cite{Graham:2015cka}. This field evolves classically in the early universe until it stops close to a critical point, defined as the field value at which the Higgs VEV is zero. A key ingredient in this picture is a potential that features periodic wiggles, similar to the one discussed in this work. Relaxion fragmentation affects this construction in a substantial way~\cite{Fonseca:2019lmc}, as we detail below
in the two main implementations of the relaxion idea which have been discussed in the literature so far.
\paragraph{Higgs dependent barriers}
In the original proposal~\cite{Graham:2015cka}, the cosine term in the relaxion potential \Eq{eq:potential} has an amplitude dependent on the Higgs VEV, $\Lambda_b^4 \propto \langle h \rangle^n$, with $n = 1, 2$. For $n=1$ (QCD relaxion), $\Lambda_b^4 \sim m_q \, \Lambda_{\textrm{QCD}}^3$ where $m_q$ is the light quark masses. For the case with $n=2$, the scale $\Lambda_b$ cannot be far from the electroweak scale, satisfying $ \Lambda_b\lesssim \mathrm{TeV}$. 
The potential contains an interaction between the Higgs and the relaxion:
\begin{equation}
V\supset -g\Lambda^3 \phi + \frac{1}{2}(\Lambda^2 - g'\Lambda\phi) h^2 +\frac{\lambda}{4}h^4 + \Lambda_b^4\cos\frac{\phi}{f} \,.
\end{equation}
Initially, the Higgs mass term $\mu_h^2\equiv \Lambda^2 - g'\Lambda\phi$ is positive, and the VEV is zero. As soon as $\phi>\Lambda/g'$, $\mu_h^2$ turns negative, a VEV develops, and the cosine term grows. In Ref.~\cite{Graham:2015cka}, it is assumed that the entire evolution takes place during a long period of inflation, and that Hubble friction is strong enough to stop the field as soon as the wiggles become larger than the average slope and the potential develop local minima, \textit{i.e.} for $g \Lambda^3 \approx \Lambda_b^4 / f$. In particular, this happens when the time that it takes to roll over one period of the cosine term is longer than one Hubble time, \textit{i.e.} for
\begin{equation}\label{eq:slow roll 1 period}
\Delta t_1 = \frac{2\pi f}{g\Lambda^3/3 H} > H^{-1} \,.
\end{equation}

Relaxion fragmentation offers an additional source of friction for the relaxion rolling. As discussed in Ref.~\cite{Fonseca:2019lmc}, this opens up two possibilities: on the one hand, the relaxion can be stopped by fragmentation even when \Eq{eq:slow roll 1 period} is not satisfied.
On the other hand, it is possible to stop the relaxion field with a much shorter period of inflation or even in the absence of an inflationary background, with a negligible Hubble friction. 
This opens new possibilities for relaxion model building, independent from constraints on the inflationary sector. If relaxation takes place after inflation, it is possible, at least in principle, to conceive a model in which this phase has observable features, most probably in gravitational waves.This study could open the way to observable relaxion models.

\paragraph{Higgs independent barriers}
An alternative relaxion construction was proposed in~\cite{Hook:2016mqo}, in which the amplitude of the cosine term is Higgs-independent, and the friction is mainly provided by gauge boson particle production.
The relaxion couples to the Chern-Simons term of the massive SM $Z$ boson, through a term
\begin{equation}
\frac{\phi}{F} Z_{\mu\nu} \widetilde Z^{\mu\nu} \,.
\end{equation}
In the presence of this coupling, the equation of motion for the transverse polarization of the $Z$ has a tachyonic instability for small mass $m_Z$
\begin{equation}
\ddot Z_\pm + \left(k^2 + m_Z^2 \pm k \frac{\dot\phi}{F} \right) Z_\pm = 0
\end{equation}
Contrarily to the case discussed above, initially the Higgs has a large VEV, and the SM particles are heavy. As the relaxion approaches the critical point and the gauge bosons become lighter, the tachyonic instability is triggered and the relaxion kinetic energy is dissipated through the production of $Z$ bosons.

Fragmentation poses a serious threat to this model \cite{Fonseca:2019lmc}. Since the amplitude of the cosine term is constant, fragmentation is always active, and the relaxion can be slowed down and stopped when the Higgs mass is large and close to the cut-off $\Lambda$, thus spoiling the successful relaxation of the Higgs VEV to its current value.
In particular,
\begin{itemize}
\item if relaxation takes place after inflation~\cite{Fonseca:2018xzp}, the parameter space is restricted by the condition of avoiding excessive fragmentation. Moreover, once cosmological constraints are taken into account, the mechanism is excluded at least for a cutoff larger than few TeV.
\item If relaxation happens during inflation, the constraints from fragmentation reduce the available parameter space but do not exclude the model. The dark matter scenario discussed in~\cite{Fonseca:2018kqf}, in particular, is not affected.
\end{itemize}

\section{Summary and outlook}\label{sec:conclusions}
In this paper, we discussed the production of quantum fluctuations 
during the evolution of  an axion-like field rolling  down a potential featuring 
wiggles,  as given in Eq.~(\ref{eq:potential}). We refer to this effect as axion fragmentation.
While the production of quanta is suppressed when an axion oscillates around the minimum of its potential, unless the initial amplitude is very large and the initial position of the field is tuned close to the maximum of the sinusoidal potential, the effect is very large in the case where the axion field crosses many of its maxima.
We studied in detail under which conditions axion fragmentation can efficiently stop the evolution of the field.
We computed the time scale needed for stopping and the corresponding field excursion.

The wavefunction of the fluctuations obeys the Mathieu equation
and the energy of the modes within the instability band Eq.~(\ref{eq:instability band}) grows exponentially.
If both the slope of the potential and the Hubble expansion rate are sufficiently small, 
this particle production effect decelerates the homogeneous mode.
The condition is given as \Eq{eq:condition 0'} and \Eq{eq:condition for no positive phiddot} in terms of the initial field velocity, the linear slope of the potential, the Hubble rate, the size of the barriers and the periodicity of the sinusoidal potential.
The corresponding acceleration $\ddot\phi$ is given in \Eq{eq:solution phiddot}.

Axion fragmentation is a generic effect 
which can have interesting phenomenological implications.
It is particularly relevant for the mechanism of cosmological relaxation of the electroweak scale.
We  dedicate a separate paper to study in details these implications in Ref.~\cite{Fonseca:2019lmc}, where we conclude that new regions of parameter space open and novel directions for relaxion model building are offered by this effect.

In the present work we study the regime in which the potential has local minima, see \Eq{eq:local minima}. It would be interesting to investigate the effect of fragmentation in the case in which the oscillating term in potential does not generate local minima. This may have implications for some relaxion models where loop effects generate small Higgs-independent barriers, so that there are constant wiggles with small amplitude during the whole scanning of the Higgs mass parameter.

Another promising direction will be to explore the impact of axion quanta on the cosmological history of the universe. As discussed in Ref.~\cite{Fonseca:2019lmc}, depending on the equation of the state of the universe during axion rolling, the produced quanta may represent  a significant fraction of the energy density of the universe. Whether they can be viable dark matter candidates,  depends on the time of fragmentation. Such quanta  may in turn induce gravitational waves. 
They may be diluted or leave observable imprints.
These effects deserve detailed studies which we postpone for future work.

\section*{Acknowledgements}
The authors thank Yohei Ema, Hyungjin Kim, Kyohei Mukaida, and Gilad Perez, Alexander Westphal for useful discussions. We are grateful to Sven Krippendorf for important discussions in the initial stages of this work.
This work is supported by the Deutsche Forschungsgemeinschaft under Germany’s Excellence Strategy - EXC 2121 ``Quantum Universe'' - 390833306. Research in Mainz is supported by the Cluster of Excellence ``Precision Physics, Fundamental Interactions, and Structure of Matter'' (PRISMA+ EXC 2118/1) funded by  the  German  Research  Foundation(DFG)  within  the German  Excellence  Strategy  (Project  ID  39083149).

\appendix

\section{Approximate solution of Eq.~(\ref{eq:fluctuation})}\label{sec:technical details}
Let us discuss the evolution of the wave functions $u_k$.
The boundary condition at $t\to-\infty$ is given by Eq.~(\ref{eq:initial}).
For the duration of the amplification process, we can neglect the Hubble friction term $3 H \dot u_k$ in the equation of motion. This is justified for $k \simeq \dot\phi/2f$ if $H \ll \Lambda_b^4/\dot\phi f$ is satisfied (see the discussion around Eq.~(\ref{eq:condition 0})).
The mode function $u_k$ satisfies the following equation of motion:
\begin{align}
\ddot u_k + \left( a^{-2} k^2 - \frac{\Lambda_b^4}{f^2}\cos\frac{\phi}{f} \right) u_k = 0. \label{eq:EOM uk}
\end{align}
Note that the Hubble expansion still has its effect via $a^{-2} k^2$ term although we have dropped $3H\dot u_k$.

\subsection{$\ddot{\phi} = 0$ and $H=0$ (Mathieu equation)}

First, let us consider the case of constant $\dot{\phi}$ and $H=0$ as in Sec.~\ref{sec:intuitive derivation}.
Taking the scale factor $a=1$, Eq.~(\ref{eq:EOM uk}) is simplified as
\begin{align}
\ddot u_k + \left( k^2 - \frac{\Lambda_b^4}{f^2} \cos \frac{\dot{\phi}}{f}t \right) u_k = 0. \label{eq:mathieu}
\end{align}
This is the Mathieu equation, and it is known that $u_k$ grows exponentially when $k$ satisfies
\begin{align}
\frac{\dot{\phi}^2}{4f^2} - \frac{\Lambda_b^4}{2 f^2} 
<
k^2
<
\frac{\dot{\phi}^2}{4f^2} + \frac{\Lambda_b^4}{2 f^2} .
\end{align}
Let us see this behavior explicitly.
We define  $\delta$ and $\epsilon$ for convenience:
\begin{align}
\delta \equiv \frac{2f^2}{\Lambda_b^4}\left( k^2 - \frac{\dot{\phi}^2}{4f^2} \right), \quad
\epsilon \equiv \frac{\Lambda_b^4}{\dot{\phi}^2}.
\end{align}
Then, Eq.~(\ref{eq:mathieu}) is rewritten as
\begin{align}
\ddot u_k + \frac{\dot{\phi}^2}{4f^2} \left[ 1 + \epsilon \left( 2\delta - 4\cos\frac{\dot{\phi}}{f}t \right) \right] u_k = 0. \label{eq:mathieu2}
\end{align}
We assume $\epsilon \ll 1$ and we will expand perturbatively in $\epsilon$.
On the other hand, we assume $\delta = {\cal O}(1)$, as we are interested in the first instability band. In the limit $\epsilon = 0$, the solution is of the form $u_k = a \cos(\dot{\phi}t/2) + b \sin(\dot{\phi}t/2)$ for constant $a$ and $b$.
This motivates the following ansatz for $u_k$:
\begin{align}
u_k = \frac{1}{\sqrt{2k}} \sum_{n=1}^\infty \left[ A_n(t) \cos\frac{2n-1}{2} \frac{\dot{\phi}}{f}t + B_n(t) \sin\frac{2n-1}{2} \frac{\dot{\phi}}{f}t \right]. \label{eq:uk mathieu}
\end{align}
Here $\cos$ and $\sin$ encode the rapid oscillations of the solution, while $A_n$'s and $B_n$'s are slowly varying coefficients.
The terms with $n\geq 2$ are introduced to maintain a consistency with Eq.~(\ref{eq:mathieu2}).
We define a dimensionless time $\tau$:
\begin{align}
\tau \equiv \frac{\Lambda_b^4 t}{2\dot{\phi}f}. \label{eq:tau}
\end{align}
By plugging Eqs.~(\ref{eq:uk mathieu}, \ref{eq:tau}) into Eq.~(\ref{eq:mathieu}), we obtain the following differential equations for $A_n$'s and $B_n$'s:
\begin{align}
(-1+\delta)A_1 - A_2 + \frac{dB_1}{d\tau} + \frac{1}{2}\epsilon \frac{d^2A_1}{d\tau^2} & = 0, \label{eq:mathieu A1 full}\\
(-1-\delta)B_1 + B_2 + \frac{dA_1}{d\tau} - \frac{1}{2}\epsilon \frac{d^2B_1}{d\tau^2} & = 0 \label{eq:mathieu B1 full}
\end{align}
for $n=1$ and, for $n\geq2$,
\begin{align}
(n-1)n A_n + \epsilon \left( -\frac{1}{2}(2n-1)\frac{dB_n}{d\tau} + \frac{1}{2}A_{n-1} - \frac{1}{2}\delta A_n + \frac{1}{2}A_{n+1}  \right) - \frac{1}{4}\epsilon^2 \frac{d^2A_n}{d\tau^2}  & = 0, \\
(n-1)n B_n + \epsilon \left( \frac{1}{2}(2n-1)\frac{dA_n}{d\tau} + \frac{1}{2}B_{n-1} - \frac{1}{2}\delta B_n + \frac{1}{2}B_{n+1}  \right) - \frac{1}{4}\epsilon^2 \frac{d^2B_n}{d\tau^2}  & = 0.
\end{align}
The differential equations above indicate a simple hierarchy between the coefficients
\begin{align}
A_n,~B_n \sim (A_1,~B_1) \times \epsilon^{n-1}.
\end{align}
At the leading order in $\epsilon$, we can neglect $A_n$ and $B_n$ with $n\geq 2$, and also the second derivative in Eqs.~(\ref{eq:mathieu A1 full}), (\ref{eq:mathieu B1 full}). We obtain
\begin{align}
\left( -1 + \delta \right) A_1 + \frac{dB_1}{d\tau} &= 0, \label{eq:mathieu A1}\\
\left( -1 - \delta \right) B_1 + \frac{dA_1}{d\tau} &= 0. \label{eq:mathieu B1}
\end{align}
For $-1<\delta < 1$, the asymptotic behavior of $A_1$ and $B_1$ are then
\begin{align}
A_1(t)  , B_1(t) \sim
\exp\left( \sqrt{ 1- \delta^2 } \tau \right)
= \exp\left( \sqrt{ 1- \delta^2 } \frac{\Lambda_b^4 t}{2\dot{\phi} f} \right).
\end{align}
For $|\delta|<1$ the solution is unstable. On the contrary, for $\delta^2 > 1$ it oscillates as $\sin ( \sqrt{\delta^2 -1} \Lambda_b^4 t/2\dot{\phi}f  ) $.

\subsection{Small non-zero $\ddot{\phi}$ and $H$}

Let us now introduce a small, constant acceleration $\ddot{\phi}$:
\begin{align}
\dot\phi = \dot\phi(0) + \ddot\phi t,
\end{align}
where $\dot\phi(0) = \dot\phi(t=0)$.
The assumption of neglecting higher derivatives is justified in the main text, see the discussion around Eq.~(\ref{eq:constant ddot phi}). Let us consider a given mode $k$.
By a simple time shift, here we define $t=0$ as the time at which $k$ is at the center of the instability band, which is now defined as
\begin{equation}
\left[ \frac{k^2}{a^2}=\left(\frac{\dot\phi(t)}{2f}\right)^2 \right]_{t=0} \,.
\end{equation}
As the velocity decreases, the instability band moves to lower modes. For a given $k$, we will solve the equations of motion from a time slightly before it enters the instability band, until slightly after it exits, and we will see how the initial oscillatory behaviour is then amplified inside the instability band until the mode exits.
We assume for $u_k$ an ansatz similar to \Eq{eq:uk mathieu}, but this time the cosine will depend on $\phi(t)$:
\begin{align}
u_k = \frac{1}{a \sqrt{2k}} \sum_{n=1}^\infty \left[ A_n(t) \cos\frac{2n-1}{2} \frac{\phi}{f} + B_n(t) \sin\frac{2n-1}{2} \frac{\phi}{f} \right].
\label{eq:uk expansion}
\end{align}
Again, we only consider the case in which the velocity is larger than the wiggles, $\epsilon \equiv \Lambda_b^4/\dot\phi^2 \ll 1$. In order to keep the evolution under perturbative control, we assume that the time spend in the instability band is short, and that the velocity changes only slightly during this time. Furthermore, we assume that the effect of Hubble friction is small:
\begin{equation}\label{eq:assumptions acceleration hubble}
\left|\frac{t \, \ddot \phi}{\dot\phi} \right| \ll 1 \,, \qquad
H t \ll 1 \,.
\end{equation}
In Eq.~(\ref{eq:uk expansion}) we make explicit the $a^{-1}$ dependence of the mode functions on the scale factor. When taking the derivatives $\dot u_k$, $\ddot u_k$ we will keep this factor as a constant, consistently with the assumption that the amplification time for any mode $k$ is much shorter than a Hubble time.
By plugging the above $u_k$ into the equation of motion Eq.~(\ref{eq:EOM uk}), we obtain the equations for the coefficients of the sine and cosine terms.
Similarly to before, the $n\geq 2$ equations show an hierarchy $A_n,~B_n \sim (A_1,~B_1) \times \epsilon^{n-1}$. This can be extracted from the equations of motion or by analogy with the constant velocity case, remembering that the acceleration gives only a small correction to the evolution according to Eq.~(\ref{eq:assumptions acceleration hubble}). Thus, we can neglect the contribution from the $n\geq 2$ terms.
For $n=1$ we obtain the equation for the coefficients of $\cos\phi/2f$ and $\sin\phi/2f$:
\begin{align}
\left( a^{-2} k^2 - \frac{\dot\phi^2}{4f^2} - \frac{\Lambda_b^4}{2f^2} \right) A_1 + \frac{d^2A_1}{dt^2} + \frac{\dot\phi}{f}\frac{dB_1}{dt} + \frac{\ddot\phi B_1}{2f} &= 0, \\
-\left( a^{-2} k^2 - \frac{\dot\phi^2}{4f^2} + \frac{\Lambda_b^4}{2f^2} \right) B_1 - \frac{d^2B_1}{dt^2} + \frac{\dot\phi}{f}\frac{dA_1}{dt} + \frac{\ddot\phi A_1}{2f} &= 0.
\end{align}
By using the results in the previous section,
we can estimate that $d^2A_1/dt^2,d^2B_1/dt^2  \sim (\Lambda_b^4/\dot\phi f)dA_1/dt, (\Lambda_b^4/\dot\phi f)dB_1/dt$.
Thus, by assuming $\Lambda_b^4 \ll \dot\phi^2$, we can drop $d^2A_1/dt^2$ and $d^2B_1/dt^2$ in the above equations.
We require the time evolution of $\dot\phi$ is sufficiently slower than that of $A_1$ and $B_1$,
\textit{i.e.}, $|\ddot\phi / \dot\phi| \ll |A_1^{-1} dA_1/dt|,~ |B_1^{-1} dB_1/dt|$.
This condition can be understood as follows.
The time evolution of the energy of zero mode is given by $d \rho_0 / dt = \dot\phi \ddot\phi = \rho_0 (\ddot\phi / \dot\phi)$.
The energy stored in the fluctuations inside the instability band is
\begin{equation}
\delta\rho_\mathrm{fluc} \sim \frac{d\rho_\mathrm{fluc}}{dk} \delta k \sim (k^4 |u_k|^2 + k^2 |\dot u_k|^2) \delta k.
\end{equation}
Thus, dropping second time derivatives, the growth of $\delta\rho_\mathrm{fluc}$ is given by
\begin{equation}
\frac{d \delta\rho_\mathrm{fluc}}{dt} \sim  \delta\rho_\mathrm{fluc} \frac{1}{(A_1+B_1)} \left(\frac{dA_1}{dt} + \frac{dB_1}{dt} \right)
\end{equation}
The growth in energy of the modes inside the instability band correspond to the slow-down of the zero mode. Thus, $d \rho_0 / dt \sim d \delta\rho_\mathrm{fluc} / dt$, and we obtain
$\ddot\phi/\dot\phi \sim (\dot A_1+\dot B_1) / (A_1+B_1) \, \delta\rho_\mathrm{fluc}/\rho_0$.
Due to the smallness of the band width $\delta k$, $\delta\rho_\mathrm{fluc}/\rho_0 \ll 1$, and hence $\ddot\phi/\dot\phi \ll (\dot A_1+\dot B_1) / (A_1+B_1)$.
The last condition can be rewritten as
\begin{equation}\label{eq:ddot phi ll Lambda_b}
\ddot\phi \ll \frac{\Lambda_b^4}{f}\,,
\end{equation}
which implies $\ddot\phi\ll\dot\phi^2/f$. Thus, we can drop the terms $\ddot\phi A_1/2f$ and $\ddot\phi B_1/2f$.
Finally, we define the auxiliary quantities
\begin{align}
\tau = \frac{\Lambda_b^4 t}{2\dot{\phi}(t) f}, \qquad
z = -\frac{2 f \dot\phi^2(0) (\ddot\phi + H\dot\phi(0))}{\Lambda_b^8}.
\label{eq:def of tau and z}
\end{align}
where $k/a = \dot\phi(0)/2f$.
Furthermore, since we assume the Hubble expansion rate is sufficiently small as explained in Eq.~(\ref{eq:assumptions acceleration hubble}),
we expand the scale factor as $a^{-2}(t) \simeq a^{-2}(0)(1 - 2Ht)$.
In terms of $\tau$ and $z$, we obtain
\begin{align}
(-1 + z\tau)A_1 + \frac{dB_1}{d\tau} &= 0\,, \label{eq:diff eq A1}\\
(-1 - z\tau)B_1 + \frac{dA_1}{d\tau} &= 0\,. \label{eq:diff eq B1}
\end{align}

\subsubsection{Generic solution for Eqs.~(\ref{eq:diff eq A1}, \ref{eq:diff eq B1})}

Let us discuss the solution of Eqs.~(\ref{eq:diff eq A1}, \ref{eq:diff eq B1}) for $z \ll 1$.
Eqs.~(\ref{eq:diff eq A1}, \ref{eq:diff eq B1}) can be rewritten as
\begin{align}
\frac{d^2 A_1}{d\tau^2} - \frac{z}{1+z\tau} \frac{dA_1}{d\tau} + (z^2\tau^2 - 1) A_1= 0, \label{eq:diff eq A1 2}\\
\frac{d^2 B_1}{d\tau^2} + \frac{z}{1-z\tau} \frac{dB_1}{d\tau} + (z^2\tau^2 - 1) B_1= 0. \label{eq:diff eq B1 2}
\end{align}
In the case of $1+z\tau,~1-z\tau \gtrsim {\cal O}(1)$, we can use WKB approximation for these differential equations.
$A_1$ can be written as $A_1 = \exp\left( \alpha_0(\tau) + z \, \alpha_1(\tau) + \ldots \right)$
where $\alpha_0',\alpha_1' = {\cal O}(1)$ and $\alpha_0'',\alpha_1'' = {\cal O}(z)$.
At leading and next to leading order in $z$, we obtain the following equations from Eq.~(\ref{eq:diff eq A1 2}):
\begin{align}
(\alpha_0')^2 + (z^2\tau^2-1) &= 0, \label{eq:WKB alpha 0} \\
2 z \alpha_0' \alpha_1' + \alpha_0'' - \frac{z}{1+z\tau} \alpha_0' &= 0. \label{eq:WKB alpha 1}
\end{align}
Solutions of the above equation are
\begin{align}
\alpha_0 = \pm i \int^\tau d\tau' \sqrt{z^2\tau'^2-1}, \qquad
\alpha_1 = \frac{1}{4z} \log \left( \frac{z\tau+1}{z\tau-1} \right) \qquad (|z\tau| > 1)\\
\alpha_0 = \pm \int^\tau d\tau' \sqrt{1-z^2\tau'^2}, \qquad
\alpha_1 = \frac{1}{4z} \log \left( \frac{1+z\tau}{1-z\tau} \right) \qquad (|z\tau| < 1)
\end{align}
Similarly, by plugging $B_1 = \exp\left( \beta_0(\tau) + z \beta_1(\tau) + \cdots \right)$ into Eq.~(\ref{eq:diff eq B1 2}),
we obtain
\begin{align}
\beta_0 = \pm i  \int^\tau d\tau' \sqrt{z^2\tau'^2-1}, \qquad
\beta_1 = \frac{1}{4z} \log \left( \frac{z\tau-1}{z\tau+1} \right) \qquad (|z\tau| > 1)\\
\beta_0 = \pm \int^\tau d\tau' \sqrt{1-z^2\tau'^2}, \qquad
\beta_1 = \frac{1}{4z} \log \left( \frac{1-z\tau}{1+z\tau} \right) \qquad (|z\tau| < 1)
\end{align}
The generic solution for $z\tau < -1$ is given as
\begin{align}
A_1 =
& \left( \frac{-z\tau-1}{-z\tau+1} \right)^{1/4}
\left[ c_+ \exp\left(  i\int_{-1/z}^\tau d\tau' \sqrt{z^2 \tau'^2 - 1} \right) 
     + c_- \exp\left( -i\int_{-1/z}^\tau d\tau' \sqrt{z^2 \tau'^2 - 1} \right) \right], \label{eq:generic A1 1}\\
B_1 =
& \left( \frac{-z\tau+1}{-z\tau-1} \right)^{1/4}
\left[ - ic_+ \exp\left(  i\int_{-1/z}^\tau d\tau' \sqrt{z^2 \tau'^2 - 1} \right) 
       + ic_- \exp\left( -i\int_{-1/z}^\tau d\tau' \sqrt{z^2 \tau'^2 - 1} \right) \right]. \label{eq:generic B1 1}
\end{align}
For $|z\tau| < 1$,
\begin{align}
A_1 =
& \left( \frac{1+z\tau}{1-z\tau} \right)^{1/4}
\left[ c'_+ \exp\left(  \int_{-1/z}^\tau d\tau' \sqrt{1-z^2 \tau'^2} \right) 
     + c'_- \exp\left( -\int_{-1/z}^\tau d\tau' \sqrt{1-z^2 \tau'^2} \right) \right], \label{eq:generic A1 3}\\
B_1 =
& \left( \frac{1-z\tau}{1+z\tau} \right)^{1/4}
\left[ c'_+ \exp\left(  \int_{-1/z}^\tau d\tau' \sqrt{1-z^2 \tau'^2} \right) 
     - c'_- \exp\left( -\int_{-1/z}^\tau d\tau' \sqrt{1-z^2 \tau'^2} \right) \right]. \label{eq:generic B1 3}
\end{align}
For $z\tau > 1$,
\begin{align}
A_1 =
& \left( \frac{z\tau+1}{z\tau-1} \right)^{1/4}
\left[ c''_+ \exp\left(  i\int_{-1/z}^\tau d\tau' \sqrt{z^2 \tau'^2 - 1} \right) 
     + c''_- \exp\left( -i\int_{-1/z}^\tau d\tau' \sqrt{z^2 \tau'^2 - 1} \right) \right], \label{eq:generic A1 2}\\
B_1 =
& \left( \frac{z\tau-1}{z\tau+1} \right)^{1/4}
\left[ ic''_+ \exp\left(  i\int_{-1/z}^\tau d\tau' \sqrt{z^2 \tau'^2 - 1} \right) 
     - ic''_- \exp\left( -i\int_{-1/z}^\tau d\tau' \sqrt{z^2 \tau'^2 - 1} \right) \right], \label{eq:generic B1 2}
\end{align}
In Eqs.~(\ref{eq:generic A1 1}-\ref{eq:generic B1 3}), $c$, $c'$, and $c''$ are constants,
and the relative factors between $A_1$ and $B_1$ are determined to satisfy Eqs.~(\ref{eq:diff eq A1}, \ref{eq:diff eq B1}).

The WKB approximation fails for $\tau \simeq \pm 1/z$, as can be seen for example from the last term in Eq.~(\ref{eq:WKB alpha 1}) becoming of order $\mathcal{O}(1)$.
For $\tau \simeq -1/z$, Eqs.~(\ref{eq:diff eq A1}, \ref{eq:diff eq B1}) can be approximated as
\begin{align}
-2 A_1 + \frac{dB_1}{d\tau} = 0 \,, \label{eq:Airy A1} \\
(-1-z\tau) B_1 + \frac{dA_1}{d\tau} = 0\, . \label{eq:Airy B1}
\end{align}
Eq.~(\ref{eq:Airy B1}) can be rewritten as an Airy/Stokes equation, the solution of which can be written in terms of the Airy functions $\mathrm{Ai}$, $\mathrm{Bi}$ as
\begin{equation}
B_1 \simeq c_A \, \mathrm{Ai}\left[(2z)^{1/3}\left(\tau+\frac{1}{z}\right)\right] + c_B \, \mathrm{Bi}\left[(2z)^{1/3}\left(\tau+\frac{1}{z}\right)\right], \label{eq:generic B1 Airy1}
\end{equation}
where $c_A$ and $c_B$ are constants. $A_i$ can be obtained using \Eq{eq:Airy A1}:
\begin{equation}
A_1  \simeq 2^{-2/3} z^{1/3}  \left\{ c_A \, \mathrm{Ai}'\left[(2z)^{1/3}\left(\tau+\frac{1}{z}\right)\right] + c_B \, \mathrm{Bi}'\left[(2z)^{1/3}\left(\tau+\frac{1}{z}\right)\right] \right\}, \label{eq:generic A1 Airy1}\\
\end{equation}
Here $\mathrm{Ai}'(x) = d\mathrm{Ai}/dx$ and $\mathrm{Bi}'(x) = d\mathrm{Bi}/dx$.
Similarly, we obtain the generic solution for $\tau \simeq 1/z$:
\begin{align}
A_1 &\simeq c'_A \, \mathrm{Ai}\left[-(2z)^{1/3}\left(\tau-\frac{1}{z}\right)\right] + c'_B \, \mathrm{Bi}\left[-(2z)^{1/3}\left(\tau-\frac{1}{z}\right)\right] \,, \label{eq:generic A1 Airy2}\\
B_1 &\simeq -2^{-2/3} z^{1/3}  \left\{ c'_A \, \mathrm{Ai}'\left[-(2z)^{1/3}\left(\tau-\frac{1}{z}\right)\right] + c'_B \, \mathrm{Bi}'\left[-(2z)^{1/3}\left(\tau-\frac{1}{z}\right)\right] \right\} \,. \label{eq:generic B1 Airy2}
\end{align}
where $c'_A$ and $c'_B$ are constants.

Now we have the generic solution of Eqs.~(\ref{eq:diff eq A1}, \ref{eq:diff eq B1})
in the following five separate region of $\tau$:
\begin{itemize}
\item $\tau < -1/z$: Eqs.~(\ref{eq:generic A1 1}, \ref{eq:generic B1 1})
\item $\tau \simeq -1/z$: Eqs.~(\ref{eq:generic A1 Airy1}, \ref{eq:generic B1 Airy1})
\item $-1/z < \tau < 1/z$: Eqs.~(\ref{eq:generic A1 3}, \ref{eq:generic B1 3})
\item $\tau \simeq 1/z$: Eqs.~(\ref{eq:generic A1 Airy2}, \ref{eq:generic B1 Airy2})
\item $1/z < \tau$: Eqs.~(\ref{eq:generic A1 2}, \ref{eq:generic B1 2})
\end{itemize}
To connect each region of $\tau$, we will need the asymptotic form of Airy functions:
\begin{equation}
\begin{aligned}
{\rm Ai}(x) & \simeq \frac{1}{2\sqrt{\pi} \, x^{1/4}} \exp\left( -\frac{2}{3} x^{3/2} \right) \\
{\rm Bi}(x) & \simeq \frac{1}{\sqrt{\pi} \, x^{1/4}} \exp\left( \frac{2}{3} x^{3/2} \right),
\end{aligned}
\qquad
\textrm{for } x\gg 1,
\end{equation}
and
\begin{equation}
\begin{aligned}
{\rm Ai}(x) & \simeq \frac{1}{\sqrt{\pi} (-x)^{1/4}} \sin\left( \frac{2}{3} (-x)^{3/2} + \frac{\pi}{4} \right) \\
{\rm Bi}(x) & \simeq \frac{1}{\sqrt{\pi} (-x)^{1/4}} \cos\left( \frac{2}{3} (-x)^{3/2}+ \frac{\pi}{4}  \right)
\end{aligned}
\qquad
\textrm{for } x\ll -1.
\end{equation}
We will also need the expansion of the phases of the WKB solutions close to the critical points $z\tau=\pm 1$:
\begin{align}
\int_{-1/z}^\tau d\tau'\sqrt{z^2\tau'^2-1} & = \frac{1}{2z}\left[z\tau \sqrt{z^2\tau^2-1} - \log\left(-z\tau-\sqrt{z^2\tau^2-1}\right)\right] \nonumber \\
& \simeq -\frac{2^{3/2}}{3z}(-1-z\tau)^{3/2} \,,\qquad z\tau\to -1^- \\
\int_{-1/z}^\tau d\tau'\sqrt{1-z^2\tau'^2} & = \frac{1}{2z}\left[z\tau \sqrt{1-z^2\tau^2} - \arcsin\left(z\tau \right)+\frac{\pi}{2} \right] \nonumber \\
& \simeq \frac{2^{3/2}}{3z}(1+z\tau)^{3/2} \,,\qquad z\tau\to -1^+ \\
\int_{-1/z}^\tau d\tau'\sqrt{1-z^2\tau'^2}
& \simeq \frac{\pi}{2z} - \frac{2^{3/2}}{3z}(1-z\tau)^{3/2} \,,\qquad z\tau\to 1^- \\
\int_{+1/z}^\tau d\tau'\sqrt{z^2\tau'^2-1} & = \frac{1}{2z}\left[z\tau \sqrt{z^2\tau^2-1} - \log\left(z\tau+\sqrt{z^2\tau^2-1}\right)\right] \nonumber \\
& \simeq \frac{2^{3/2}}{3z}(z\tau-1)^{3/2} \,,\qquad z\tau\to 1^+
\end{align}

\subsubsection{Matching with the initial condition Eq.~(\ref{eq:initial})}

The solutions described have free coefficients that can be obtained from matching at the intersection of their regime of validity and with the initial condition
\begin{equation}\label{eq:initial appendix A}
u_k \simeq \frac{e^{-i(k/a) t }}{a\sqrt{2\,k}}\,,
\end{equation}
where Eq.~(\ref{eq:initial appendix A}) is specified up to a phase, which we choose in order to make the phases of $A_1$ and $B_1$ real and positive for $\tau>-1/z$.
The initial conditions for $A_1(\tau)$ and $B_1(\tau)$ are given as
\begin{align}
A_1(\tau) \simeq \exp\left( -\frac{i}{2}z \tau^2 + i\theta\right), \quad
B_1(\tau) \simeq -i \exp\left( -\frac{i}{2}z \tau^2 + i\theta\right), \label{eq:initial A1 B1}
\end{align}
where $\theta$ is a phase which we do not specify here.

Matching Eqs.~(\ref{eq:generic A1 1}, \ref{eq:generic B1 1}) with the initial condition Eq.~(\ref{eq:initial A1 B1}) at $\tau < -1/z$, we have
\begin{align}
A_1(\tau) &\simeq i\left( \frac{-1-z\tau}{1-z\tau} \right)^{1/4} \exp\left( i\int^\tau_{-1/z} d\tau' \sqrt{z^2 \tau'^2 -1} - \frac{i\pi}{4}  \right), \label{eq:A1 t negative}\\
B_1(\tau) &\simeq \left( \frac{1-z\tau}{-1-z\tau} \right)^{1/4} \exp\left( i\int^\tau_{-1/z} d\tau' \sqrt{z^2 \tau'^2 -1} - \frac{i\pi}{4} \right). \label{eq:B1 t negative}
\end{align}
For $\tau \simeq -1/z$, $A_1$ can be written as a linear combination of Airy functions.
The solution which is consistent with Eqs.~(\ref{eq:A1 t negative}, \ref{eq:B1 t negative}) is
\begin{align}
A_1(\tau) &\simeq 2^{-1/3} z^{1/6} \sqrt{\pi} [-i {\rm Ai}'( (2z)^{1/3}(\tau+1/z) ) + {\rm Bi}'( (2z)^{1/3}(\tau+1/z) )], \label{eq:A1 transition1}\\ 
B_1(\tau) &\simeq 2^{1/3} z^{-1/6} \sqrt{\pi} [-i{\rm Ai}( (2z)^{1/3}(\tau+1/z) ) + {\rm Bi}( (2z)^{1/3}(\tau+1/z) )]. \label{eq:B1 transition1}
\end{align}
For $-1/z < \tau < 1/z$, we can use WKB approximation again.
The solution consistent with Eqs.~(\ref{eq:A1 transition1}, \ref{eq:B1 transition1}) is given as
\begin{align}
A_1(\tau) &\simeq \left( \frac{1+z\tau}{1-z\tau} \right)^{1/4} \exp\left( \int^\tau_{-1/z} d\tau' \sqrt{1-z^2 \tau'^2} \right), \label{eq:A1 t middle}\\
B_1(\tau) &\simeq \left( \frac{1-z\tau}{1+z\tau} \right)^{1/4} \exp\left( \int^\tau_{-1/z} d\tau' \sqrt{1-z^2 \tau'^2} \right). \label{eq:B1 t middle}
\end{align}
Here we dropped exponentially suppressed term $\exp(-\int^\tau_{-1/z} d\tau' \sqrt{1-z^2\tau'^2})$.
For $\tau\sim 1/z$, $B_1$ can be written as a linear combination of Airy functions.
Matching onto Eq.~(\ref{eq:A1 t middle}, \ref{eq:B1 t middle}) gives
\begin{align}
A_1(\tau) &\simeq 2^{4/3} z^{-1/6} \sqrt{\pi} \exp\left(\frac{\pi}{2z} \right) {\rm Ai}( -(2z)^{1/3}(\tau-1/z) ), \label{eq:A1 transition2}\\
B_1(\tau) &\simeq -2^{2/3} z^{1/6} \sqrt{\pi} \exp\left(\frac{\pi}{2z} \right) {\rm Ai}'( -(2z)^{1/3}(\tau-1/z) ). \label{eq:B1 transition2}
\end{align}
Finally, for $\tau > 1/z$, we can use WKB approximation again.
Matching with Eqs.~(\ref{eq:A1 transition2}, \ref{eq:B1 transition2}) gives
\begin{align}
A_1(\tau) &= 2 \exp\left(\frac{\pi}{2z}\right) \left( \frac{z\tau+1}{z\tau-1}\right)^{1/4} \sin\left( \frac{\pi}{4} + \int_{1/z}^\tau d\tau' \sqrt{z^2 \tau'^2-1} \right), \label{eq:A1 t later}\\
B_1(\tau) &=  2 \exp\left(\frac{\pi}{2z}\right) \left( \frac{z\tau-1}{z\tau+1}\right)^{1/4} \cos\left( \frac{\pi}{4} + \int_{1/z}^\tau d\tau' \sqrt{z^2 \tau'^2-1} \right). \label{eq:B1 t later}
\end{align}
Therefore, after amplification has ended, the asymptotic behavior of $u_k$ is
\begin{align}
u_k \to  \frac{1}{a}\sqrt{\frac{2}{k}} \exp\left(\frac{\pi}{2z}\right) \sin\left( \frac{\pi}{4} + \frac{\phi}{2f} + \int_{1/z}^{\frac{\Lambda_b^4 t}{2\dot{\phi}_0 f^2}} d\tau' \sqrt{z^2\tau'^2-1} \right). \label{eq:asymptotic form}
\end{align}
\begin{figure}
\centering
\includegraphics[width=0.8\hsize]{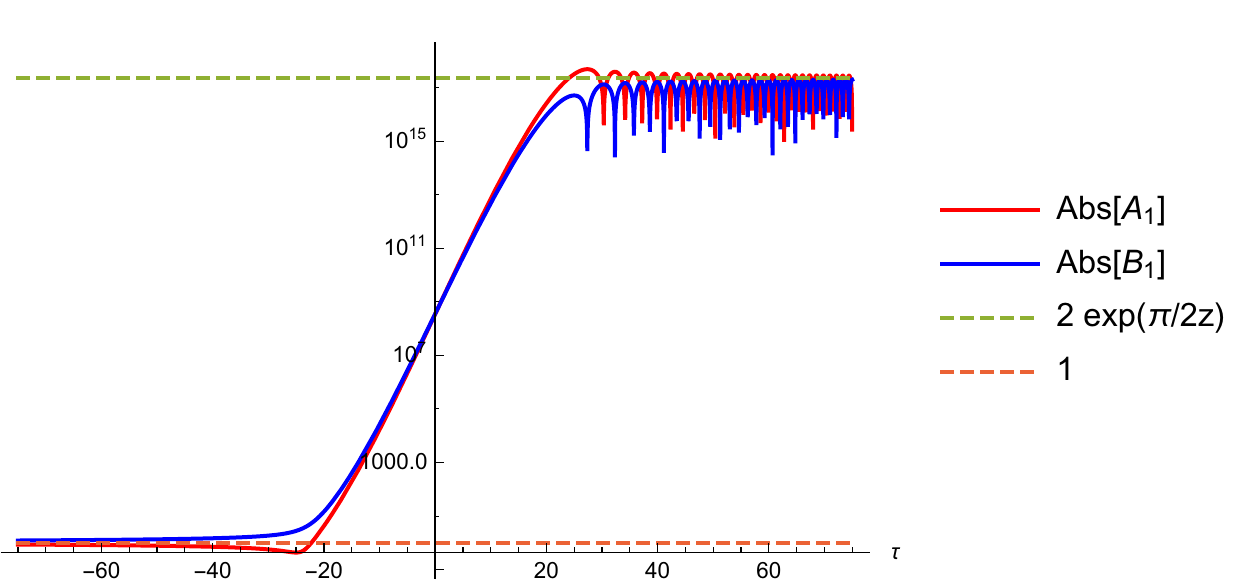}\\[2cm]
\includegraphics[width=0.8\hsize]{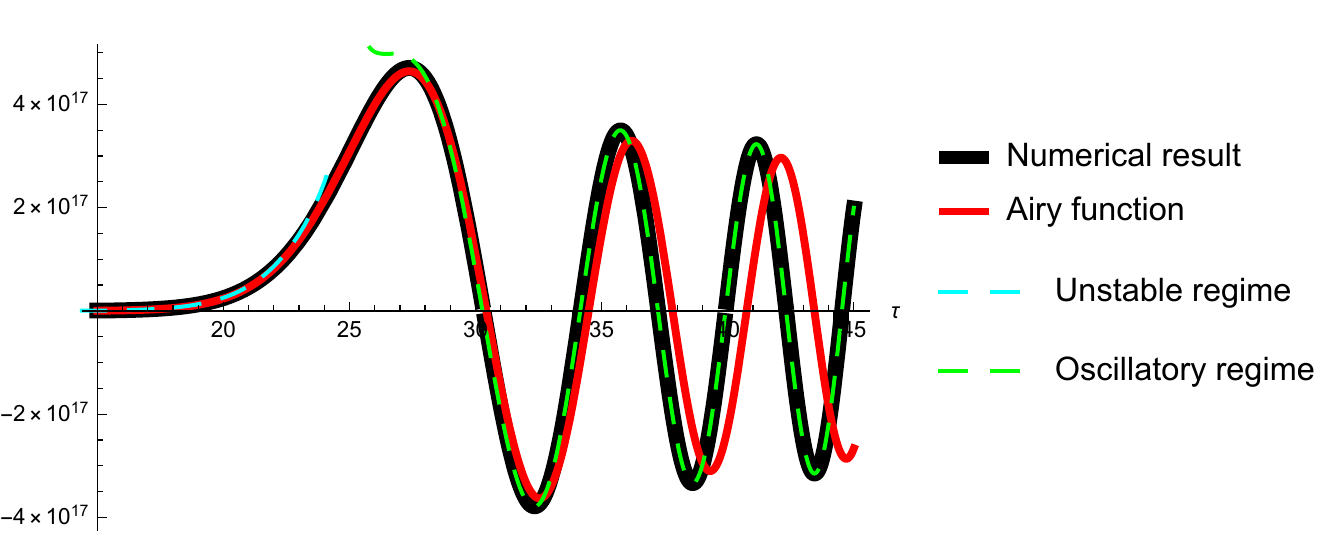}\\[2cm]
\includegraphics[width=0.8\hsize]{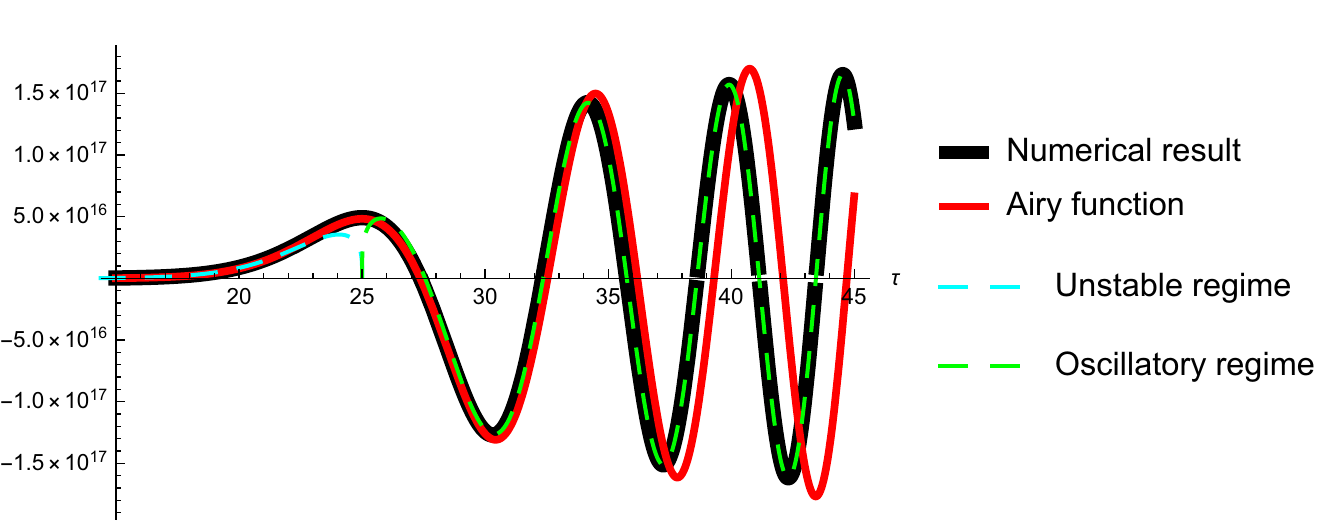}
\caption{
Solution of Eqs.~(\ref{eq:diff eq A1}, \ref{eq:diff eq B1}) with $z=0.04$.
The top panel shows $|A_1|$ and $|B_1|$.
The middle (bottom) panel shows comparison between numerical result of $A_1$ ($B_1$) at $\tau \simeq 1/z = 25$
with 
Eq.~(\ref{eq:A1 transition2}, \ref{eq:B1 transition2}) (red lines),
Eq.~(\ref{eq:A1 t later}, \ref{eq:B1 t later}) (green dashed lines),
and Eq.~(\ref{eq:A1 t middle}, \ref{eq:B1 t middle}) (cyan dashed lines).
}\label{fig:wkb}
\end{figure}
Figure~\ref{fig:wkb} shows the good agreement of a numerical solution of Eqs.~(\ref{eq:diff eq A1}, \ref{eq:diff eq B1}) with the analytic formulae above.
By using the definition of $z$ given in Eq.~(\ref{eq:def of tau and z}), we obtain Eq.~(\ref{eq:asymptotic uk 2}).

So far, we have discussed the case with $z>0$, \textit{i.e.,} $\ddot\phi < -H\dot\phi$.
In closing this section, let us briefly look at the solutions of Eqs.~(\ref{eq:diff eq A1}, \ref{eq:diff eq B1}) for $z<0$,
\textit{i.e.,} $\ddot\phi > -H\dot\phi$.
This solution is required to discuss positive $\ddot\phi$ solution and $\ddot\phi = 0$ solution in Eq.~(\ref{eq:eq for phi double dot}).
For $z<0$, the initial conditions for $A_1$ and $B_1$ is 
\begin{align}
A_1(\tau) \simeq \exp\left( \frac{i}{2}|z| \tau^2 + i\theta\right), \quad
B_1(\tau) \simeq -i \exp\left( \frac{i}{2}|z| \tau^2 + i\theta\right), \label{eq:initial A1 B1 for negative z}
\end{align}
Eq.~(\ref{eq:diff eq A1}, \ref{eq:diff eq B1}) are symmetric under the transformation of $z \to -z$ and $A_1 \leftrightarrow B_1$. Also, complex conjugation is a symmetry of the equation of motion.
The initial conditions in Eq.~(\ref{eq:initial A1 B1 for negative z}) can be obtained by taking complex conjugate of Eq.~(\ref{eq:initial A1 B1}) after exchanging $A_1 \leftrightarrow B_1$.
Thus, $A_1$ and $B_1$ for $z<0$ can be obtained from Eqs.~(\ref{eq:A1 t negative}--\ref{eq:B1 t later}) by replacing $z\to |z|$ and
\begin{equation}
\left. A_1 \right|_{z<0} = (\left. B_1 \right|_{z>0})^* \, , \qquad
\left. B_1 \right|_{z<0} = (\left. A_1 \right|_{z>0})^* \, .
\end{equation}
The coefficients $A_1$, $B_1$ for $\tau < -1/|z|$ are given as
\begin{align}
A_1(\tau) &\simeq \left( \frac{1-|z|\tau}{-1-|z|\tau} \right)^{1/4} \exp\left( -i\int^\tau_{-1/|z|} d\tau' \sqrt{z^2 \tau'^2 -1} + \frac{i\pi}{4} \right), \\
B_1(\tau) &\simeq -i\left( \frac{-1-|z|\tau}{1-|z|\tau} \right)^{1/4} \exp\left( -i\int^\tau_{-1/|z|} d\tau' \sqrt{z^2 \tau'^2 -1} + \frac{i\pi}{4}  \right).
\end{align}
$A_1$ and $B_1$ for $\tau > 1/|z|$ are given as
\begin{align}
A_1(\tau) &=  2 \exp\left(\frac{\pi}{2|z|}\right) \left( \frac{|z|\tau-1}{|z|\tau+1}\right)^{1/4} \cos\left( \frac{\pi}{4} + \int_{1/|z|}^\tau d\tau' \sqrt{z^2 \tau'^2-1} \right), \\
B_1(\tau) &= 2 \exp\left(\frac{\pi}{2|z|}\right) \left( \frac{|z|\tau+1}{|z|\tau-1}\right)^{1/4} \sin\left( \frac{\pi}{4} + \int_{1/|z|}^\tau d\tau' \sqrt{z^2 \tau'^2-1} \right).
\end{align}
Therefore, after amplification has ended, the asymptotic behavior of $u_k$ is
\begin{align}
u_k \to  \frac{1}{a}\sqrt{\frac{2}{k}} \exp\left(\frac{\pi}{2|z|}\right) \cos\left( -\frac{\pi}{4} + \frac{\phi}{2f} - \int_{1/|z|}^{\frac{\Lambda_b^4 t}{2\dot{\phi}_0 f^2}} d\tau' \sqrt{z^2\tau'^2-1} \right).
\end{align}
By using the definition of $z$ given in Eq.~(\ref{eq:def of tau and z}), we obtain Eq.~(\ref{eq:asymptotic uk 2}).

\section{Detailed analysis on Eq.~(\ref{eq:eq for phi double dot})}\label{sec:detail on phiddot}
In this section, we discuss Eq.~(\ref{eq:eq for phi double dot}) in detail.

\subsection{The solutions of Eq.~(\ref{eq:eq for phi double dot})} \label{sec:solution phiddot}
Let us discuss the solutions of Eq.~(\ref{eq:eq for phi double dot}), which we report here for simplicity:
\begin{align}
\dot\phi\ddot\phi
=
- 3 H \dot\phi^2
+ \mu^3 \dot\phi
- \frac{1}{32\pi^2 f^4} \dot\phi^3 |\ddot\phi + H\dot\phi| \exp\left( \frac{\pi \Lambda_b^8}{2f\dot\phi^2|\ddot\phi + H\dot\phi|} \right)\,.
\label{eq:eq for phi double dot appendix}
\end{align}
By defining
\begin{align}
x \equiv \frac{\pi \Lambda_b^8}{2f\dot\phi^2(\ddot\phi + H\dot\phi)},
\end{align}
Eq.~(\ref{eq:eq for phi double dot}) can be rewritten as
\begin{align}
\frac{1}{x} = \frac{1}{b} - \frac{a}{|x|} e^{|x|}. \label{eq:deq for a,b,x}
\end{align}
where $a$ and $b$ are dimensionless parameters which are defined as
\begin{align}\label{eq:prod log equation}
a \equiv \frac{\dot\phi^2}{32\pi^2 f^4},\quad
b \equiv \frac{\pi\Lambda_b^8}{2f\dot\phi^2(\mu^3-2H\dot\phi)}.
\end{align}
Eq.~(\ref{eq:deq for a,b,x}) can be rewritten as
\begin{equation}
\begin{cases}
(x-b) e^{x-b} = -abe^{-b} & (x<0) \\
(b-x) e^{b-x} = -abe^b & (x>0)
\end{cases}
\end{equation}
which has the following solutions:
\begin{equation}
\begin{cases}
x = b + W_{-1}(-abe^{-b}) & (b>0)  \\
x = b + W_{0}(-abe^{-b}) & (b<0)  \\
x = b - W_{-1}(-abe^{b}) & (0<abe^b<e^{-1})  \\
x = b - W_{0}(-abe^{b}) & (0<abe^b<e^{-1})
\end{cases}
\end{equation}
where $W_0$, $W_{-1}$ are the two main branches of the Lambert function (product logarithm). The first two solutions have $x<0$, the other two have $x>0$. Formally, the first solution only exists for $a<1$.
By using the consistency condition of EFT, we can assume $\dot\phi^2 < 32\pi^2 f^4$, \textit{i.e.,} $0<a<1$.
In terms of the acceleration $\ddot\phi$, the analytic solutions of Eq.~(\ref{eq:eq for phi double dot appendix}) are given as
\begin{align}
\ddot\phi_1 &=
\begin{cases}
-H\dot\phi + \displaystyle\frac{\pi \Lambda_b^8}{2f\dot\phi^2}\left[ b + W_{-1}(-ab e^{-b} ) \right]^{-1} & (b>0) \\
-H\dot\phi + \displaystyle\frac{\pi \Lambda_b^8}{2f\dot\phi^2}\left[ b + W_0(-ab e^{-b} ) \right]^{-1} & (b<0) \\
\end{cases}
, \label{eq:phiddot 1}\\
\ddot\phi_2 &= -H\dot\phi + \frac{\pi \Lambda_b^8}{2f\dot\phi^2}\left[ b - W_{-1}(-ab e^b ) \right]^{-1}, \label{eq:phiddot 2}\\
\ddot\phi_3 &= -H\dot\phi + \frac{\pi \Lambda_b^8}{2f\dot\phi^2}\left[ b - W_0(-ab e^b ) \right]^{-1}. \label{eq:phiddot 3}
\end{align}
Again, $\ddot\phi_2$ and $\ddot\phi_3$ exist only if $0<ab e^b < e^{-1}$.
Moreover, $\ddot\phi_1 < -H\dot\phi$, and $- H\dot\phi < \ddot\phi_2 \leq \ddot\phi_3 < \mu^3 - 3H\dot\phi$ if they exist.
The solution $\ddot\phi_1$ has two different rappresentations for positive or negative $b$, but it is continuous in $\dot\phi$ at $\mu^3-2H\dot\phi = 0$ when $b$ diverges.
Finally, by looking at Eq.~(\ref{eq:eq for phi double dot appendix}), one notices that $\ddot\phi_{1,2,3} < \mu^3 - 3H\dot\phi$, as it is expected since the particle production effect always takes away energy from the zero mode.

\begin{figure}[t]
\centering
\includegraphics[width=0.7\hsize]{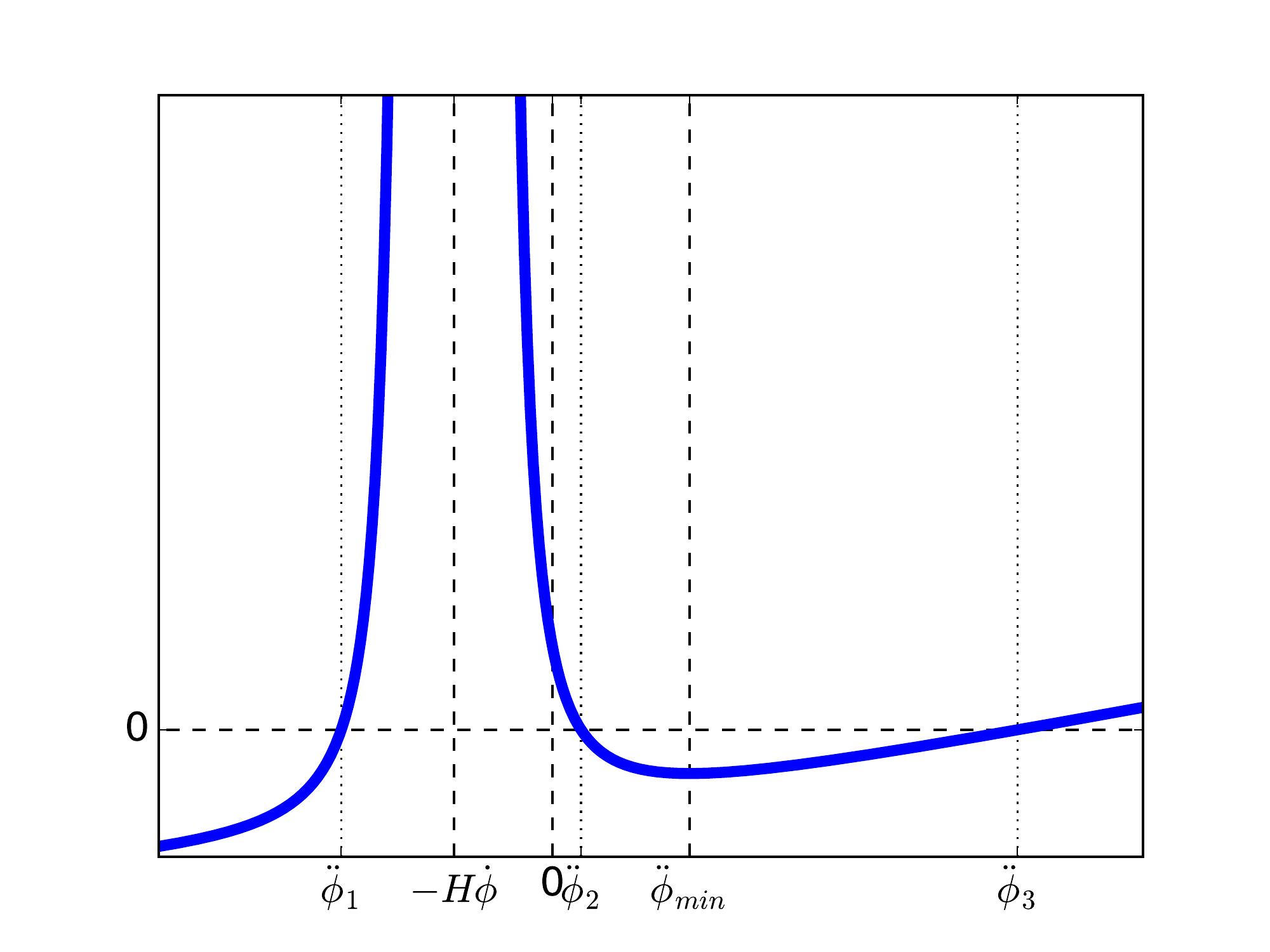}
\caption{A typical behavior of $f(\dot\phi, \ddot\phi)$ as a function of $\ddot\phi$.
}\label{fig:ffunc}
\end{figure}

\subsection{The condition not to have positive $\ddot\phi$ solution}\label{sec:detail on no positive phiddot condition}

\begin{figure}[t]
\centering
\includegraphics[width=0.9\hsize]{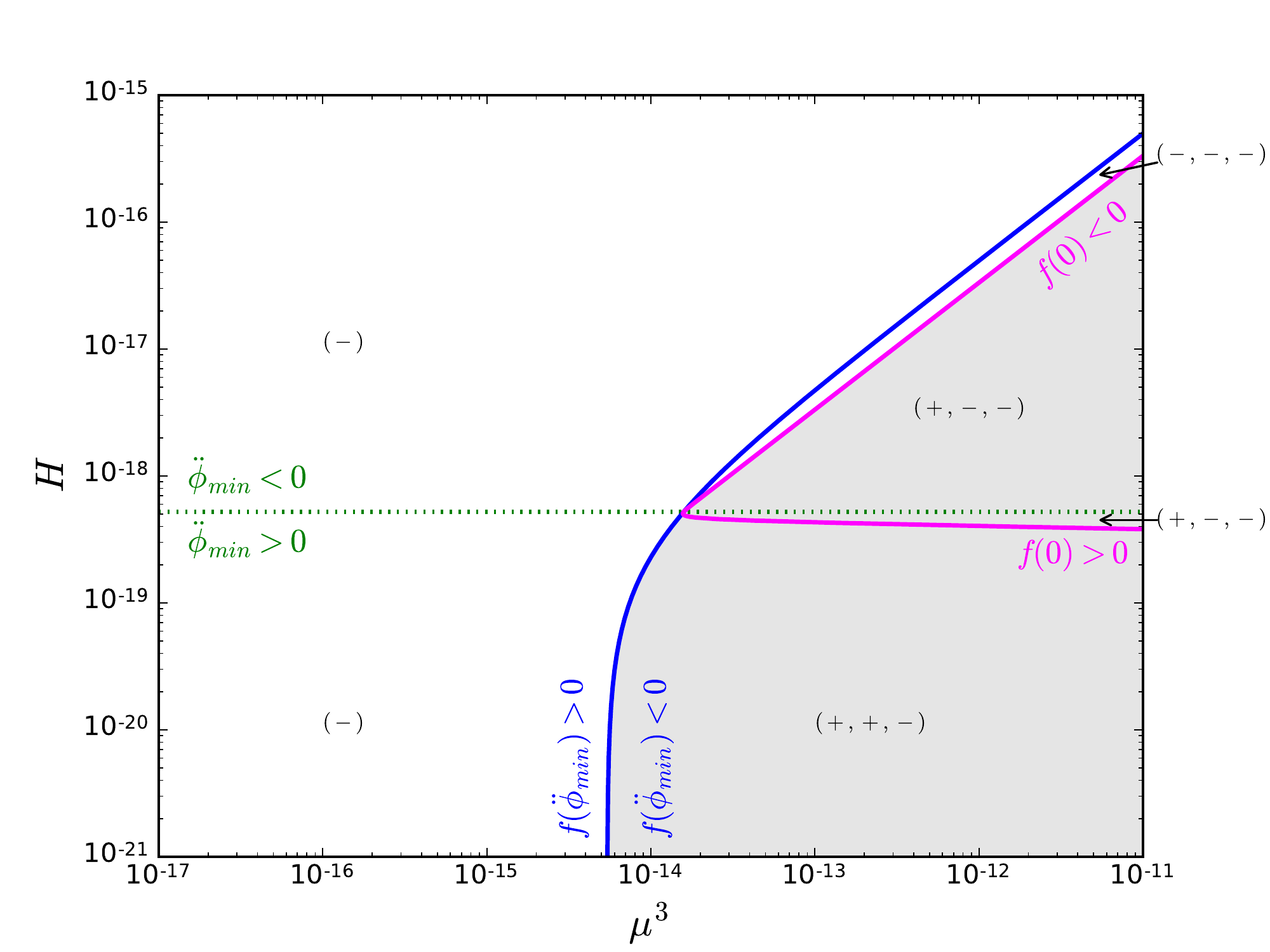}
\caption{
The solutions of Eq.~(\ref{eq:eq for phi double dot}) are classified in the $\mu^3$-$H$ plane.
We take $\dot\phi= 10^4$, $f= 10^5$ in units of $\Lambda_b=1$.
$\ddot\phi_\mathrm{min}<0$ ($\ddot\phi_\mathrm{min}>0$) is satisfied above (below) the green line.
$f(\dot\phi,\ddot\phi_\mathrm{min})<0$ ($f(\dot\phi,\ddot\phi_\mathrm{min})>0$) is satisfied on the right (left) of the blue line.
$f(\dot\phi,0)<0$ ($f(\dot\phi,0)>0$) is satisfied inside (outside) the magenta line.
There exists at least one positive $\ddot\phi$ solution in the grey region.
There is no positive $\ddot\phi$ solution in the white region.
$(-)$ indicates one negative solution for $f(\dot\phi,\ddot\phi) = 0$.
$(-,-,-)$ three negative solution, 
$(+,-,-)$ one positive and two negative solution, 
and $(+,+,-)$ two positive and one negative solution.
}\label{fig:glam-H-plane}
\end{figure}

\begin{figure}[t]
\centering
\includegraphics[width=0.6\hsize]{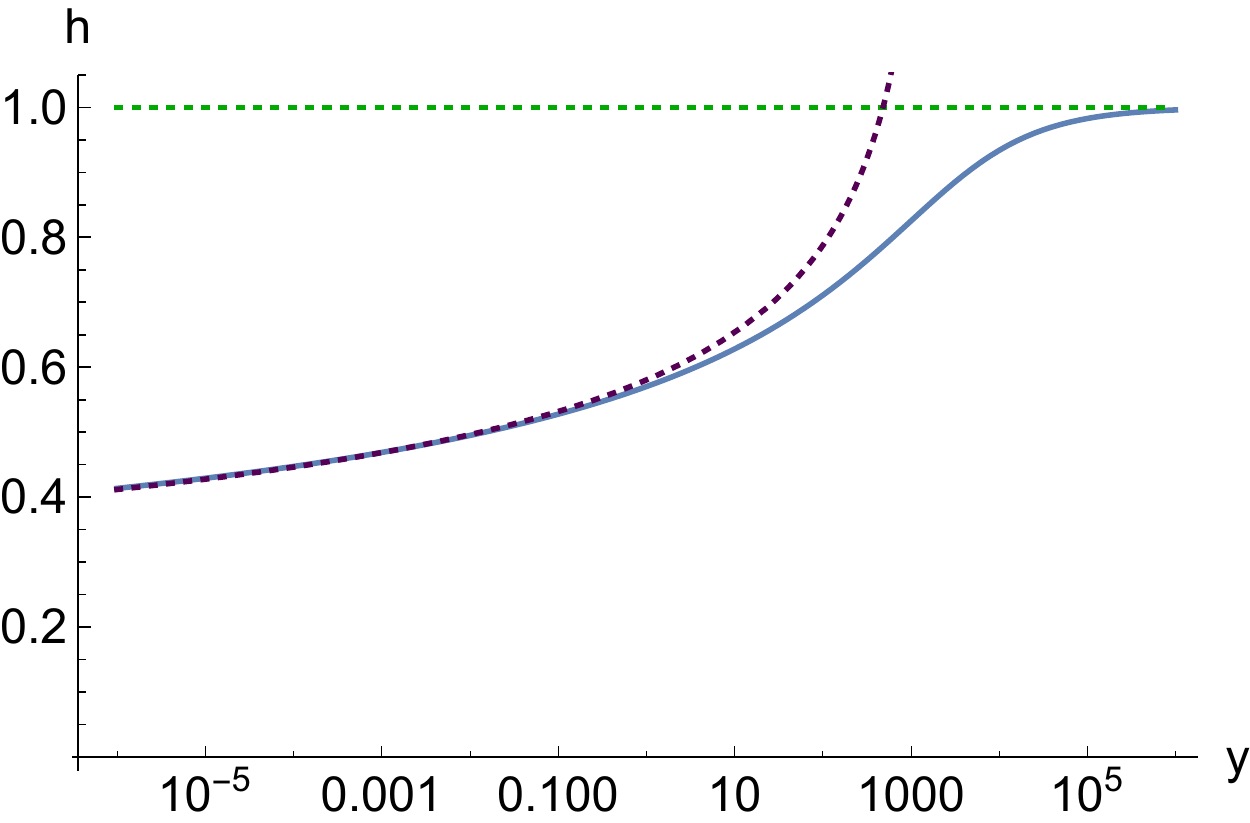}
\caption{
The function $h(y)$ as the solution of Eq.~(\ref{eq:deq for hfunc}) is shown as a solid line.
The green and purple dotted lines correspond to $1$ and $(3/\log(98304\pi^6/e^3 y^2) )^{1/3}$, and show the behaviour of $h(y)$ for small and large $y$, respectively. }
\label{fig:hfunc}
\end{figure}

In order to stop the axion rolling, the acceleration $\ddot\phi$ should always be negative. Before fragmentation starts, the field is only subject to the slope and Hubble friction, $\ddot\phi = \mu^3 - 3H\dot\phi$. Let us start by deriving a condition not to have a solution $0<\ddot\phi<\mu^3-3H\dot\phi$.
Let us define $f(\dot\phi, \ddot\phi)$ as
\begin{align}
f(\dot\phi, \ddot\phi) \equiv
\dot\phi \ddot\phi + 3H\dot\phi^2 - \mu^3 \dot\phi + \frac{1}{32\pi^2 f^4} \dot\phi^3 |\ddot\phi + H\dot\phi| \exp\left( \frac{\pi\Lambda_b^8}{2f\dot\phi^2|\ddot\phi + H\dot\phi|}\right).
\end{align}
Eq.~(\ref{eq:eq for phi double dot}) is equivalent to $f(\dot\phi,\ddot\phi) = 0$.
We can easily see the following property of $f(\dot\phi, \ddot\phi)$:
\begin{align}
\frac{\partial^2f}{\partial\ddot\phi^2} &> 0, \\
\lim_{\ddot\phi \to -\infty} \frac{\partial f}{\partial \ddot\phi} &= \dot\phi - \frac{1}{32\pi^2 f^4} \dot\phi^3, \\
\lim_{\ddot\phi \to (-H\dot\phi) }f &= +\infty, \\
\lim_{\ddot\phi \to (-H\dot\phi)^-} \frac{\partial f}{\partial \ddot\phi} &= +\infty, \\
\lim_{\ddot\phi \to (-H\dot\phi)^+} \frac{\partial f}{\partial \ddot\phi} &= -\infty, \\
\lim_{\ddot\phi \to +\infty} \frac{\partial f}{\partial \ddot\phi} &= \dot\phi + \frac{1}{32\pi^2 f^4} \dot\phi^3.
\end{align}
The function $f(\dot\phi, \ddot\phi)$ monotonously increases for $\ddot\phi < -H\dot\phi$ because of $\dot\phi^2 / 32\pi^2 f^4 < 1$.
There exists one local minimum at $\ddot\phi = \ddot\phi_\mathrm{min}> -H\dot\phi$.
A typical behavior of $f(\dot\phi,\ddot\phi)$ is shown in Fig.~\ref{fig:ffunc}.
The solutions of $f(\dot\phi,\ddot\phi)=0$ are classified depending on the sign of
$\ddot\phi_\mathrm{min}(\dot\phi)$, $f(\dot\phi, \ddot\phi_\mathrm{min}(\dot\phi))$, and $f(\dot\phi,0)$,
which are given as
\begin{align}
\ddot\phi_\mathrm{min}(\dot\phi) &= - H\dot\phi + \frac{\pi\Lambda_b^8}{2f\dot\phi^2} \left( 1+W_0\left( \frac{32\pi^2f^4}{e\dot\phi^2} \right) \right)^{-1}, \\
f(\dot\phi, \ddot\phi_\mathrm{min}(\dot\phi)) &= 2H\dot\phi^2 - \mu^3 \dot\phi + \frac{\pi\Lambda_b^8}{2f\dot\phi}\left( W_0\left(\frac{32\pi^2 f^4}{e\dot\phi^2} \right) \right)^{-1}, \\
f(\dot\phi,0) &= 3H\dot\phi^2 - \mu^3 \dot\phi + \frac{1}{32\pi^2 f^4} H \dot\phi^4 \exp\left( \frac{\pi\Lambda_b^8}{2fH \dot\phi^3}\right).
\end{align}
After having discussed the behaviour of the function $f$, it is clear that the equation $f(\dot\phi,\ddot\phi)=0$ has up to three solutions, which correspond to $\ddot\phi_1$, $\ddot\phi_2$, $\ddot\phi_3$ of Eqs.~(\ref{eq:phiddot 1}-\ref{eq:phiddot 3}). These are summarized in Tab.~\ref{tab:solution}.
We are interested in a situation in which the field is slown-down by fragmentation, thus we want to avoid solutions with $\ddot\phi>0$.
This condition, depending on the sign of $\ddot\phi_\mathrm{min}$, can be written as
\begin{align}
f(\dot\phi,\ddot\phi_\mathrm{min})>0 & \quad \textrm{for}\quad \ddot\phi_\mathrm{min} >0 \,, \label{eq:condition on f fdd>0}\\
f(\dot\phi,0)>0 & \quad \textrm{for}\quad \ddot\phi_\mathrm{min} <0 \,. \label{eq:condition on f fdd<0}
\end{align}
Eqauations~(\ref{eq:condition on f fdd>0}, \ref{eq:condition on f fdd<0}) cover, respectively, a situation in which the solution $\ddot\phi_{2,3}$ do not exist and one in which they do exist but they are negative.
It is convenient to rewrite this condition as an upperbound on $\mu^3$:
\begin{align}
\mu^3 < 
\begin{cases}
\mu_1^3 & (H < H_{\dot\phi}) \\
\mu_2^3 & (H \geq H_{\dot\phi}) \\
\end{cases}
\label{eq:bound on glambda 2}
\end{align}
where we have defined
\begin{align}
H_{\dot\phi} &\equiv \frac{\pi\Lambda_b^8}{2f\dot\phi^3}\left( 1 + W_0\left( \frac{32\pi^2 f^4}{e\dot\phi^2} \right) \right)^{-1}, \label{eq:Hphi}\\
\mu_1^3 &\equiv 2H\dot\phi + \frac{\pi\Lambda_b^8}{2f\dot\phi^2} \left( W_0\left( \frac{32\pi^2f^4}{e\dot\phi^2} \right) \right)^{-1}, \label{eq:g1Lam3}\\
\mu_2^3 &\equiv 3H\dot\phi + \frac{H\dot\phi^3}{32\pi^2 f^4} \exp \left( \frac{\pi\Lambda_b^8}{2fH\dot\phi^3} \right). \label{eq:g2Lam3}
\end{align}
Note that $\ddot\phi_\mathrm{min} \lessgtr 0$ corresponds to $H \gtrless H_{\dot\phi}$,
$\mu^3 \lessgtr \mu_1^3$ corresponds to $f(\dot\phi,\ddot\phi_\mathrm{min}(\dot\phi)) \gtrless 0$,
and $\mu^3 \lessgtr \mu_2^3$ corresponds to $f(\dot\phi,0) \gtrless 0$.
Moreover, $\mu_1^3=\mu_2^3$ for $H=H_{\dot\phi}$.
Figure~\ref{fig:glam-H-plane} shows the parameter region which is excluded by the condition Eq.~(\ref{eq:bound on glambda 2}).
$H_{\dot\phi}$ is a monotonously decreasing function of $\dot\phi$;
thus, Eq.~(\ref{eq:bound on glambda 2}) are equivalent to
\begin{align}
\mu^3 < 
\begin{cases}
\mu_1^3 & (\dot\phi < \dot\phi_H) \\
\mu_2^3 & (\dot\phi \geq \dot\phi_H) \\
\end{cases}
\label{eq:bound on glambda 3}
\end{align}
Here $\dot\phi_H$ is defined as the solution of $H = (\pi \Lambda_b^8/2f\dot\phi_H^3) ( 1 + W_0( 32\pi^2f^4/e\dot\phi_H^2 ))^{-1}$, so that $\dot\phi=\dot\phi_H$ is equivalent to $H=H_{\dot\phi}$,
and it is explicitly written as
\begin{align}
\dot\phi_H = \left( \frac{\pi \Lambda_b^8}{2 f H} \right)^{1/3} h\left( \frac{\pi \Lambda_b^8}{f^7 H} \right),
\end{align}
where $h(y) \sim {\cal O}(1)$ satisfies
\begin{align}
h^3(y) \left[ 1 + W_0\left( \frac{32\pi^2}{e \, 2^{-2/3} y^{2/3} h^2(y)} \right) \right] = 1. \label{eq:deq for hfunc}
\end{align}
The numerical value of $h(y)$ is shown in Fig.~\ref{fig:hfunc}.

\begin{table}
\centering
\begin{tabular}{|l|l|}
\hline
&\\
$f(\dot\phi,\ddot\phi_\mathrm{min})>0$                                           & $\ddot\phi_1 < -H\dot\phi$ ($\ddot\phi_2, \ddot\phi_3$ do not exist.)\\
$f(\dot\phi,\ddot\phi_\mathrm{min})<0$, $f(\dot\phi,0)<0$                        & $\ddot\phi_1 < -H\dot\phi < \ddot\phi_2 < 0 < \ddot\phi_3$\\
$f(\dot\phi,\ddot\phi_\mathrm{min})<0$, $f(\dot\phi,0)>0$, $\ddot\phi_\mathrm{min} > 0$ & $\ddot\phi_1 < -H\dot\phi < 0 < \ddot\phi_2 < \ddot\phi_3$\\
$f(\dot\phi,\ddot\phi_\mathrm{min})<0$, $f(\dot\phi,0)>0$, $\ddot\phi_\mathrm{min} < 0$ & $\ddot\phi_1 < -H\dot\phi < \ddot\phi_2 < \ddot\phi_3 < 0$\\
&\\
\hline
\end{tabular}
\caption{Solutions of $f(\dot\phi,\ddot\phi)=0$ for given $\dot\phi$ and their classification.
Note that $f(\dot\phi, \ddot\phi_\mathrm{min}(\dot\phi))>0$ is equivalent to $0<ab e^b<e^{-1}$.}
\label{tab:solution}
\end{table}

\subsection{The stopping condition}\label{sec:stopping condition}
Before fragmentation is active, $\ddot\phi\approx \mu^3-3 H \dot\phi$. This is not a solution of Eq.~(\ref{eq:eq for phi double dot appendix}), and does not result as the $\Lambda_b\to 0$ limit of $\ddot\phi_{1,2,3}$. The failure of $\ddot\phi_{1,2,3}$ to reproduce this initial condition is due to our assumption Eq.~(\ref{eq:ddot phi ll Lambda_b}), which gives a minimal value of $\Lambda_b$ (and thus a minimal efficiency of the fragmentation effect) below which our calculation is not reliable. Still, in this regime, we can assume that $\ddot\phi$ varies continuously. When fragmentation turns on and Eq.~(\ref{eq:eq for phi double dot appendix}) becomes valid, $\ddot\phi$ will smoothly decrease from its initial value until it reaches one of the solutions in Eqs.~(\ref{eq:phiddot 1}-\ref{eq:phiddot 3}), and it will stick to it for the rest of the evolution.
In order to understand the behaviour during this phase, it is also useful to notice that, if $f(\dot\phi,\ddot\phi)>0$, $\ddot\phi$ will decrease. Oppositely, if $f(\dot\phi,\ddot\phi)<0$, $\ddot\phi$ will increase. Hence, $\ddot\phi_1$ is a stable solution, as well as $\ddot\phi_3$ (if it exists). On the contrary, $\ddot\phi_2$, if it exists, is an unstable solution.

Let us now discuss the necessary and sufficient conditions in order to guarantee that the field slows-down until it stops. 
The conditions in Eqs.~(\ref{eq:condition on f fdd>0}, \ref{eq:condition on f fdd<0}), or equivalently in Eq.~(\ref{eq:bound on glambda 3}), give a bound on the slope $\mu^3$ that depends on the velocity $\dot\phi$.
If this condition is satisfied for $\dot\phi = \dot\phi_0$, initially the solutions to Eq.~(\ref{eq:eq for phi double dot appendix}) have $\ddot\phi<0$:
\begin{align}
\mu^3 <
\begin{cases}
\mu_1^3(\dot\phi_0) & (\dot\phi_0 < \dot\phi_H) \\
\mu_2^3(\dot\phi_0) & (\dot\phi_0 \geq \dot\phi_H) \\
\end{cases}.
\end{align}
This is a necessary but not sufficient condition to stop the rolling of the field $\phi$. Indeed, $\ddot\phi$ should be negative until the axion will be stopped.
In the case $\mu^3 < \mu_1^3(\dot\phi_0)$, the only solution of $f(\dot\phi_0,\ddot\phi)=0$ is $\ddot\phi_1$.
Thus, at the beginning, $\ddot\phi$ becomes $\ddot\phi_1(\dot\phi_0)$ and $\dot\phi$ starts to decrease.
As $\dot\phi$ decreases, $\ddot\phi$ changes continuously and $\ddot\phi = \ddot\phi_1(\dot\phi)$ is always satisfied
even if $\ddot\phi_2$ or $\ddot\phi_3$ solution appear at some velocity $\dot\phi < \dot\phi_0$.
As we have seen in App.~\ref{sec:solution phiddot}, $\ddot\phi_1$ is always negative and the field will be stopped in the end.

\begin{figure}[t]
\centering
\includegraphics[width=0.8\hsize]{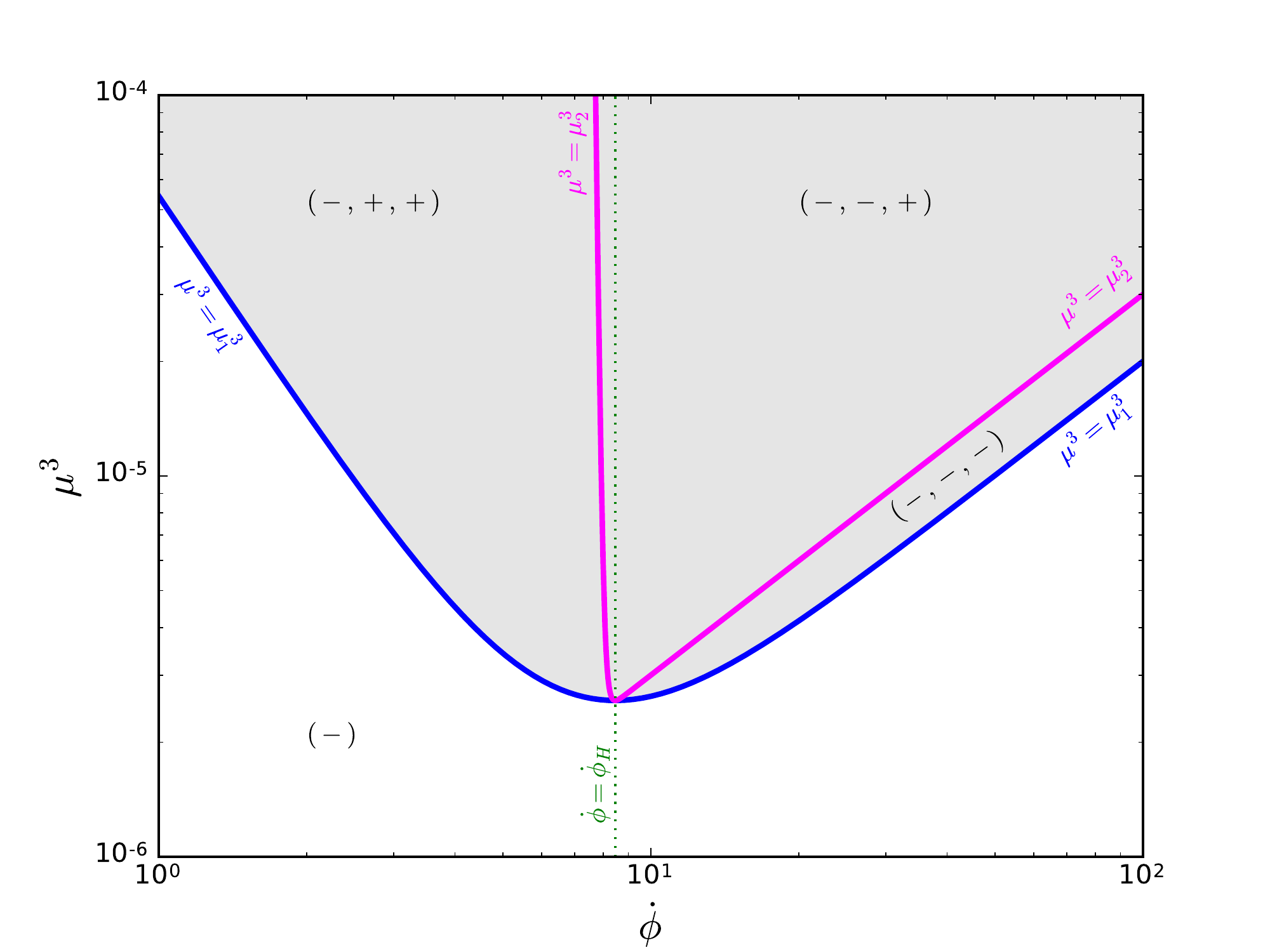}
\caption{
$\mu_1^3$ (the blue line) and $\mu_2^3$ (the magenta line) are shown as a function of $\dot\phi$.
Here we take $f = 10^3$ and $H = 10^{-7}$ in the unit of $\Lambda_b = 1$.
$(-)$, $(-,+,+)$, $(-,-,+)$, and $(-,-,-)$ are same as Fig.~\ref{fig:glam-H-plane}.
The gray region does not satisfy Eq.~(\ref{eq:stopping condition appendix}).
}\label{fig:phidot-mu3 plane}
\end{figure}

On the other hand, in the case $\mu^3 > \mu_1^3(\dot\phi_0)$,
$f(\dot\phi_0,\ddot\phi) = 0$ has three solutions.
The acceleration decreases from the initial value $\mu^3 -3H\dot\phi_0$ until it reaches $\ddot\phi_3$, because $\ddot\phi_1 < \ddot\phi_2 \leq \ddot\phi_3 < \mu^3 - 3H\dot\phi_0$.
Following the discussion in Sec.~\ref{sec:detail on no positive phiddot condition}, the solution $\ddot\phi_3$ exists and is negative if both $\mu_1^3 < \mu^3 < \mu_2^3$ and $\dot\phi > \dot\phi_H$ are satisfied.
In this case, the field starts to decelerate.
As $\dot\phi$ decreases, $\ddot\phi$ changes continuously and $\ddot\phi = \ddot\phi_3(\dot\phi)$ is always satisfied.
However, $\ddot\phi_3$ is not always negative, and in particular, $\ddot\phi_3(\dot\phi_H) > 0$, since, as we discussed below the definitions Eqs.~(\ref{eq:Hphi}, \ref{eq:g1Lam3}, \ref{eq:g2Lam3}), $\ddot\phi_\mathrm{min}=0$ for $\dot\phi=\dot\phi_H$, and $\ddot\phi_3>\ddot\phi_\mathrm{min}$.\footnote{
Alternatively, by using the definition Eqs.~(\ref{eq:Hphi}, \ref{eq:g1Lam3}, \ref{eq:g2Lam3}),
one can show that $\mu_1^3$ takes its minimum at $\dot\phi = \dot\phi_H$,
and then, $\mu_1^3(\dot\phi_H) \leq \mu_1^3(\dot\phi_0)$.
Also, we can show $\mu_1^3(\dot\phi_H) = \mu_2^3(\dot\phi_H)$.
Thus, $\mu_2^3(\dot\phi_H) < \mu^2$ holds and $\ddot\phi_3$ for $\dot\phi = \dot\phi_H$ is positive. This is depicted in Fig.~\ref{fig:phidot-mu3 plane}, which shows that for $\dot\phi = \dot\phi_H$ $\mu_1^3$ and $\mu_2^3$ coincide, leaving space only for positive $
\ddot\phi_3$.
}
This means that there exists a velocity such that $\ddot\phi_3 = 0$ for $\dot\phi_H < \dot\phi < \dot\phi_0$.
(For an existence condition of $\ddot\phi = 0$ solution, see also the Appendix \ref{sec:modified slow roll velocity}.)
If the axion evolution is governed by the solution $\ddot\phi_3$, the field will approach that constant velocity, and its rolling cannot be stopped.
Thus, in order to stop the axion rolling, we need to assume that the solutions $\ddot\phi_{2,3}$ do not exist, \textit{i.e.}, that
\begin{align}
\mu^3 < \mu_1^3(\dot\phi_0). \label{eq:stopping condition appendix}
\end{align}

\medskip

\begin{figure}[t]
\centering
\includegraphics[width=0.8\hsize]{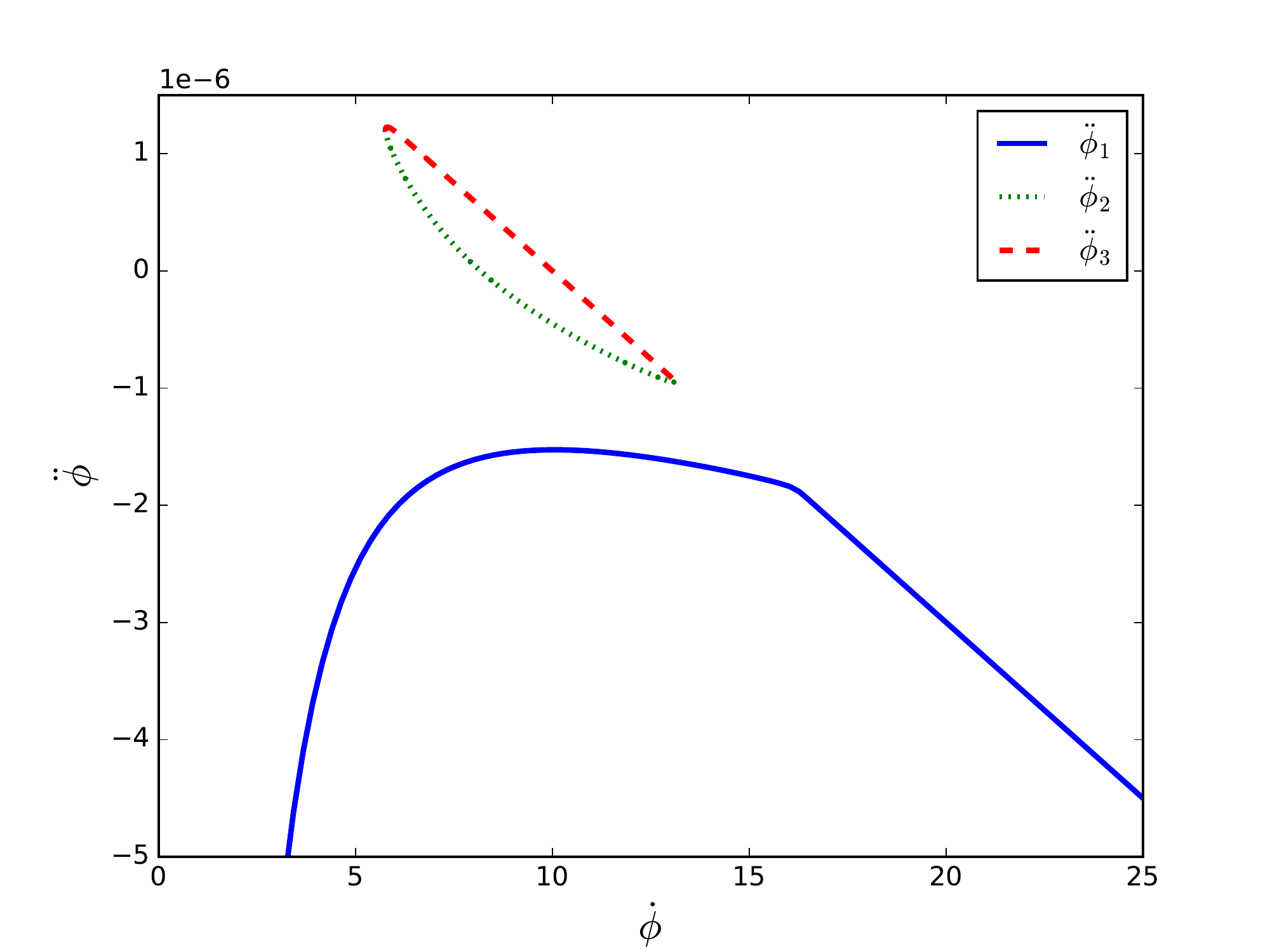}
\caption{
$\ddot\phi$ as a function of $\dot\phi$.
Here we take $f = 10^3$, $H = 10^{-7}$, and $\mu^3 = 3 \times 10^{-6}$ in the unit of $\Lambda_b = 1$.
}\label{fig:phiddot as a function of phidot}
\end{figure}

\begin{figure}[t]
\centering
\includegraphics[width=0.8\hsize]{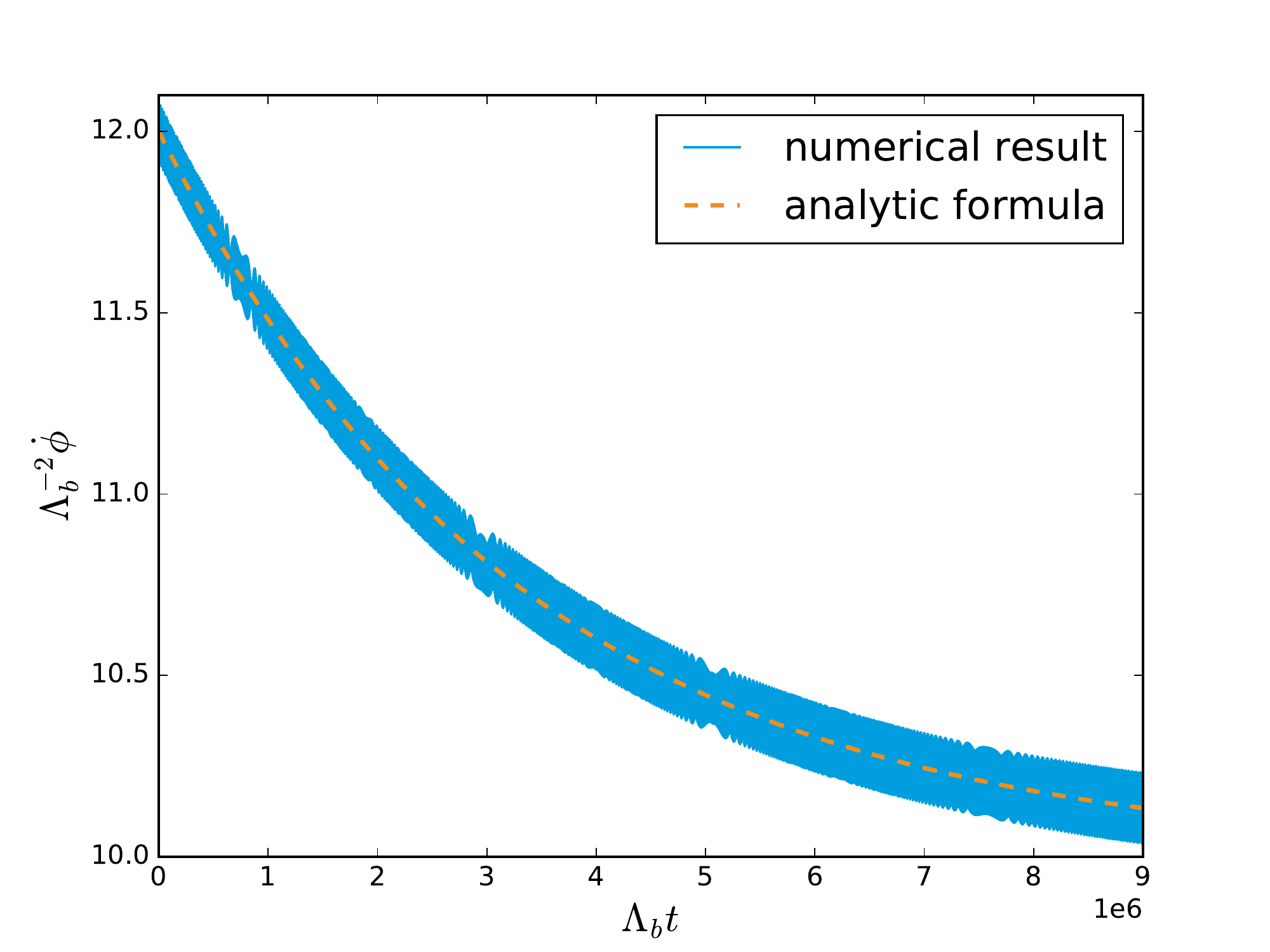}
\includegraphics[width=0.8\hsize]{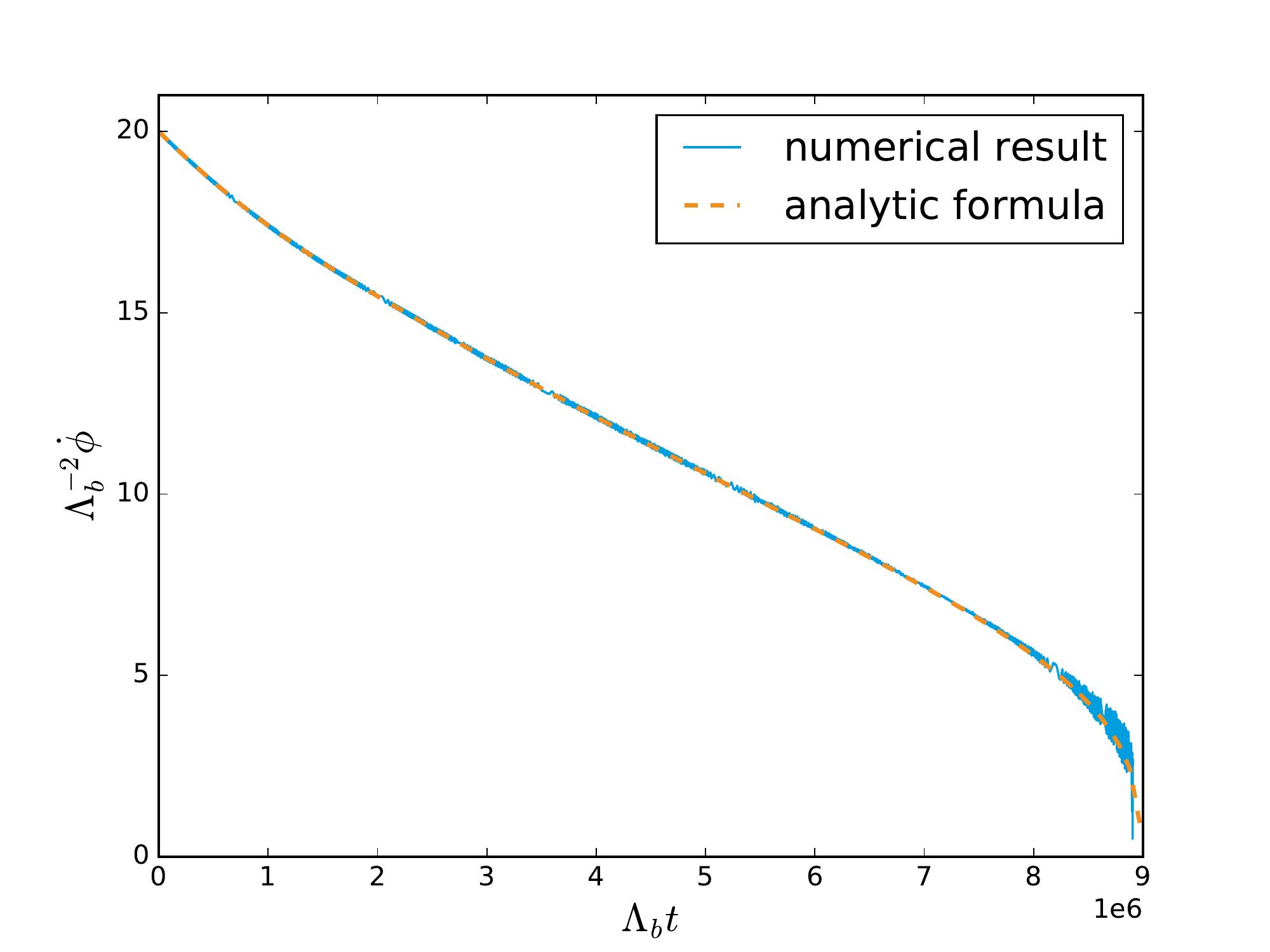}
\caption{
Time evolution of $\dot\phi$.  Here we take $f = 10^3$, $H = 10^{-7}$, and $\mu^3 = 3 \times 10^{-6}$ in the unit of $\Lambda_b = 1$.
Also, we take $\dot\phi_0 = 12 \Lambda_b^2$ in the upper panel and $20 \Lambda_b^2$ in the lower panel.
The blue lines are results of numerical calculations.
The orange lines are obtained by integrating $\ddot\phi_3$ in the upper panel and $\ddot\phi_1$ in the lower panel.
}\label{fig:time evolution of phidot (appendix)}
\end{figure}

To conclude this section, let us show a numerical example that supports the analytical discussion above.
In Fig.~\ref{fig:phiddot as a function of phidot}, we show $\ddot\phi_1$, $\ddot\phi_2$, and $\ddot\phi_3$ as a function of $\dot\phi$.
We take $f = 10^3$, $H = 10^{-7}$, and $\mu^3 = 3 \times 10^{-6}$ in units of $\Lambda_b = 1$.
In this example, $\ddot\phi$ at the beginning of fragmentation becomes $\ddot\phi_3$ if $6 \lesssim \dot\phi_0/\Lambda_b^2 \lesssim 13$,
and $\ddot\phi_1$ if $\dot\phi_0/\Lambda_b^2 \gtrsim 13$.
In Fig.~\ref{fig:time evolution of phidot (appendix)}, we show the time evolution of $\dot\phi$ for $\dot\phi_0 = 12 \Lambda_b^2$ and $20 \Lambda_b^2$.
The numerical result is consistent with $\ddot\phi = \ddot\phi_3$ for $\dot\phi_0 = 12\Lambda_b^2$,
and $\ddot\phi = \ddot\phi_1$ for $\dot\phi_0 = 20\Lambda_b^2$, confirming our understanding.

\subsection{Modified slow roll velocity} \label{sec:modified slow roll velocity}

In this section, we want to determine the existence conditions and the value of a constant-velocity solution of the equation of motion, under the effect of fragmentation. Fragmentation acts as an additional source friction, thus it is clear that this velocity, which we call $\dot\phi_{\textrm{SR(frag)}}$, will be smaller than the usual slow-roll velocity $\dot\phi_{\textrm{SR}} \equiv \mu^3 / 3H$.
As we discussed in App.~\ref{sec:stopping condition}, if such a solution exists the field will decelerate until it reaches $\dot\phi_{\textrm{SR(frag)}}$, and its evolution cannot be stopped. 

First, let us discuss the existence condition of the solution of $f(\dot\phi,0)=0$.
Let us define $\zeta \equiv \dot\phi/\dot\phi_{\textrm{SR}} > 0$.
Then, we can rewrite $f(\dot\phi,0)$ as
\begin{align}\label{eq: f ddphi=0 zeta app B}
- \frac{3H}{\mu^6}f(\dot\phi,0) =
	\zeta
	- \zeta^2
	- \frac{\dot\phi_{\textrm{SR}}^2}{96\pi^2 f^4} \zeta^4 \exp\left( \frac{\pi \Lambda_b^8}{2fH\dot\phi_{\textrm{SR}}^3} \frac{1}{\zeta^3}  \right) = 0.
\end{align}
For given $\dot\phi_{\textrm{SR}}$, $f$, and $\Lambda_b$, there exists a critical value of the Hubble expansion rate $H_{\rm cr}$, such that $f(\dot\phi,0)=0$ has two solution if $H>H_{\rm cr}$, one solution if $H=H_{\rm cr}$, and no solution if $H<H_{\rm cr}$. In order to avoid constant velocity solutions, we need to impose $H<H_{\rm cr}$.
It should be noticed that this condion is an upper bound on $H$ if $\dot\phi_\mathrm{SR}$ is fixed, but it can be rewritten as a lower bound on $H$ for fixed $\mu^3$.
If $\dot\phi_{\textrm{SR(frag)}} < \dot\phi$, $f(\dot\phi,0)=0$ should not have solution tostop the axion.
For the consistency of the EFT description we assume that $\dot\phi_\mathrm{SR} < f^2$.
Thus, we can treat $\dot\phi_{\textrm{SR}}^2/32\pi^2 f^4$ as a small parameter.
In this case, the solution of \Eq{eq: f ddphi=0 zeta app B} with $H = H_{\rm cr}$ is close to 1, since we expect the deviations from the slow roll velocity to be at least of order $\dot\phi_{\textrm{SR}}^2/32\pi^2 f^4$.
Equation \Eq{eq: f ddphi=0 zeta app B} can then be expressed as
\begin{equation}
\frac{\dot\phi_{\textrm{SR}}^2}{96\pi^2 f^4} \exp\left( \frac{\pi \Lambda_b^8}{2fH\dot\phi_{\textrm{SR}}^3} (1+3(1-\zeta) ) \right) \approx 1-\zeta \,, \label{eq:B32}
\end{equation}
where, after simplifying a factor $\zeta$, we expanded separately the right-hand side and the exponent of the left-hand side for $\zeta\simeq 1$. This procedure is justified by the observation that the exponent is very large and thus the LHS changes much more rapidly than the RHS with $\zeta$.
By solving $f(\dot\phi,0) = 0$ and $(\partial f/\partial\dot\phi)(\dot\phi,0)=0$ simultaneously, we obtain
\begin{align}
\frac{\dot\phi_{\textrm{SR(frag)}}}{\dot\phi_{\textrm{SR}}} \simeq 1 - \frac{1}{3} \left( W_0\left( \frac{32\pi^2 f^4}{e \dot\phi_{\textrm{SR}}^2} \right) \right)^{-1}, \qquad
H_{\rm cr} \simeq \frac{\pi \Lambda_b^8}{2f\dot\phi_{\textrm{SR}}^3} \left( W_0\left( \frac{32\pi^2 f^4}{e \dot\phi_{\textrm{SR}}^2} \right) \right)^{-1}. \label{eq:modifield slow roll velocity}
\end{align}
For $H>H_{\rm cr}$, Eq.~(\ref{eq: f ddphi=0 zeta app B}) admits two solutions. The smallest of the two has $d\ddot\phi/d\dot\phi > 0$ and is unstable. The largest solution is instead stable, and can be regarded the modified slow roll velocity, given by
\begin{align}
\frac{\dot\phi_{\textrm{SR(frag)}}}{\dot\phi_{\textrm{SR}}}
\simeq
1 + \frac{2fH\dot\phi_{\textrm{SR}}^3}{3\pi\Lambda_b^8} W_0\left( -\frac{\pi\Lambda_b^8}{ 64 \pi^2 f^5 H \dot\phi_{\textrm{SR}}} \exp\left( \frac{\pi\Lambda_b^8}{2fH\dot\phi_{\textrm{SR}}^3} \right)  \right).
\end{align}

\bibliography{ref}
\bibliographystyle{JHEP}

\end{document}